\documentclass{eptcs}

\usepackage[latin1]{inputenc}

\usepackage[T1]{fontenc}

\usepackage[english]{babel}

\usepackage{hyperref}

\usepackage{cite}

\usepackage[final]{microtype}

\hypersetup
	{
	pdfsubject	=	{},
	pdfkeywords	=	{},
	pdfproducer	=	{},
	pdfcreator	=	{}
	}

\usepackage{etoolbox}

\AtEndPreamble
	{

	\usepackage{changebar}

	\usepackage{fixltx2e}

	}

\usepackage{xargs}

\usepackage{xspace}

\usepackage{xstring}

\usepackage{boolexpr}

\usepackage[cmex10]{mathtools}
\usepackage{amssymb}
\usepackage{amsfonts}
\usepackage{mathrsfs}
\usepackage{latexsym}
\usepackage{textcomp}
\usepackage{pifont}
\usepackage{bbold}

\newcommand{\argemp}[2]
	{\if&#1&\else#2\fi}

\newcommand{\argdef}[2]
	{\if&#1&#2\else#1\fi}

\newcommand{\argint}[3]
	{\if&#2&\else#1#2#3\fi}

\newcommand{\argext}[3]
	{\if&#1&#3\else#1\if&#3&\else#2#3\fi\fi}

\newcommandx{\argsubsup}[3][2=, 3=]
	{\def\argsubscript{{#2}}\def\argsuperscript{{#3}}#1}

\newcommandx{\argind}[9][2=, 3=, 4=, 5=, 6=, 7=, 8=, 9=]
	{%
	\switch[#1=]%
		\case{0}#2%
		\case{1}#3%
		\case{2}#4%
		\case{3}#5%
		\case{4}#6%
		\case{5}#7%
		\case{6}#8%
		\case{7}#9%
		\otherwise\ensuremath{\clubsuit}%
	\endswitch%
	}

\newcommand{\arga}[1]
	{#1}
\newcommand{\argb}[2]
	{\argext{\arga{#1}}{, \allowbreak}{#2}}
\newcommand{\argc}[3]
	{\argext{\argb{#1}{#2}}{, \allowbreak}{#3}}
\newcommand{\argd}[4]
	{\argext{\argc{#1}{#2}{#3}}{, \allowbreak}{#4}}
\newcommand{\arge}[5]
	{\argext{\argd{#1}{#2}{#3}{#4}}{, \allowbreak}{#5}}
\newcommand{\argf}[6]
	{\argext{\arge{#1}{#2}{#3}{#4}{#5}}{, \allowbreak}{#6}}
\newcommand{\argg}[7]
	{\argext{\argf{#1}{#2}{#3}{#4}{#5}{#6}}{, \allowbreak}{#7}}
\newcommand{\argh}[8]
	{\argext{\argg{#1}{#2}{#3}{#4}{#5}{#6}{#7}}{, \allowbreak}{#8}}
\newcommand{\argi}[9]
	{\argext{\argh{#1}{#2}{#3}{#4}{#5}{#6}{#7}{#8}}{, \allowbreak}{#9}}

\newcommand{\argj}[9]
	{%
	\def\valarga{#1}%
	\def\valargb{#2}%
	\def\valargc{#3}%
	\def\valargd{#4}%
	\def\valarge{#5}%
	\def\valargf{#6}%
	\def\valargg{#7}%
	\def\valargh{#8}%
	\def\valargi{#9}%
	\argauxj%
	}
\newcommand{\argk}[9]
	{%
	\def\valarga{#1}%
	\def\valargb{#2}%
	\def\valargc{#3}%
	\def\valargd{#4}%
	\def\valarge{#5}%
	\def\valargf{#6}%
	\def\valargg{#7}%
	\def\valargh{#8}%
	\def\valargi{#9}%
	\argauxk%
	}
\newcommand{\argl}[9]
	{%
	\def\valarga{#1}%
	\def\valargb{#2}%
	\def\valargc{#3}%
	\def\valargd{#4}%
	\def\valarge{#5}%
	\def\valargf{#6}%
	\def\valargg{#7}%
	\def\valargh{#8}%
	\def\valargi{#9}%
	\argauxl%
	}
\newcommand{\argm}[9]
	{%
	\def\valarga{#1}%
	\def\valargb{#2}%
	\def\valargc{#3}%
	\def\valargd{#4}%
	\def\valarge{#5}%
	\def\valargf{#6}%
	\def\valargg{#7}%
	\def\valargh{#8}%
	\def\valargi{#9}%
	\argauxm%
	}
\newcommand{\argn}[9]
	{%
	\def\valarga{#1}%
	\def\valargb{#2}%
	\def\valargc{#3}%
	\def\valargd{#4}%
	\def\valarge{#5}%
	\def\valargf{#6}%
	\def\valargg{#7}%
	\def\valargh{#8}%
	\def\valargi{#9}%
	\argauxn%
	}
\newcommand{\argo}[9]
	{%
	\def\valarga{#1}%
	\def\valargb{#2}%
	\def\valargc{#3}%
	\def\valargd{#4}%
	\def\valarge{#5}%
	\def\valargf{#6}%
	\def\valargg{#7}%
	\def\valargh{#8}%
	\def\valargi{#9}%
	\argauxo%
	}
\newcommand{\argp}[9]
	{%
	\def\valarga{#1}%
	\def\valargb{#2}%
	\def\valargc{#3}%
	\def\valargd{#4}%
	\def\valarge{#5}%
	\def\valargf{#6}%
	\def\valargg{#7}%
	\def\valargh{#8}%
	\def\valargi{#9}%
	\argauxp%
	}
\newcommand{\argq}[9]
	{%
	\def\valarga{#1}%
	\def\valargb{#2}%
	\def\valargc{#3}%
	\def\valargd{#4}%
	\def\valarge{#5}%
	\def\valargf{#6}%
	\def\valargg{#7}%
	\def\valargh{#8}%
	\def\valargi{#9}%
	\argauxq%
	}
\newcommand{\argr}[9]
	{%
	\def\valarga{#1}%
	\def\valargb{#2}%
	\def\valargc{#3}%
	\def\valargd{#4}%
	\def\valarge{#5}%
	\def\valargf{#6}%
	\def\valargg{#7}%
	\def\valargh{#8}%
	\def\valargi{#9}%
	\argauxr%
	}

\newcommand{\argauxj}[1]
	{%
	\argext%
		{\argi{\valarga}{\valargb}{\valargc}{\valargd}{\valarge}{\valargf}{\valargg}
			{\valargh}{\valargi}}
		{, \allowbreak}{#1}%
	}
\newcommand{\argauxk}[2]
	{\argext{\argauxj{#1}}{, \allowbreak}{#2}}
\newcommand{\argauxl}[3]
	{\argext{\argauxk{#1}{#2}}{, \allowbreak}{#3}}
\newcommand{\argauxm}[4]
	{\argext{\argauxl{#1}{#2}{#3}}{, \allowbreak}{#4}}
\newcommand{\argauxn}[5]
	{\argext{\argauxm{#1}{#2}{#3}{#4}}{, \allowbreak}{#5}}
\newcommand{\argauxo}[6]
	{\argext{\argauxn{#1}{#2}{#3}{#4}{#5}}{, \allowbreak}{#6}}
\newcommand{\argauxp}[7]
	{\argext{\argauxo{#1}{#2}{#3}{#4}{#5}{#6}}{, \allowbreak}{#7}}
\newcommand{\argauxq}[8]
	{\argext{\argauxp{#1}{#2}{#3}{#4}{#5}{#6}{#7}}{, \allowbreak}{#8}}
\newcommand{\argauxr}[9]
	{\argext{\argauxq{#1}{#2}{#3}{#4}{#5}{#6}{#7}{#8}}{, \allowbreak}{#9}}

\newcommand{\txtfnt}[2][]
	{{%
	\IfStrEq{#1}{}
		{#2}
		{%
		\StrLeft{#1}{2}[\optbgn]%
		\StrGobbleLeft{#1}{2}[\optend]%
		\IfStrEqCase{\optbgn}
			{%
			{Rm}{\rmfamily\txtfnt[\optend]{#2}}%
			{Sf}{\sffamily\txtfnt[\optend]{#2}}%
			{Tt}{\ttfamily\txtfnt[\optend]{#2}}%
			{Up}{\upshape\txtfnt[\optend]{#2}}%
			{It}{\itshape\txtfnt[\optend]{#2}}%
			{Sl}{\slshape\txtfnt[\optend]{#2}}%
			{Sc}{\scshape\txtfnt[\optend]{#2}}%
			{Md}{\mdseries\txtfnt[\optend]{#2}}%
			{Bf}{\bfseries\txtfnt[\optend]{#2}}%
			{Em}{\emph{\txtfnt[\optend]{#2}}}%
			}
			[\ensuremath{\clubsuit}]%
		}%
	}}

\newcommand{\txtsub}[2][]
	{\argemp{#2}{\ensuremath{_{\text{\txtfnt[#1]{#2}}}}}}

\newcommand{\txtsup}[2][]
	{\argemp{#2}{\ensuremath{^{\text{\txtfnt[#1]{#2}}}}}}

\newcommandx{\txt}[4][1=, 3=, 4=]
	{\text{\txtfnt[#1]{#2}\ensuremath{\txtsub[#1]{#3}\txtsup[#1]{#4}}}}

\newcommandx{\txtarg}[5][1=, 3=, 4=]
	{{\txt[#1]{#2}[#3][#4]\argint{(}{#5}{)}}}

\newcommand{\txtstyname}{RmScMd}
\newcommand{\txtname}[1][]
	{\txt[\argdef{#1}{\txtstyname}]}
\newcommand{\txtargname}[1][]
	{\txtarg[\argdef{#1}{\txtstyname}]}

\newcommand{\txtstyabr}{Em}
\newcommand{\txtabr}[1][]
	{\txt[\argdef{#1}{\txtstyabr}]}

\newcommandx{\mthfnt}[3][1=, 2=0]
	{{%
	\IfStrEqCase{#1}
		{%
		{}%
			{#3}%
		{Name}%
			{%
			\IfStrEqCase{#2}
				{%
				{0}{\mathcal{#3}}%
				{1}{\mathscr{#3}}%
				{2}{\mathfrak{#3}}%
				{3}{\mathbf{#3}}%
				}
				[\ensuremath{\clubsuit}]%
			}%
		{Set}%
			{%
			\IfStrEqCase{#2}
				{%
				{0}{\mathrm{#3}}%
				{1}{\mathsf{#3}}%
				{2}{\mathbb{#3}}%
				{3}{\mathtt{#3}}%
				}
				[\ensuremath{\clubsuit}]%
			}%
		{Fun}%
			{%
			\IfStrEqCase{#2}
				{%
				{0}{\mathsf{#3}}%
				{1}{\mathrm{#3}}%
				}
				[\ensuremath{\clubsuit}]%
			}%
		{Rel}%
			{%
			\IfStrEqCase{#2}
				{%
				{0}{\mathit{#3}}%
				{1}{\mathtt{#3}}%
				}
				[\ensuremath{\clubsuit}]%
			}%
		{Sym}%
			{%
			\IfStrEqCase{#2}
				{%
				{0}{\mathtt{#3}}%
				{1}{\mathbf{#3}}%
				}
				[\ensuremath{\clubsuit}]%
			}%
		{Elm}%
			{\mathnormal{#3}}
		}
		[\ensuremath{\clubsuit}]%
	}}

\newcommand{\mthsub}[1]
	{\argemp{#1}{\ensuremath{_{\mathnormal{#1}}}}}

\newcommand{\mthsup}[1]
	{\argemp{#1}{\ensuremath{^{\mathnormal{#1}}}}}

\newcommandx{\mth}[5][1=, 2=0, 4=, 5=]
	{{\ensuremath{\mthfnt[#1][#2]{#3}\mthsub{#4}\mthsup{#5}}}}

\newcommandx{\mtharg}[6][1=, 2=0, 4=, 5=]
	{{\mth[#1][#2]{#3}[#4][#5]\ensuremath{\argint{(}{#6}{)}}}}

\newcommand{\mthempty}
	{\mth[][]}

\newcommand{\mthstyname}{0}
\newcommand{\mthname}[1][]
	{\mth[Name][\argdef{#1}{\mthstyname}]}

\newcommand{\mthstyset}{0}
\newcommand{\mthset}[1][]
	{\mth[Set][\argdef{#1}{\mthstyset}]}
\newcommand{\mthargset}[1][]
	{\mtharg[Set][\argdef{#1}{\mthstyset}]}

\newcommand{\mthstyfun}{0}
\newcommand{\mthfun}[1][]
	{\mth[Fun][\argdef{#1}{\mthstyfun}]}
\newcommand{\mthargfun}[1][]
	{\mtharg[Fun][\argdef{#1}{\mthstyfun}]}

\newcommand{\mthstyrel}{0}
\newcommand{\mthrel}[1][]
	{\mth[Rel][\argdef{#1}{\mthstyrel}]}

\newcommand{\mthstysym}{0}
\newcommand{\mthsym}[1][]
	{\mth[Sym][\argdef{#1}{\mthstysym}]}

\newcommand{\mthstyelm}{0}
\newcommand{\mthelm}[1][]
	{\mth[Elm][\argdef{#1}{\mthstyelm}]}

\newcommandx{\AName}[4][1=, 2=, 3=, 4=]{\mthname[#4]{A#3}[#1][#2]}
\newcommandx{\BName}[4][1=, 2=, 3=, 4=]{\mthname[#4]{B#3}[#1][#2]}
\newcommandx{\CName}[4][1=, 2=, 3=, 4=]{\mthname[#4]{C#3}[#1][#2]}
\newcommandx{\DName}[4][1=, 2=, 3=, 4=]{\mthname[#4]{D#3}[#1][#2]}
\newcommandx{\EName}[4][1=, 2=, 3=, 4=]{\mthname[#4]{E#3}[#1][#2]}
\newcommandx{\FName}[4][1=, 2=, 3=, 4=]{\mthname[#4]{F#3}[#1][#2]}
\newcommandx{\GName}[4][1=, 2=, 3=, 4=]{\mthname[#4]{G#3}[#1][#2]}
\newcommandx{\HName}[4][1=, 2=, 3=, 4=]{\mthname[#4]{H#3}[#1][#2]}
\newcommandx{\IName}[4][1=, 2=, 3=, 4=]{\mthname[#4]{I#3}[#1][#2]}
\newcommandx{\JName}[4][1=, 2=, 3=, 4=]{\mthname[#4]{J#3}[#1][#2]}
\newcommandx{\KName}[4][1=, 2=, 3=, 4=]{\mthname[#4]{K#3}[#1][#2]}
\newcommandx{\LName}[4][1=, 2=, 3=, 4=]{\mthname[#4]{L#3}[#1][#2]}
\newcommandx{\MName}[4][1=, 2=, 3=, 4=]{\mthname[#4]{M#3}[#1][#2]}
\newcommandx{\NName}[4][1=, 2=, 3=, 4=]{\mthname[#4]{N#3}[#1][#2]}
\newcommandx{\OName}[4][1=, 2=, 3=, 4=]{\mthname[#4]{O#3}[#1][#2]}
\newcommandx{\PName}[4][1=, 2=, 3=, 4=]{\mthname[#4]{P#3}[#1][#2]}
\newcommandx{\QName}[4][1=, 2=, 3=, 4=]{\mthname[#4]{Q#3}[#1][#2]}
\newcommandx{\RName}[4][1=, 2=, 3=, 4=]{\mthname[#4]{R#3}[#1][#2]}
\newcommandx{\SName}[4][1=, 2=, 3=, 4=]{\mthname[#4]{S#3}[#1][#2]}
\newcommandx{\TName}[4][1=, 2=, 3=, 4=]{\mthname[#4]{T#3}[#1][#2]}
\newcommandx{\UName}[4][1=, 2=, 3=, 4=]{\mthname[#4]{U#3}[#1][#2]}
\newcommandx{\VName}[4][1=, 2=, 3=, 4=]{\mthname[#4]{V#3}[#1][#2]}
\newcommandx{\WName}[4][1=, 2=, 3=, 4=]{\mthname[#4]{W#3}[#1][#2]}
\newcommandx{\XName}[4][1=, 2=, 3=, 4=]{\mthname[#4]{X#3}[#1][#2]}
\newcommandx{\YName}[4][1=, 2=, 3=, 4=]{\mthname[#4]{Y#3}[#1][#2]}
\newcommandx{\ZName}[4][1=, 2=, 3=, 4=]{\mthname[#4]{Z#3}[#1][#2]}

\newcommandx{\ASet}[4][1=, 2=, 3=, 4=]{\mthset[#4]{A#3}[#1][#2]}
\newcommandx{\BSet}[4][1=, 2=, 3=, 4=]{\mthset[#4]{B#3}[#1][#2]}
\newcommandx{\CSet}[4][1=, 2=, 3=, 4=]{\mthset[#4]{C#3}[#1][#2]}
\newcommandx{\DSet}[4][1=, 2=, 3=, 4=]{\mthset[#4]{D#3}[#1][#2]}
\newcommandx{\ESet}[4][1=, 2=, 3=, 4=]{\mthset[#4]{E#3}[#1][#2]}
\newcommandx{\FSet}[4][1=, 2=, 3=, 4=]{\mthset[#4]{F#3}[#1][#2]}
\newcommandx{\GSet}[4][1=, 2=, 3=, 4=]{\mthset[#4]{G#3}[#1][#2]}
\newcommandx{\HSet}[4][1=, 2=, 3=, 4=]{\mthset[#4]{H#3}[#1][#2]}
\newcommandx{\ISet}[4][1=, 2=, 3=, 4=]{\mthset[#4]{I#3}[#1][#2]}
\newcommandx{\JSet}[4][1=, 2=, 3=, 4=]{\mthset[#4]{J#3}[#1][#2]}
\newcommandx{\KSet}[4][1=, 2=, 3=, 4=]{\mthset[#4]{K#3}[#1][#2]}
\newcommandx{\LSet}[4][1=, 2=, 3=, 4=]{\mthset[#4]{L#3}[#1][#2]}
\newcommandx{\MSet}[4][1=, 2=, 3=, 4=]{\mthset[#4]{M#3}[#1][#2]}
\newcommandx{\NSet}[4][1=, 2=, 3=, 4=]{\mthset[#4]{N#3}[#1][#2]}
\newcommandx{\OSet}[4][1=, 2=, 3=, 4=]{\mthset[#4]{O#3}[#1][#2]}
\newcommandx{\PSet}[4][1=, 2=, 3=, 4=]{\mthset[#4]{P#3}[#1][#2]}
\newcommandx{\QSet}[4][1=, 2=, 3=, 4=]{\mthset[#4]{Q#3}[#1][#2]}
\newcommandx{\RSet}[4][1=, 2=, 3=, 4=]{\mthset[#4]{R#3}[#1][#2]}
\newcommandx{\SSet}[4][1=, 2=, 3=, 4=]{\mthset[#4]{S#3}[#1][#2]}
\newcommandx{\TSet}[4][1=, 2=, 3=, 4=]{\mthset[#4]{T#3}[#1][#2]}
\newcommandx{\USet}[4][1=, 2=, 3=, 4=]{\mthset[#4]{U#3}[#1][#2]}
\newcommandx{\VSet}[4][1=, 2=, 3=, 4=]{\mthset[#4]{V#3}[#1][#2]}
\newcommandx{\WSet}[4][1=, 2=, 3=, 4=]{\mthset[#4]{W#3}[#1][#2]}
\newcommandx{\XSet}[4][1=, 2=, 3=, 4=]{\mthset[#4]{X#3}[#1][#2]}
\newcommandx{\YSet}[4][1=, 2=, 3=, 4=]{\mthset[#4]{Y#3}[#1][#2]}
\newcommandx{\ZSet}[4][1=, 2=, 3=, 4=]{\mthset[#4]{Z#3}[#1][#2]}

\newcommandx{\aSet}[4][1=, 2=, 3=, 4=]{\mthset[#4]{a#3}[#1][#2]}
\newcommandx{\bSet}[4][1=, 2=, 3=, 4=]{\mthset[#4]{b#3}[#1][#2]}
\newcommandx{\cSet}[4][1=, 2=, 3=, 4=]{\mthset[#4]{c#3}[#1][#2]}
\newcommandx{\dSet}[4][1=, 2=, 3=, 4=]{\mthset[#4]{d#3}[#1][#2]}
\newcommandx{\eSet}[4][1=, 2=, 3=, 4=]{\mthset[#4]{e#3}[#1][#2]}
\newcommandx{\fSet}[4][1=, 2=, 3=, 4=]{\mthset[#4]{f#3}[#1][#2]}
\newcommandx{\gSet}[4][1=, 2=, 3=, 4=]{\mthset[#4]{g#3}[#1][#2]}
\newcommandx{\hSet}[4][1=, 2=, 3=, 4=]{\mthset[#4]{h#3}[#1][#2]}
\newcommandx{\iSet}[4][1=, 2=, 3=, 4=]{\mthset[#4]{i#3}[#1][#2]}
\newcommandx{\jSet}[4][1=, 2=, 3=, 4=]{\mthset[#4]{j#3}[#1][#2]}
\newcommandx{\kSet}[4][1=, 2=, 3=, 4=]{\mthset[#4]{k#3}[#1][#2]}
\newcommandx{\lSet}[4][1=, 2=, 3=, 4=]{\mthset[#4]{l#3}[#1][#2]}
\newcommandx{\mSet}[4][1=, 2=, 3=, 4=]{\mthset[#4]{m#3}[#1][#2]}
\newcommandx{\nSet}[4][1=, 2=, 3=, 4=]{\mthset[#4]{n#3}[#1][#2]}
\newcommandx{\oSet}[4][1=, 2=, 3=, 4=]{\mthset[#4]{o#3}[#1][#2]}
\newcommandx{\pSet}[4][1=, 2=, 3=, 4=]{\mthset[#4]{p#3}[#1][#2]}
\newcommandx{\qSet}[4][1=, 2=, 3=, 4=]{\mthset[#4]{q#3}[#1][#2]}
\newcommandx{\rSet}[4][1=, 2=, 3=, 4=]{\mthset[#4]{r#3}[#1][#2]}
\newcommandx{\sSet}[4][1=, 2=, 3=, 4=]{\mthset[#4]{s#3}[#1][#2]}
\newcommandx{\tSet}[4][1=, 2=, 3=, 4=]{\mthset[#4]{t#3}[#1][#2]}
\newcommandx{\uSet}[4][1=, 2=, 3=, 4=]{\mthset[#4]{u#3}[#1][#2]}
\newcommandx{\vSet}[4][1=, 2=, 3=, 4=]{\mthset[#4]{v#3}[#1][#2]}
\newcommandx{\wSet}[4][1=, 2=, 3=, 4=]{\mthset[#4]{w#3}[#1][#2]}
\newcommandx{\xSet}[4][1=, 2=, 3=, 4=]{\mthset[#4]{x#3}[#1][#2]}
\newcommandx{\ySet}[4][1=, 2=, 3=, 4=]{\mthset[#4]{y#3}[#1][#2]}
\newcommandx{\zSet}[4][1=, 2=, 3=, 4=]{\mthset[#4]{z#3}[#1][#2]}

\newcommandx{\AFun}[4][1=, 2=, 3=, 4=]{\mthfun[#4]{A#3}[#1][#2]}
\newcommandx{\BFun}[4][1=, 2=, 3=, 4=]{\mthfun[#4]{B#3}[#1][#2]}
\newcommandx{\CFun}[4][1=, 2=, 3=, 4=]{\mthfun[#4]{C#3}[#1][#2]}
\newcommandx{\DFun}[4][1=, 2=, 3=, 4=]{\mthfun[#4]{D#3}[#1][#2]}
\newcommandx{\EFun}[4][1=, 2=, 3=, 4=]{\mthfun[#4]{E#3}[#1][#2]}
\newcommandx{\FFun}[4][1=, 2=, 3=, 4=]{\mthfun[#4]{F#3}[#1][#2]}
\newcommandx{\GFun}[4][1=, 2=, 3=, 4=]{\mthfun[#4]{G#3}[#1][#2]}
\newcommandx{\HFun}[4][1=, 2=, 3=, 4=]{\mthfun[#4]{H#3}[#1][#2]}
\newcommandx{\IFun}[4][1=, 2=, 3=, 4=]{\mthfun[#4]{I#3}[#1][#2]}
\newcommandx{\JFun}[4][1=, 2=, 3=, 4=]{\mthfun[#4]{J#3}[#1][#2]}
\newcommandx{\KFun}[4][1=, 2=, 3=, 4=]{\mthfun[#4]{K#3}[#1][#2]}
\newcommandx{\LFun}[4][1=, 2=, 3=, 4=]{\mthfun[#4]{L#3}[#1][#2]}
\newcommandx{\MFun}[4][1=, 2=, 3=, 4=]{\mthfun[#4]{M#3}[#1][#2]}
\newcommandx{\NFun}[4][1=, 2=, 3=, 4=]{\mthfun[#4]{N#3}[#1][#2]}
\newcommandx{\OFun}[4][1=, 2=, 3=, 4=]{\mthfun[#4]{O#3}[#1][#2]}
\newcommandx{\PFun}[4][1=, 2=, 3=, 4=]{\mthfun[#4]{P#3}[#1][#2]}
\newcommandx{\QFun}[4][1=, 2=, 3=, 4=]{\mthfun[#4]{Q#3}[#1][#2]}
\newcommandx{\RFun}[4][1=, 2=, 3=, 4=]{\mthfun[#4]{R#3}[#1][#2]}
\newcommandx{\SFun}[4][1=, 2=, 3=, 4=]{\mthfun[#4]{S#3}[#1][#2]}
\newcommandx{\TFun}[4][1=, 2=, 3=, 4=]{\mthfun[#4]{T#3}[#1][#2]}
\newcommandx{\UFun}[4][1=, 2=, 3=, 4=]{\mthfun[#4]{U#3}[#1][#2]}
\newcommandx{\VFun}[4][1=, 2=, 3=, 4=]{\mthfun[#4]{V#3}[#1][#2]}
\newcommandx{\WFun}[4][1=, 2=, 3=, 4=]{\mthfun[#4]{W#3}[#1][#2]}
\newcommandx{\XFun}[4][1=, 2=, 3=, 4=]{\mthfun[#4]{X#3}[#1][#2]}
\newcommandx{\YFun}[4][1=, 2=, 3=, 4=]{\mthfun[#4]{Y#3}[#1][#2]}
\newcommandx{\ZFun}[4][1=, 2=, 3=, 4=]{\mthfun[#4]{Z#3}[#1][#2]}

\newcommandx{\aFun}[4][1=, 2=, 3=, 4=]{\mthfun[#4]{a#3}[#1][#2]}
\newcommandx{\bFun}[4][1=, 2=, 3=, 4=]{\mthfun[#4]{b#3}[#1][#2]}
\newcommandx{\cFun}[4][1=, 2=, 3=, 4=]{\mthfun[#4]{c#3}[#1][#2]}
\newcommandx{\dFun}[4][1=, 2=, 3=, 4=]{\mthfun[#4]{d#3}[#1][#2]}
\newcommandx{\eFun}[4][1=, 2=, 3=, 4=]{\mthfun[#4]{e#3}[#1][#2]}
\newcommandx{\fFun}[4][1=, 2=, 3=, 4=]{\mthfun[#4]{f#3}[#1][#2]}
\newcommandx{\gFun}[4][1=, 2=, 3=, 4=]{\mthfun[#4]{g#3}[#1][#2]}
\newcommandx{\hFun}[4][1=, 2=, 3=, 4=]{\mthfun[#4]{h#3}[#1][#2]}
\newcommandx{\iFun}[4][1=, 2=, 3=, 4=]{\mthfun[#4]{i#3}[#1][#2]}
\newcommandx{\jFun}[4][1=, 2=, 3=, 4=]{\mthfun[#4]{j#3}[#1][#2]}
\newcommandx{\kFun}[4][1=, 2=, 3=, 4=]{\mthfun[#4]{k#3}[#1][#2]}
\newcommandx{\lFun}[4][1=, 2=, 3=, 4=]{\mthfun[#4]{l#3}[#1][#2]}
\newcommandx{\mFun}[4][1=, 2=, 3=, 4=]{\mthfun[#4]{m#3}[#1][#2]}
\newcommandx{\nFun}[4][1=, 2=, 3=, 4=]{\mthfun[#4]{n#3}[#1][#2]}
\newcommandx{\oFun}[4][1=, 2=, 3=, 4=]{\mthfun[#4]{o#3}[#1][#2]}
\newcommandx{\pFun}[4][1=, 2=, 3=, 4=]{\mthfun[#4]{p#3}[#1][#2]}
\newcommandx{\qFun}[4][1=, 2=, 3=, 4=]{\mthfun[#4]{q#3}[#1][#2]}
\newcommandx{\rFun}[4][1=, 2=, 3=, 4=]{\mthfun[#4]{r#3}[#1][#2]}
\newcommandx{\sFun}[4][1=, 2=, 3=, 4=]{\mthfun[#4]{s#3}[#1][#2]}
\newcommandx{\tFun}[4][1=, 2=, 3=, 4=]{\mthfun[#4]{t#3}[#1][#2]}
\newcommandx{\uFun}[4][1=, 2=, 3=, 4=]{\mthfun[#4]{u#3}[#1][#2]}
\newcommandx{\vFun}[4][1=, 2=, 3=, 4=]{\mthfun[#4]{v#3}[#1][#2]}
\newcommandx{\wFun}[4][1=, 2=, 3=, 4=]{\mthfun[#4]{w#3}[#1][#2]}
\newcommandx{\xFun}[4][1=, 2=, 3=, 4=]{\mthfun[#4]{x#3}[#1][#2]}
\newcommandx{\yFun}[4][1=, 2=, 3=, 4=]{\mthfun[#4]{y#3}[#1][#2]}
\newcommandx{\zFun}[4][1=, 2=, 3=, 4=]{\mthfun[#4]{z#3}[#1][#2]}

\newcommandx{\ARel}[4][1=, 2=, 3=, 4=]{\mthrel[#4]{A#3}[#1][#2]}
\newcommandx{\BRel}[4][1=, 2=, 3=, 4=]{\mthrel[#4]{B#3}[#1][#2]}
\newcommandx{\CRel}[4][1=, 2=, 3=, 4=]{\mthrel[#4]{C#3}[#1][#2]}
\newcommandx{\DRel}[4][1=, 2=, 3=, 4=]{\mthrel[#4]{D#3}[#1][#2]}
\newcommandx{\ERel}[4][1=, 2=, 3=, 4=]{\mthrel[#4]{E#3}[#1][#2]}
\newcommandx{\FRel}[4][1=, 2=, 3=, 4=]{\mthrel[#4]{F#3}[#1][#2]}
\newcommandx{\GRel}[4][1=, 2=, 3=, 4=]{\mthrel[#4]{G#3}[#1][#2]}
\newcommandx{\HRel}[4][1=, 2=, 3=, 4=]{\mthrel[#4]{H#3}[#1][#2]}
\newcommandx{\IRel}[4][1=, 2=, 3=, 4=]{\mthrel[#4]{I#3}[#1][#2]}
\newcommandx{\JRel}[4][1=, 2=, 3=, 4=]{\mthrel[#4]{J#3}[#1][#2]}
\newcommandx{\KRel}[4][1=, 2=, 3=, 4=]{\mthrel[#4]{K#3}[#1][#2]}
\newcommandx{\LRel}[4][1=, 2=, 3=, 4=]{\mthrel[#4]{L#3}[#1][#2]}
\newcommandx{\MRel}[4][1=, 2=, 3=, 4=]{\mthrel[#4]{M#3}[#1][#2]}
\newcommandx{\NRel}[4][1=, 2=, 3=, 4=]{\mthrel[#4]{N#3}[#1][#2]}
\newcommandx{\ORel}[4][1=, 2=, 3=, 4=]{\mthrel[#4]{O#3}[#1][#2]}
\newcommandx{\PRel}[4][1=, 2=, 3=, 4=]{\mthrel[#4]{P#3}[#1][#2]}
\newcommandx{\QRel}[4][1=, 2=, 3=, 4=]{\mthrel[#4]{Q#3}[#1][#2]}
\newcommandx{\RRel}[4][1=, 2=, 3=, 4=]{\mthrel[#4]{R#3}[#1][#2]}
\newcommandx{\SRel}[4][1=, 2=, 3=, 4=]{\mthrel[#4]{S#3}[#1][#2]}
\newcommandx{\TRel}[4][1=, 2=, 3=, 4=]{\mthrel[#4]{T#3}[#1][#2]}
\newcommandx{\URel}[4][1=, 2=, 3=, 4=]{\mthrel[#4]{U#3}[#1][#2]}
\newcommandx{\VRel}[4][1=, 2=, 3=, 4=]{\mthrel[#4]{V#3}[#1][#2]}
\newcommandx{\WRel}[4][1=, 2=, 3=, 4=]{\mthrel[#4]{W#3}[#1][#2]}
\newcommandx{\XRel}[4][1=, 2=, 3=, 4=]{\mthrel[#4]{X#3}[#1][#2]}
\newcommandx{\YRel}[4][1=, 2=, 3=, 4=]{\mthrel[#4]{Y#3}[#1][#2]}
\newcommandx{\ZRel}[4][1=, 2=, 3=, 4=]{\mthrel[#4]{Z#3}[#1][#2]}

\newcommandx{\aRel}[4][1=, 2=, 3=, 4=]{\mthrel[#4]{a#3}[#1][#2]}
\newcommandx{\bRel}[4][1=, 2=, 3=, 4=]{\mthrel[#4]{b#3}[#1][#2]}
\newcommandx{\cRel}[4][1=, 2=, 3=, 4=]{\mthrel[#4]{c#3}[#1][#2]}
\newcommandx{\dRel}[4][1=, 2=, 3=, 4=]{\mthrel[#4]{d#3}[#1][#2]}
\newcommandx{\eRel}[4][1=, 2=, 3=, 4=]{\mthrel[#4]{e#3}[#1][#2]}
\newcommandx{\fRel}[4][1=, 2=, 3=, 4=]{\mthrel[#4]{f#3}[#1][#2]}
\newcommandx{\gRel}[4][1=, 2=, 3=, 4=]{\mthrel[#4]{g#3}[#1][#2]}
\newcommandx{\hRel}[4][1=, 2=, 3=, 4=]{\mthrel[#4]{h#3}[#1][#2]}
\newcommandx{\iRel}[4][1=, 2=, 3=, 4=]{\mthrel[#4]{i#3}[#1][#2]}
\newcommandx{\jRel}[4][1=, 2=, 3=, 4=]{\mthrel[#4]{j#3}[#1][#2]}
\newcommandx{\kRel}[4][1=, 2=, 3=, 4=]{\mthrel[#4]{k#3}[#1][#2]}
\newcommandx{\lRel}[4][1=, 2=, 3=, 4=]{\mthrel[#4]{l#3}[#1][#2]}
\newcommandx{\mRel}[4][1=, 2=, 3=, 4=]{\mthrel[#4]{m#3}[#1][#2]}
\newcommandx{\nRel}[4][1=, 2=, 3=, 4=]{\mthrel[#4]{n#3}[#1][#2]}
\newcommandx{\oRel}[4][1=, 2=, 3=, 4=]{\mthrel[#4]{o#3}[#1][#2]}
\newcommandx{\pRel}[4][1=, 2=, 3=, 4=]{\mthrel[#4]{p#3}[#1][#2]}
\newcommandx{\qRel}[4][1=, 2=, 3=, 4=]{\mthrel[#4]{q#3}[#1][#2]}
\newcommandx{\rRel}[4][1=, 2=, 3=, 4=]{\mthrel[#4]{r#3}[#1][#2]}
\newcommandx{\sRel}[4][1=, 2=, 3=, 4=]{\mthrel[#4]{s#3}[#1][#2]}
\newcommandx{\tRel}[4][1=, 2=, 3=, 4=]{\mthrel[#4]{t#3}[#1][#2]}
\newcommandx{\uRel}[4][1=, 2=, 3=, 4=]{\mthrel[#4]{u#3}[#1][#2]}
\newcommandx{\vRel}[4][1=, 2=, 3=, 4=]{\mthrel[#4]{v#3}[#1][#2]}
\newcommandx{\wRel}[4][1=, 2=, 3=, 4=]{\mthrel[#4]{w#3}[#1][#2]}
\newcommandx{\xRel}[4][1=, 2=, 3=, 4=]{\mthrel[#4]{x#3}[#1][#2]}
\newcommandx{\yRel}[4][1=, 2=, 3=, 4=]{\mthrel[#4]{y#3}[#1][#2]}
\newcommandx{\zRel}[4][1=, 2=, 3=, 4=]{\mthrel[#4]{z#3}[#1][#2]}

\newcommandx{\ASym}[4][1=, 2=, 3=, 4=]{\mthsym[#4]{A#3}[#1][#2]}
\newcommandx{\BSym}[4][1=, 2=, 3=, 4=]{\mthsym[#4]{B#3}[#1][#2]}
\newcommandx{\CSym}[4][1=, 2=, 3=, 4=]{\mthsym[#4]{C#3}[#1][#2]}
\newcommandx{\DSym}[4][1=, 2=, 3=, 4=]{\mthsym[#4]{D#3}[#1][#2]}
\newcommandx{\ESym}[4][1=, 2=, 3=, 4=]{\mthsym[#4]{E#3}[#1][#2]}
\newcommandx{\FSym}[4][1=, 2=, 3=, 4=]{\mthsym[#4]{F#3}[#1][#2]}
\newcommandx{\GSym}[4][1=, 2=, 3=, 4=]{\mthsym[#4]{G#3}[#1][#2]}
\newcommandx{\HSym}[4][1=, 2=, 3=, 4=]{\mthsym[#4]{H#3}[#1][#2]}
\newcommandx{\ISym}[4][1=, 2=, 3=, 4=]{\mthsym[#4]{I#3}[#1][#2]}
\newcommandx{\JSym}[4][1=, 2=, 3=, 4=]{\mthsym[#4]{J#3}[#1][#2]}
\newcommandx{\KSym}[4][1=, 2=, 3=, 4=]{\mthsym[#4]{K#3}[#1][#2]}
\newcommandx{\LSym}[4][1=, 2=, 3=, 4=]{\mthsym[#4]{L#3}[#1][#2]}
\newcommandx{\MSym}[4][1=, 2=, 3=, 4=]{\mthsym[#4]{M#3}[#1][#2]}
\newcommandx{\NSym}[4][1=, 2=, 3=, 4=]{\mthsym[#4]{N#3}[#1][#2]}
\newcommandx{\OSym}[4][1=, 2=, 3=, 4=]{\mthsym[#4]{O#3}[#1][#2]}
\newcommandx{\PSym}[4][1=, 2=, 3=, 4=]{\mthsym[#4]{P#3}[#1][#2]}
\newcommandx{\QSym}[4][1=, 2=, 3=, 4=]{\mthsym[#4]{Q#3}[#1][#2]}
\newcommandx{\RSym}[4][1=, 2=, 3=, 4=]{\mthsym[#4]{R#3}[#1][#2]}
\newcommandx{\SSym}[4][1=, 2=, 3=, 4=]{\mthsym[#4]{S#3}[#1][#2]}
\newcommandx{\TSym}[4][1=, 2=, 3=, 4=]{\mthsym[#4]{T#3}[#1][#2]}
\newcommandx{\USym}[4][1=, 2=, 3=, 4=]{\mthsym[#4]{U#3}[#1][#2]}
\newcommandx{\VSym}[4][1=, 2=, 3=, 4=]{\mthsym[#4]{V#3}[#1][#2]}
\newcommandx{\WSym}[4][1=, 2=, 3=, 4=]{\mthsym[#4]{W#3}[#1][#2]}
\newcommandx{\XSym}[4][1=, 2=, 3=, 4=]{\mthsym[#4]{X#3}[#1][#2]}
\newcommandx{\YSym}[4][1=, 2=, 3=, 4=]{\mthsym[#4]{Y#3}[#1][#2]}
\newcommandx{\ZSym}[4][1=, 2=, 3=, 4=]{\mthsym[#4]{Z#3}[#1][#2]}

\newcommandx{\aSym}[4][1=, 2=, 3=, 4=]{\mthsym[#4]{a#3}[#1][#2]}
\newcommandx{\bSym}[4][1=, 2=, 3=, 4=]{\mthsym[#4]{b#3}[#1][#2]}
\newcommandx{\cSym}[4][1=, 2=, 3=, 4=]{\mthsym[#4]{c#3}[#1][#2]}
\newcommandx{\dSym}[4][1=, 2=, 3=, 4=]{\mthsym[#4]{d#3}[#1][#2]}
\newcommandx{\eSym}[4][1=, 2=, 3=, 4=]{\mthsym[#4]{e#3}[#1][#2]}
\newcommandx{\fSym}[4][1=, 2=, 3=, 4=]{\mthsym[#4]{f#3}[#1][#2]}
\newcommandx{\gSym}[4][1=, 2=, 3=, 4=]{\mthsym[#4]{g#3}[#1][#2]}
\newcommandx{\hSym}[4][1=, 2=, 3=, 4=]{\mthsym[#4]{h#3}[#1][#2]}
\newcommandx{\iSym}[4][1=, 2=, 3=, 4=]{\mthsym[#4]{i#3}[#1][#2]}
\newcommandx{\jSym}[4][1=, 2=, 3=, 4=]{\mthsym[#4]{j#3}[#1][#2]}
\newcommandx{\kSym}[4][1=, 2=, 3=, 4=]{\mthsym[#4]{k#3}[#1][#2]}
\newcommandx{\lSym}[4][1=, 2=, 3=, 4=]{\mthsym[#4]{l#3}[#1][#2]}
\newcommandx{\mSym}[4][1=, 2=, 3=, 4=]{\mthsym[#4]{m#3}[#1][#2]}
\newcommandx{\nSym}[4][1=, 2=, 3=, 4=]{\mthsym[#4]{n#3}[#1][#2]}
\newcommandx{\oSym}[4][1=, 2=, 3=, 4=]{\mthsym[#4]{o#3}[#1][#2]}
\newcommandx{\pSym}[4][1=, 2=, 3=, 4=]{\mthsym[#4]{p#3}[#1][#2]}
\newcommandx{\qSym}[4][1=, 2=, 3=, 4=]{\mthsym[#4]{q#3}[#1][#2]}
\newcommandx{\rSym}[4][1=, 2=, 3=, 4=]{\mthsym[#4]{r#3}[#1][#2]}
\newcommandx{\sSym}[4][1=, 2=, 3=, 4=]{\mthsym[#4]{s#3}[#1][#2]}
\newcommandx{\tSym}[4][1=, 2=, 3=, 4=]{\mthsym[#4]{t#3}[#1][#2]}
\newcommandx{\uSym}[4][1=, 2=, 3=, 4=]{\mthsym[#4]{u#3}[#1][#2]}
\newcommandx{\vSym}[4][1=, 2=, 3=, 4=]{\mthsym[#4]{v#3}[#1][#2]}
\newcommandx{\wSym}[4][1=, 2=, 3=, 4=]{\mthsym[#4]{w#3}[#1][#2]}
\newcommandx{\xSym}[4][1=, 2=, 3=, 4=]{\mthsym[#4]{x#3}[#1][#2]}
\newcommandx{\ySym}[4][1=, 2=, 3=, 4=]{\mthsym[#4]{y#3}[#1][#2]}
\newcommandx{\zSym}[4][1=, 2=, 3=, 4=]{\mthsym[#4]{z#3}[#1][#2]}

\newcommandx{\AElm}[4][1=, 2=, 3=, 4=]{\mthelm[#4]{A#3}[#1][#2]}
\newcommandx{\BElm}[4][1=, 2=, 3=, 4=]{\mthelm[#4]{B#3}[#1][#2]}
\newcommandx{\CElm}[4][1=, 2=, 3=, 4=]{\mthelm[#4]{C#3}[#1][#2]}
\newcommandx{\DElm}[4][1=, 2=, 3=, 4=]{\mthelm[#4]{D#3}[#1][#2]}
\newcommandx{\EElm}[4][1=, 2=, 3=, 4=]{\mthelm[#4]{E#3}[#1][#2]}
\newcommandx{\FElm}[4][1=, 2=, 3=, 4=]{\mthelm[#4]{F#3}[#1][#2]}
\newcommandx{\GElm}[4][1=, 2=, 3=, 4=]{\mthelm[#4]{G#3}[#1][#2]}
\newcommandx{\HElm}[4][1=, 2=, 3=, 4=]{\mthelm[#4]{H#3}[#1][#2]}
\newcommandx{\IElm}[4][1=, 2=, 3=, 4=]{\mthelm[#4]{I#3}[#1][#2]}
\newcommandx{\JElm}[4][1=, 2=, 3=, 4=]{\mthelm[#4]{J#3}[#1][#2]}
\newcommandx{\KElm}[4][1=, 2=, 3=, 4=]{\mthelm[#4]{K#3}[#1][#2]}
\newcommandx{\LElm}[4][1=, 2=, 3=, 4=]{\mthelm[#4]{L#3}[#1][#2]}
\newcommandx{\MElm}[4][1=, 2=, 3=, 4=]{\mthelm[#4]{M#3}[#1][#2]}
\newcommandx{\NElm}[4][1=, 2=, 3=, 4=]{\mthelm[#4]{N#3}[#1][#2]}
\newcommandx{\OElm}[4][1=, 2=, 3=, 4=]{\mthelm[#4]{O#3}[#1][#2]}
\newcommandx{\PElm}[4][1=, 2=, 3=, 4=]{\mthelm[#4]{P#3}[#1][#2]}
\newcommandx{\QElm}[4][1=, 2=, 3=, 4=]{\mthelm[#4]{Q#3}[#1][#2]}
\newcommandx{\RElm}[4][1=, 2=, 3=, 4=]{\mthelm[#4]{R#3}[#1][#2]}
\newcommandx{\SElm}[4][1=, 2=, 3=, 4=]{\mthelm[#4]{S#3}[#1][#2]}
\newcommandx{\TElm}[4][1=, 2=, 3=, 4=]{\mthelm[#4]{T#3}[#1][#2]}
\newcommandx{\UElm}[4][1=, 2=, 3=, 4=]{\mthelm[#4]{U#3}[#1][#2]}
\newcommandx{\VElm}[4][1=, 2=, 3=, 4=]{\mthelm[#4]{V#3}[#1][#2]}
\newcommandx{\WElm}[4][1=, 2=, 3=, 4=]{\mthelm[#4]{W#3}[#1][#2]}
\newcommandx{\XElm}[4][1=, 2=, 3=, 4=]{\mthelm[#4]{X#3}[#1][#2]}
\newcommandx{\YElm}[4][1=, 2=, 3=, 4=]{\mthelm[#4]{Y#3}[#1][#2]}
\newcommandx{\ZElm}[4][1=, 2=, 3=, 4=]{\mthelm[#4]{Z#3}[#1][#2]}

\newcommandx{\aElm}[4][1=, 2=, 3=, 4=]{\mthelm[#4]{a#3}[#1][#2]}
\newcommandx{\bElm}[4][1=, 2=, 3=, 4=]{\mthelm[#4]{b#3}[#1][#2]}
\newcommandx{\cElm}[4][1=, 2=, 3=, 4=]{\mthelm[#4]{c#3}[#1][#2]}
\newcommandx{\dElm}[4][1=, 2=, 3=, 4=]{\mthelm[#4]{d#3}[#1][#2]}
\newcommandx{\eElm}[4][1=, 2=, 3=, 4=]{\mthelm[#4]{e#3}[#1][#2]}
\newcommandx{\fElm}[4][1=, 2=, 3=, 4=]{\mthelm[#4]{f#3}[#1][#2]}
\newcommandx{\gElm}[4][1=, 2=, 3=, 4=]{\mthelm[#4]{g#3}[#1][#2]}
\newcommandx{\hElm}[4][1=, 2=, 3=, 4=]{\mthelm[#4]{h#3}[#1][#2]}
\newcommandx{\iElm}[4][1=, 2=, 3=, 4=]{\mthelm[#4]{i#3}[#1][#2]}
\newcommandx{\jElm}[4][1=, 2=, 3=, 4=]{\mthelm[#4]{j#3}[#1][#2]}
\newcommandx{\kElm}[4][1=, 2=, 3=, 4=]{\mthelm[#4]{k#3}[#1][#2]}
\newcommandx{\lElm}[4][1=, 2=, 3=, 4=]{\mthelm[#4]{l#3}[#1][#2]}
\newcommandx{\mElm}[4][1=, 2=, 3=, 4=]{\mthelm[#4]{m#3}[#1][#2]}
\newcommandx{\nElm}[4][1=, 2=, 3=, 4=]{\mthelm[#4]{n#3}[#1][#2]}
\newcommandx{\oElm}[4][1=, 2=, 3=, 4=]{\mthelm[#4]{o#3}[#1][#2]}
\newcommandx{\pElm}[4][1=, 2=, 3=, 4=]{\mthelm[#4]{p#3}[#1][#2]}
\newcommandx{\qElm}[4][1=, 2=, 3=, 4=]{\mthelm[#4]{q#3}[#1][#2]}
\newcommandx{\rElm}[4][1=, 2=, 3=, 4=]{\mthelm[#4]{r#3}[#1][#2]}
\newcommandx{\sElm}[4][1=, 2=, 3=, 4=]{\mthelm[#4]{s#3}[#1][#2]}
\newcommandx{\tElm}[4][1=, 2=, 3=, 4=]{\mthelm[#4]{t#3}[#1][#2]}
\newcommandx{\uElm}[4][1=, 2=, 3=, 4=]{\mthelm[#4]{u#3}[#1][#2]}
\newcommandx{\vElm}[4][1=, 2=, 3=, 4=]{\mthelm[#4]{v#3}[#1][#2]}
\newcommandx{\wElm}[4][1=, 2=, 3=, 4=]{\mthelm[#4]{w#3}[#1][#2]}
\newcommandx{\xElm}[4][1=, 2=, 3=, 4=]{\mthelm[#4]{x#3}[#1][#2]}
\newcommandx{\yElm}[4][1=, 2=, 3=, 4=]{\mthelm[#4]{y#3}[#1][#2]}
\newcommandx{\zElm}[4][1=, 2=, 3=, 4=]{\mthelm[#4]{z#3}[#1][#2]}

\newcommand{\ie}
	{\txtabr{i.e.}\xspace}

\newcommand{\wrt}
	{\txtabr{w.r.t.}\xspace}

\newcommand{\defeq}
	{\ensuremath{\triangleq}}
\newcommand{\seteq}
	{\ensuremath{:=}}

\newcommand{\tuple}[1]
	{\ensuremath{\!\argint{\langle}{#1}{\rangle}}}

\newcommand{\tupleb}[2]
	{\tuple{\argb{#1}{#2}}}
\newcommand{\tuplec}[3]
	{\tuple{\argc{#1}{#2}{#3}}}
\newcommand{\tupled}[4]
	{\tuple{\argd{#1}{#2}{#3}{#4}}}
\newcommand{\tuplee}[5]
	{\tuple{\arge{#1}{#2}{#3}{#4}{#5}}}
\newcommand{\tuplef}[6]
	{\tuple{\argf{#1}{#2}{#3}{#4}{#5}{#6}}}
\newcommand{\tupleg}[7]
	{\tuple{\argg{#1}{#2}{#3}{#4}{#5}{#6}{#7}}}
\newcommand{\tupleh}[8]
	{\tuple{\argh{#1}{#2}{#3}{#4}{#5}{#6}{#7}{#8}}}
\newcommand{\tuplei}[9]
	{\tuple{\argi{#1}{#2}{#3}{#4}{#5}{#6}{#7}{#8}{#9}}}

\newcommand{\tuplej}[9]
	{%
	\def\defarga{#1}%
	\def\defargb{#2}%
	\def\defargc{#3}%
	\def\defargd{#4}%
	\def\defarge{#5}%
	\def\defargf{#6}%
	\def\defargg{#7}%
	\def\defargh{#8}%
	\def\defargi{#9}%
	\tupleauxj%
	}
\newcommand{\tuplek}[9]
	{%
	\def\defarga{#1}%
	\def\defargb{#2}%
	\def\defargc{#3}%
	\def\defargd{#4}%
	\def\defarge{#5}%
	\def\defargf{#6}%
	\def\defargg{#7}%
	\def\defargh{#8}%
	\def\defargi{#9}%
	\tupleauxk%
	}
\newcommand{\tuplel}[9]
	{%
	\def\defarga{#1}%
	\def\defargb{#2}%
	\def\defargc{#3}%
	\def\defargd{#4}%
	\def\defarge{#5}%
	\def\defargf{#6}%
	\def\defargg{#7}%
	\def\defargh{#8}%
	\def\defargi{#9}%
	\tupleauxl%
	}
\newcommand{\tuplem}[9]
	{%
	\def\defarga{#1}%
	\def\defargb{#2}%
	\def\defargc{#3}%
	\def\defargd{#4}%
	\def\defarge{#5}%
	\def\defargf{#6}%
	\def\defargg{#7}%
	\def\defargh{#8}%
	\def\defargi{#9}%
	\tupleauxm%
	}
\newcommand{\tuplen}[9]
	{%
	\def\defarga{#1}%
	\def\defargb{#2}%
	\def\defargc{#3}%
	\def\defargd{#4}%
	\def\defarge{#5}%
	\def\defargf{#6}%
	\def\defargg{#7}%
	\def\defargh{#8}%
	\def\defargi{#9}%
	\tupleauxn%
	}
\newcommand{\tupleo}[9]
	{%
	\def\defarga{#1}%
	\def\defargb{#2}%
	\def\defargc{#3}%
	\def\defargd{#4}%
	\def\defarge{#5}%
	\def\defargf{#6}%
	\def\defargg{#7}%
	\def\defargh{#8}%
	\def\defargi{#9}%
	\tupleauxo%
	}
\newcommand{\tuplep}[9]
	{%
	\def\defarga{#1}%
	\def\defargb{#2}%
	\def\defargc{#3}%
	\def\defargd{#4}%
	\def\defarge{#5}%
	\def\defargf{#6}%
	\def\defargg{#7}%
	\def\defargh{#8}%
	\def\defargi{#9}%
	\tupleauxp%
	}
\newcommand{\tupleq}[9]
	{%
	\def\defarga{#1}%
	\def\defargb{#2}%
	\def\defargc{#3}%
	\def\defargd{#4}%
	\def\defarge{#5}%
	\def\defargf{#6}%
	\def\defargg{#7}%
	\def\defargh{#8}%
	\def\defargi{#9}%
	\tupleauxq%
	}
\newcommand{\tupler}[9]
	{%
	\def\defarga{#1}%
	\def\defargb{#2}%
	\def\defargc{#3}%
	\def\defargd{#4}%
	\def\defarge{#5}%
	\def\defargf{#6}%
	\def\defargg{#7}%
	\def\defargh{#8}%
	\def\defargi{#9}%
	\tupleauxr%
	}

\newcommand{\tupleauxj}[1]
	{%
	\tuple{\argj{\defarga}{\defargb}{\defargc}{\defargd}{\defarge}{\defargf}%
		{\defargg}{\defargh}{\defargi}{#1}}%
	}
\newcommand{\tupleauxk}[2]
	{%
	\tuple{\argk{\defarga}{\defargb}{\defargc}{\defargd}{\defarge}{\defargf}%
		{\defargg}{\defargh}{\defargi}{#1}{#2}}%
	}
\newcommand{\tupleauxl}[3]
	{%
	\tuple{\argl{\defarga}{\defargb}{\defargc}{\defargd}{\defarge}{\defargf}%
		{\defargg}{\defargh}{\defargi}{#1}{#2}{#3}}%
	}
\newcommand{\tupleauxm}[4]
	{%
	\tuple{\argm{\defarga}{\defargb}{\defargc}{\defargd}{\defarge}{\defargf}%
		{\defargg}{\defargh}{\defargi}{#1}{#2}{#3}{#4}}%
	}
\newcommand{\tupleauxn}[5]
	{%
	\tuple{\argn{\defarga}{\defargb}{\defargc}{\defargd}{\defarge}{\defargf}%
		{\defargg}{\defargh}{\defargi}{#1}{#2}{#3}{#4}{#5}}%
	}
\newcommand{\tupleauxo}[6]
	{%
	\tuple{\argo{\defarga}{\defargb}{\defargc}{\defargd}{\defarge}{\defargf}%
		{\defargg}{\defargh}{\defargi}{#1}{#2}{#3}{#4}{#5}{#6}}%
	}
\newcommand{\tupleauxp}[7]
	{%
	\tuple{\argp{\defarga}{\defargb}{\defargc}{\defargd}{\defarge}{\defargf}%
		{\defargg}{\defargh}{\defargi}{#1}{#2}{#3}{#4}{#5}{#6}{#7}}%
	}
\newcommand{\tupleauxq}[8]
	{%
	\tuple{\argq{\defarga}{\defargb}{\defargc}{\defargd}{\defarge}{\defargf}%
		{\defargg}{\defargh}{\defargi}{#1}{#2}{#3}{#4}{#5}{#6}{#7}{#8}}%
	}
\newcommand{\tupleauxr}[9]
	{%
	\tuple{\argr{\defarga}{\defargb}{\defargc}{\defargd}{\defarge}{\defargf}%
		{\defargg}{\defargh}{\defargi}{#1}{#2}{#3}{#4}{#5}{#6}{#7}{#8}{#9}}%
	}

\newcommand{\tuplecx}[3]
	{%
	\def\defarga{#1}%
	\def\defargb{#2}%
	\def\defargc{#3}%
	\argsubsup{\tupleauxcx}%
	}
\newcommand{\tupledx}[4]
	{%
	\def\defarga{#1}%
	\def\defargb{#2}%
	\def\defargc{#3}%
	\def\defargd{#4}%
	\argsubsup{\tupleauxdx}%
	}
\newcommand{\tupleex}[5]
	{%
	\def\defarga{#1}%
	\def\defargb{#2}%
	\def\defargc{#3}%
	\def\defargd{#4}%
	\def\defarge{#5}%
	\argsubsup{\tupleauxex}%
	}
\newcommand{\tuplefx}[6]
	{%
	\def\defarga{#1}%
	\def\defargb{#2}%
	\def\defargc{#3}%
	\def\defargd{#4}%
	\def\defarge{#5}%
	\def\defargf{#6}%
	\argsubsup{\tupleauxfx}%
	}
\newcommand{\tuplegx}[7]
	{%
	\def\defarga{#1}%
	\def\defargb{#2}%
	\def\defargc{#3}%
	\def\defargd{#4}%
	\def\defarge{#5}%
	\def\defargf{#6}%
	\def\defargg{#7}%
	\argsubsup{\tupleauxgx}%
	}
\newcommand{\tuplehx}[8]
	{%
	\def\defarga{#1}%
	\def\defargb{#2}%
	\def\defargc{#3}%
	\def\defargd{#4}%
	\def\defarge{#5}%
	\def\defargf{#6}%
	\def\defargg{#7}%
	\def\defargh{#8}%
	\argsubsup{\tupleauxhx}%
	}
\newcommand{\tupleix}[9]
	{%
	\def\defarga{#1}%
	\def\defargb{#2}%
	\def\defargc{#3}%
	\def\defargd{#4}%
	\def\defarge{#5}%
	\def\defargf{#6}%
	\def\defargg{#7}%
	\def\defargh{#8}%
	\def\defargi{#9}%
	\argsubsup{\tupleauxix}%
	}

\newcommandx{\tupleauxbx}[2][1=, 2=]
	{%
	\tupleb
		{\argdef{#1}{\defarga[\argsubscript][\argsuperscript]}}
		{\argdef{#2}{\defargb[\argsubscript][\argsuperscript]}}%
	}
\newcommandx{\tupleauxcx}[3][1=, 2=, 3=]
	{%
	\tuplec
		{\argdef{#1}{\defarga[\argsubscript][\argsuperscript]}}
		{\argdef{#2}{\defargb[\argsubscript][\argsuperscript]}}
		{\argdef{#3}{\defargc[\argsubscript][\argsuperscript]}}%
	}
\newcommandx{\tupleauxdx}[4][1=, 2=, 3=, 4=]
	{%
	\tupled
		{\argdef{#1}{\defarga[\argsubscript][\argsuperscript]}}
		{\argdef{#2}{\defargb[\argsubscript][\argsuperscript]}}
		{\argdef{#3}{\defargc[\argsubscript][\argsuperscript]}}
		{\argdef{#4}{\defargd[\argsubscript][\argsuperscript]}}%
	}
\newcommandx{\tupleauxex}[5][1=, 2=, 3=, 4=, 5=]
	{%
	\tuplee
		{\argdef{#1}{\defarga[\argsubscript][\argsuperscript]}}
		{\argdef{#2}{\defargb[\argsubscript][\argsuperscript]}}
		{\argdef{#3}{\defargc[\argsubscript][\argsuperscript]}}
		{\argdef{#4}{\defargd[\argsubscript][\argsuperscript]}}
		{\argdef{#5}{\defarge[\argsubscript][\argsuperscript]}}%
	}
\newcommandx{\tupleauxfx}[6][1=, 2=, 3=, 4=, 5=, 6=]
	{%
	\tuplef
		{\argdef{#1}{\defarga[\argsubscript][\argsuperscript]}}
		{\argdef{#2}{\defargb[\argsubscript][\argsuperscript]}}
		{\argdef{#3}{\defargc[\argsubscript][\argsuperscript]}}
		{\argdef{#4}{\defargd[\argsubscript][\argsuperscript]}}
		{\argdef{#5}{\defarge[\argsubscript][\argsuperscript]}}
		{\argdef{#6}{\defargf[\argsubscript][\argsuperscript]}}%
	}
\newcommandx{\tupleauxgx}[7][1=, 2=, 3=, 4=, 5=, 6=, 7=]
	{%
	\tupleg
		{\argdef{#1}{\defarga[\argsubscript][\argsuperscript]}}
		{\argdef{#2}{\defargb[\argsubscript][\argsuperscript]}}
		{\argdef{#3}{\defargc[\argsubscript][\argsuperscript]}}
		{\argdef{#4}{\defargd[\argsubscript][\argsuperscript]}}
		{\argdef{#5}{\defarge[\argsubscript][\argsuperscript]}}
		{\argdef{#6}{\defargf[\argsubscript][\argsuperscript]}}
		{\argdef{#7}{\defargg[\argsubscript][\argsuperscript]}}%
	}
\newcommandx{\tupleauxhx}[8][1=, 2=, 3=, 4=, 5=, 6=, 7=, 8=]
	{%
	\tupleh
		{\argdef{#1}{\defarga[\argsubscript][\argsuperscript]}}
		{\argdef{#2}{\defargb[\argsubscript][\argsuperscript]}}
		{\argdef{#3}{\defargc[\argsubscript][\argsuperscript]}}
		{\argdef{#4}{\defargd[\argsubscript][\argsuperscript]}}
		{\argdef{#5}{\defarge[\argsubscript][\argsuperscript]}}
		{\argdef{#6}{\defargf[\argsubscript][\argsuperscript]}}
		{\argdef{#7}{\defargg[\argsubscript][\argsuperscript]}}
		{\argdef{#8}{\defargh[\argsubscript][\argsuperscript]}}%
	}
\newcommandx{\tupleauxix}[9][1=, 2=, 3=, 4=, 5=, 6=, 7=, 8=, 9=]
	{%
	\tuplei
		{\argdef{#1}{\defarga[\argsubscript][\argsuperscript]}}
		{\argdef{#2}{\defargb[\argsubscript][\argsuperscript]}}
		{\argdef{#3}{\defargc[\argsubscript][\argsuperscript]}}
		{\argdef{#4}{\defargd[\argsubscript][\argsuperscript]}}
		{\argdef{#5}{\defarge[\argsubscript][\argsuperscript]}}
		{\argdef{#6}{\defargf[\argsubscript][\argsuperscript]}}
		{\argdef{#7}{\defargg[\argsubscript][\argsuperscript]}}
		{\argdef{#8}{\defargh[\argsubscript][\argsuperscript]}}
		{\argdef{#9}{\defargi[\argsubscript][\argsuperscript]}}%
	}

\newcommand{\tuplejx}[9]
	{%
	\def\tuplearga{#1}%
	\def\tupleargb{#2}%
	\def\tupleargc{#3}%
	\def\tupleargd{#4}%
	\def\tuplearge{#5}%
	\def\tupleargf{#6}%
	\def\tupleargg{#7}%
	\def\tupleargh{#8}%
	\def\tupleargi{#9}%
	\argsubsup{\tupleauxjx}%
	}

\newcommand{\tupleauxjx}[1]
	{%
	\def\tupleargj{#1}%
	\argsubsup{\tupleauxxjx}%
	}

\newcommandx{\tupleauxxjx}[9][1=, 2=, 3=, 4=, 5=, 6=, 7=, 8=, 9=]
	{%
	\def\optarga{#1}%
	\def\optargb{#2}%
	\def\optargc{#3}%
	\def\optargd{#4}%
	\def\optarge{#5}%
	\def\optargf{#6}%
	\def\optargg{#7}%
	\def\optargh{#8}%
	\def\optargi{#9}%
	\tupleauxxxjx%
	}
\newcommandx{\tupleauxxkx}[9][1=, 2=, 3=, 4=, 5=, 6=, 7=, 8=, 9=]
	{%
	\def\optarga{#1}%
	\def\optargb{#2}%
	\def\optargc{#3}%
	\def\optargd{#4}%
	\def\optarge{#5}%
	\def\optargf{#6}%
	\def\optargg{#7}%
	\def\optargh{#8}%
	\def\optargi{#9}%
	\tupleauxxxkx%
	}
\newcommandx{\tupleauxxlx}[9][1=, 2=, 3=, 4=, 5=, 6=, 7=, 8=, 9=]
	{%
	\def\optarga{#1}%
	\def\optargb{#2}%
	\def\optargc{#3}%
	\def\optargd{#4}%
	\def\optarge{#5}%
	\def\optargf{#6}%
	\def\optargg{#7}%
	\def\optargh{#8}%
	\def\optargi{#9}%
	\tupleauxxxlx%
	}
\newcommandx{\tupleauxxmx}[9][1=, 2=, 3=, 4=, 5=, 6=, 7=, 8=, 9=]
	{%
	\def\optarga{#1}%
	\def\optargb{#2}%
	\def\optargc{#3}%
	\def\optargd{#4}%
	\def\optarge{#5}%
	\def\optargf{#6}%
	\def\optargg{#7}%
	\def\optargh{#8}%
	\def\optargi{#9}%
	\tupleauxxxmx%
	}
\newcommandx{\tupleauxxnx}[9][1=, 2=, 3=, 4=, 5=, 6=, 7=, 8=, 9=]
	{%
	\def\optarga{#1}%
	\def\optargb{#2}%
	\def\optargc{#3}%
	\def\optargd{#4}%
	\def\optarge{#5}%
	\def\optargf{#6}%
	\def\optargg{#7}%
	\def\optargh{#8}%
	\def\optargi{#9}%
	\tupleauxxxnx%
	}
\newcommandx{\tupleauxxox}[9][1=, 2=, 3=, 4=, 5=, 6=, 7=, 8=, 9=]
	{%
	\def\optarga{#1}%
	\def\optargb{#2}%
	\def\optargc{#3}%
	\def\optargd{#4}%
	\def\optarge{#5}%
	\def\optargf{#6}%
	\def\optargg{#7}%
	\def\optargh{#8}%
	\def\optargi{#9}%
	\tupleauxxxox%
	}
\newcommandx{\tupleauxxpx}[9][1=, 2=, 3=, 4=, 5=, 6=, 7=, 8=, 9=]
	{%
	\def\optarga{#1}%
	\def\optargb{#2}%
	\def\optargc{#3}%
	\def\optargd{#4}%
	\def\optarge{#5}%
	\def\optargf{#6}%
	\def\optargg{#7}%
	\def\optargh{#8}%
	\def\optargi{#9}%
	\tupleauxxxpx%
	}
\newcommandx{\tupleauxxqx}[9][1=, 2=, 3=, 4=, 5=, 6=, 7=, 8=, 9=]
	{%
	\def\optarga{#1}%
	\def\optargb{#2}%
	\def\optargc{#3}%
	\def\optargd{#4}%
	\def\optarge{#5}%
	\def\optargf{#6}%
	\def\optargg{#7}%
	\def\optargh{#8}%
	\def\optargi{#9}%
	\tupleauxxxqx%
	}
\newcommandx{\tupleauxxrx}[9][1=, 2=, 3=, 4=, 5=, 6=, 7=, 8=, 9=]
	{%
	\def\optarga{#1}%
	\def\optargb{#2}%
	\def\optargc{#3}%
	\def\optargd{#4}%
	\def\optarge{#5}%
	\def\optargf{#6}%
	\def\optargg{#7}%
	\def\optargh{#8}%
	\def\optargi{#9}%
	\tupleauxxxrx%
	}

\newcommandx{\tupleauxxxjx}[1][1=]
	{%
	\tuplej
		{\argdef{\optarga}{\tuplearga[\argsubscript][\argsuperscript]}}
		{\argdef{\optargb}{\tupleargb[\argsubscript][\argsuperscript]}}
		{\argdef{\optargc}{\tupleargc[\argsubscript][\argsuperscript]}}
		{\argdef{\optargd}{\tupleargd[\argsubscript][\argsuperscript]}}
		{\argdef{\optarge}{\tuplearge[\argsubscript][\argsuperscript]}}
		{\argdef{\optargf}{\tupleargf[\argsubscript][\argsuperscript]}}
		{\argdef{\optargg}{\tupleargg[\argsubscript][\argsuperscript]}}
		{\argdef{\optargh}{\tupleargh[\argsubscript][\argsuperscript]}}
		{\argdef{\optargi}{\tupleargi[\argsubscript][\argsuperscript]}}
		{\argdef{#1}{\tupleargj[\argsubscript][\argsuperscript]}}%
	}
\newcommandx{\tupleauxxxkx}[2][1=, 2=]
	{%
	\tuplek
		{\argdef{\optarga}{\tuplearga[\argsubscript][\argsuperscript]}}
		{\argdef{\optargb}{\tupleargb[\argsubscript][\argsuperscript]}}
		{\argdef{\optargc}{\tupleargc[\argsubscript][\argsuperscript]}}
		{\argdef{\optargd}{\tupleargd[\argsubscript][\argsuperscript]}}
		{\argdef{\optarge}{\tuplearge[\argsubscript][\argsuperscript]}}
		{\argdef{\optargf}{\tupleargf[\argsubscript][\argsuperscript]}}
		{\argdef{\optargg}{\tupleargg[\argsubscript][\argsuperscript]}}
		{\argdef{\optargh}{\tupleargh[\argsubscript][\argsuperscript]}}
		{\argdef{\optargi}{\tupleargi[\argsubscript][\argsuperscript]}}
		{\argdef{#1}{\tupleargj[\argsubscript][\argsuperscript]}}
		{\argdef{#2}{\tupleargk[\argsubscript][\argsuperscript]}}
	}
\newcommandx{\tupleauxxxlx}[3][1=, 2=, 3=]
	{%
	\tuplel
		{\argdef{\optarga}{\tuplearga[\argsubscript][\argsuperscript]}}
		{\argdef{\optargb}{\tupleargb[\argsubscript][\argsuperscript]}}
		{\argdef{\optargc}{\tupleargc[\argsubscript][\argsuperscript]}}
		{\argdef{\optargd}{\tupleargd[\argsubscript][\argsuperscript]}}
		{\argdef{\optarge}{\tuplearge[\argsubscript][\argsuperscript]}}
		{\argdef{\optargf}{\tupleargf[\argsubscript][\argsuperscript]}}
		{\argdef{\optargg}{\tupleargg[\argsubscript][\argsuperscript]}}
		{\argdef{\optargh}{\tupleargh[\argsubscript][\argsuperscript]}}
		{\argdef{\optargi}{\tupleargi[\argsubscript][\argsuperscript]}}
		{\argdef{#1}{\tupleargj[\argsubscript][\argsuperscript]}}
		{\argdef{#2}{\tupleargk[\argsubscript][\argsuperscript]}}
		{\argdef{#3}{\tupleargl[\argsubscript][\argsuperscript]}}
	}
\newcommandx{\tupleauxxxmx}[4][1=, 2=, 3=, 4=]
	{%
	\tuplem
		{\argdef{\optarga}{\tuplearga[\argsubscript][\argsuperscript]}}
		{\argdef{\optargb}{\tupleargb[\argsubscript][\argsuperscript]}}
		{\argdef{\optargc}{\tupleargc[\argsubscript][\argsuperscript]}}
		{\argdef{\optargd}{\tupleargd[\argsubscript][\argsuperscript]}}
		{\argdef{\optarge}{\tuplearge[\argsubscript][\argsuperscript]}}
		{\argdef{\optargf}{\tupleargf[\argsubscript][\argsuperscript]}}
		{\argdef{\optargg}{\tupleargg[\argsubscript][\argsuperscript]}}
		{\argdef{\optargh}{\tupleargh[\argsubscript][\argsuperscript]}}
		{\argdef{\optargi}{\tupleargi[\argsubscript][\argsuperscript]}}
		{\argdef{#1}{\tupleargj[\argsubscript][\argsuperscript]}}
		{\argdef{#2}{\tupleargk[\argsubscript][\argsuperscript]}}
		{\argdef{#3}{\tupleargl[\argsubscript][\argsuperscript]}}
		{\argdef{#4}{\tupleargm[\argsubscript][\argsuperscript]}}
	}
\newcommandx{\tupleauxxxnx}[5][1=, 2=, 3=, 4=, 5=]
	{%
	\tuplen
		{\argdef{\optarga}{\tuplearga[\argsubscript][\argsuperscript]}}
		{\argdef{\optargb}{\tupleargb[\argsubscript][\argsuperscript]}}
		{\argdef{\optargc}{\tupleargc[\argsubscript][\argsuperscript]}}
		{\argdef{\optargd}{\tupleargd[\argsubscript][\argsuperscript]}}
		{\argdef{\optarge}{\tuplearge[\argsubscript][\argsuperscript]}}
		{\argdef{\optargf}{\tupleargf[\argsubscript][\argsuperscript]}}
		{\argdef{\optargg}{\tupleargg[\argsubscript][\argsuperscript]}}
		{\argdef{\optargh}{\tupleargh[\argsubscript][\argsuperscript]}}
		{\argdef{\optargi}{\tupleargi[\argsubscript][\argsuperscript]}}
		{\argdef{#1}{\tupleargj[\argsubscript][\argsuperscript]}}
		{\argdef{#2}{\tupleargk[\argsubscript][\argsuperscript]}}
		{\argdef{#3}{\tupleargl[\argsubscript][\argsuperscript]}}
		{\argdef{#4}{\tupleargm[\argsubscript][\argsuperscript]}}
		{\argdef{#5}{\tupleargn[\argsubscript][\argsuperscript]}}
	}
\newcommandx{\tupleauxxxox}[6][1=, 2=, 3=, 4=, 5=, 6=]
	{%
	\tupleo
		{\argdef{\optarga}{\tuplearga[\argsubscript][\argsuperscript]}}
		{\argdef{\optargb}{\tupleargb[\argsubscript][\argsuperscript]}}
		{\argdef{\optargc}{\tupleargc[\argsubscript][\argsuperscript]}}
		{\argdef{\optargd}{\tupleargd[\argsubscript][\argsuperscript]}}
		{\argdef{\optarge}{\tuplearge[\argsubscript][\argsuperscript]}}
		{\argdef{\optargf}{\tupleargf[\argsubscript][\argsuperscript]}}
		{\argdef{\optargg}{\tupleargg[\argsubscript][\argsuperscript]}}
		{\argdef{\optargh}{\tupleargh[\argsubscript][\argsuperscript]}}
		{\argdef{\optargi}{\tupleargi[\argsubscript][\argsuperscript]}}
		{\argdef{#1}{\tupleargj[\argsubscript][\argsuperscript]}}
		{\argdef{#2}{\tupleargk[\argsubscript][\argsuperscript]}}
		{\argdef{#3}{\tupleargl[\argsubscript][\argsuperscript]}}
		{\argdef{#4}{\tupleargm[\argsubscript][\argsuperscript]}}
		{\argdef{#5}{\tupleargn[\argsubscript][\argsuperscript]}}
		{\argdef{#6}{\tupleargo[\argsubscript][\argsuperscript]}}
	}
\newcommandx{\tupleauxxxpx}[7][1=, 2=, 3=, 4=, 5=, 6=, 7=]
	{%
	\tuplep
		{\argdef{\optarga}{\tuplearga[\argsubscript][\argsuperscript]}}
		{\argdef{\optargb}{\tupleargb[\argsubscript][\argsuperscript]}}
		{\argdef{\optargc}{\tupleargc[\argsubscript][\argsuperscript]}}
		{\argdef{\optargd}{\tupleargd[\argsubscript][\argsuperscript]}}
		{\argdef{\optarge}{\tuplearge[\argsubscript][\argsuperscript]}}
		{\argdef{\optargf}{\tupleargf[\argsubscript][\argsuperscript]}}
		{\argdef{\optargg}{\tupleargg[\argsubscript][\argsuperscript]}}
		{\argdef{\optargh}{\tupleargh[\argsubscript][\argsuperscript]}}
		{\argdef{\optargi}{\tupleargi[\argsubscript][\argsuperscript]}}
		{\argdef{#1}{\tupleargj[\argsubscript][\argsuperscript]}}
		{\argdef{#2}{\tupleargk[\argsubscript][\argsuperscript]}}
		{\argdef{#3}{\tupleargl[\argsubscript][\argsuperscript]}}
		{\argdef{#4}{\tupleargm[\argsubscript][\argsuperscript]}}
		{\argdef{#5}{\tupleargn[\argsubscript][\argsuperscript]}}
		{\argdef{#6}{\tupleargo[\argsubscript][\argsuperscript]}}
		{\argdef{#7}{\tupleargp[\argsubscript][\argsuperscript]}}
	}
\newcommandx{\tupleauxxxqx}[8][1=, 2=, 3=, 4=, 5=, 6=, 7=, 8=]
	{%
	\tupleq
		{\argdef{\optarga}{\tuplearga[\argsubscript][\argsuperscript]}}
		{\argdef{\optargb}{\tupleargb[\argsubscript][\argsuperscript]}}
		{\argdef{\optargc}{\tupleargc[\argsubscript][\argsuperscript]}}
		{\argdef{\optargd}{\tupleargd[\argsubscript][\argsuperscript]}}
		{\argdef{\optarge}{\tuplearge[\argsubscript][\argsuperscript]}}
		{\argdef{\optargf}{\tupleargf[\argsubscript][\argsuperscript]}}
		{\argdef{\optargg}{\tupleargg[\argsubscript][\argsuperscript]}}
		{\argdef{\optargh}{\tupleargh[\argsubscript][\argsuperscript]}}
		{\argdef{\optargi}{\tupleargi[\argsubscript][\argsuperscript]}}
		{\argdef{#1}{\tupleargj[\argsubscript][\argsuperscript]}}
		{\argdef{#2}{\tupleargk[\argsubscript][\argsuperscript]}}
		{\argdef{#3}{\tupleargl[\argsubscript][\argsuperscript]}}
		{\argdef{#4}{\tupleargm[\argsubscript][\argsuperscript]}}
		{\argdef{#5}{\tupleargn[\argsubscript][\argsuperscript]}}
		{\argdef{#6}{\tupleargo[\argsubscript][\argsuperscript]}}
		{\argdef{#7}{\tupleargp[\argsubscript][\argsuperscript]}}
		{\argdef{#8}{\tupleargq[\argsubscript][\argsuperscript]}}
	}
\newcommandx{\tupleauxxxrx}[9][1=, 2=, 3=, 4=, 5=, 6=, 7=, 8=, 9=]
	{%
	\tupler
		{\argdef{\optarga}{\tuplearga[\argsubscript][\argsuperscript]}}
		{\argdef{\optargb}{\tupleargb[\argsubscript][\argsuperscript]}}
		{\argdef{\optargc}{\tupleargc[\argsubscript][\argsuperscript]}}
		{\argdef{\optargd}{\tupleargd[\argsubscript][\argsuperscript]}}
		{\argdef{\optarge}{\tuplearge[\argsubscript][\argsuperscript]}}
		{\argdef{\optargf}{\tupleargf[\argsubscript][\argsuperscript]}}
		{\argdef{\optargg}{\tupleargg[\argsubscript][\argsuperscript]}}
		{\argdef{\optargh}{\tupleargh[\argsubscript][\argsuperscript]}}
		{\argdef{\optargi}{\tupleargi[\argsubscript][\argsuperscript]}}
		{\argdef{#1}{\tupleargj[\argsubscript][\argsuperscript]}}
		{\argdef{#2}{\tupleargk[\argsubscript][\argsuperscript]}}
		{\argdef{#3}{\tupleargl[\argsubscript][\argsuperscript]}}
		{\argdef{#4}{\tupleargm[\argsubscript][\argsuperscript]}}
		{\argdef{#5}{\tupleargn[\argsubscript][\argsuperscript]}}
		{\argdef{#6}{\tupleargo[\argsubscript][\argsuperscript]}}
		{\argdef{#7}{\tupleargp[\argsubscript][\argsuperscript]}}
		{\argdef{#8}{\tupleargq[\argsubscript][\argsuperscript]}}
		{\argdef{#9}{\tupleargr[\argsubscript][\argsuperscript]}}%
	}

\newcommand{\set}[2]
	{\ensuremath{\argint{\{}{\argext{#1}{\allowbreak:\allowbreak}{#2}}{\}}}}

\newcommand{\pow}[1]
	{\ensuremath{2^{#1}}}

\newcommand{\card}[1]
	{\mthempty{\argint{\vert}{#1}{\vert}}}

\newcommand{\dom}
	{\mthargfun{dom}}

\newcommandx{\pto}[2][1=, 2=]
	{\ensuremath{\rightharpoonup}}

\newcommandx{\cto}[2][1=, 2=]
	{\:\mthempty{\to}[#1][#2]\:}
\newcommandx{\cpto}[2][1=, 2=]
	{\:\mthempty{\pto}[#1][#2]\:}

\newcommand{\SetN}
	{\mthset[2]{N}}

\newcommand{\numcc}[2]
	{\mthempty{[\argb{#1}{#2}]}}

\newcommand{\argset}{Ar}
\newcommandx{\ArgSet}[3][1=, 2=, 3=]
	{\mthset{\argset#3}[#1][#2]}

\newcommand{\argsym}{a}
\newcommandx{\argSym}[3][1=, 2=, 3=]
	{\mthsym{\argsym#3}[#1][#2]}

\newcommand{\argelm}{a}
\newcommandx{\argElm}[3][1=, 2=, 3=]
	{\mthelm{\argelm#3}[#1][#2]}

\newcommand{\relset}{Rl}
\newcommandx{\RelSet}[3][1=, 2=, 3=]
	{\mthset{\relset#3}[#1][#2]}

\newcommand{\relsym}{r}
\newcommandx{\relSym}[3][1=, 2=, 3=]
	{\mthsym{\relsym#3}[#1][#2]}

\newcommand{\relelm}{r}
\newcommandx{\relElm}[3][1=, 2=, 3=]
	{\mthelm{\relelm#3}[#1][#2]}

\newcommand{\argfun}{ar}
\newcommandx{\argFun}[4][1=, 2=, 3=, 4=]
	{\mthargfun{\argfun#4}[#1][#2]{#3}}

\newcommand{\lansig}{LS}
\newcommandx{\LanSig}[5][1=, 2=, 3=, 4=, 5=]
	{\txtargname{\lansig#5{\small\argint{$[$}{#1}{$]$}}}[#2][#3]{#4}\xspace}

\newcommand{\lansigcls}{LS}
\newcommandx{\LanSigCls}[5][1=, 2=, 3=, 4=, 5=]
	{\mthset[#5]{\lansigcls#4\text{\txtname{\small\argint{$[$}{#1}{$]$}}}}[#2]%
	[#3]}

\newcommand{\domset}{Dm}
\newcommandx{\DomSet}[3][1=, 2=, 3=]
	{\mthset{\domset#3}[#1][#2]}

\newcommand{\domsym}{d}
\newcommandx{\domSym}[3][1=, 2=, 3=]
	{\mthsym{\domsym#3}[#1][#2]}

\newcommand{\domelm}{d}
\newcommandx{\domElm}[3][1=, 2=, 3=]
	{\mthelm{\domelm#3}[#1][#2]}

\newcommand{\relfun}{rl}
\newcommandx{\relFun}[4][1=, 2=, 3=, 4=]
	{\mthargfun{\relfun#4}[#1][#2]{#3}}

\newcommand{\relstr}{RS}
\newcommandx{\RelStr}[5][1=, 2=, 3=, 4=, 5=]
	{\txtargname{\relstr#5{\small\argint{$[$}{#1}{$]$}}}[#2][#3]{#4}\xspace}

\newcommand{\relstrcls}{RS}
\newcommandx{\RelStrCls}[5][1=, 2=, 3=, 4=, 5=]
	{\mthset[#5]{\relstrcls#4\text{\txtname{\small\argint{$[$}{#1}{$]$}}}}[#2]%
	[#3]}

\newcommandx{\ordFun}[3][1=, 2=, 3=]
	{\mthempty{\argint{\left\vert}{#3}{\right\vert}}[#1][#2]}

\newcommandx{\sizFun}[3][1=, 2=, 3=]
	{\mthempty{\argint{\left\Vert}{#3}{\right\Vert}}[#1][#2]}

\newcommand{\verset}{Vr}
\newcommandx{\VerSet}[3][1=, 2=, 3=]
	{\mthset{\verset#3}[#1][#2]}

\newcommand{\versym}{v}
\newcommandx{\verSym}[3][1=, 2=, 3=]
	{\mthsym{\versym#3}[#1][#2]}

\newcommand{\verelm}{v}
\newcommandx{\verElm}[3][1=, 2=, 3=]
	{\mthelm{\verelm#3}[#1][#2]}

\newcommand{\edgrel}{Ed}
\newcommandx{\EdgRel}[3][1=, 2=, 3=]
	{\mthrel{\edgrel#3}[#1][#2]}

\newcommand{\edgsym}{e}
\newcommandx{\edgSym}[3][1=, 2=, 3=]
	{\mthsym{\edgsym#3}[#1][#2]}

\newcommand{\edgelm}{e}
\newcommandx{\edgElm}[3][1=, 2=, 3=]
	{\mthelm{\edgelm#3}[#1][#2]}

\newcommand{\orgfun}{or}
\newcommandx{\orgFun}[4][1=, 2=, 3=, 4=]
	{\mthargfun{\orgfun#4}[#1][#2]{#3}}

\newcommand{\desfun}{ds}
\newcommandx{\desFun}[4][1=, 2=, 3=, 4=]
	{\mthargfun{\desfun#4}[#1][#2]{#3}}

\newcommand{\grp}{Gr}
\newcommandx{\Grp}[5][1=, 2=, 3=, 4=, 5=]
	{\txtargname{\grp#5{\small\argint{$[$}{#1}{$]$}}}[#2][#3]{#4}\xspace}

\newcommand{\grpcls}{Gr}
\newcommandx{\GrpCls}[5][1=, 2=, 3=, 4=, 5=]
	{\mthset[#5]{\grpcls#4\text{\small\txtname{\argint{$[$}{#1}{$]$}}}}[#2][#3]}

\newcommand{\pthset}{Pth}
\newcommandx{\PthSet}[3][1=, 2=, 3=]
	{\mthset{\pthset#3}[#1][#2]}

\newcommand{\pthsym}{\pi}
\newcommandx{\pthSym}[3][1=, 2=, 3=]
	{\mthsym{\pthsym#3}[#1][#2]}

\newcommand{\pthelm}{\pi}
\newcommandx{\pthElm}[3][1=, 2=, 3=]
	{\mthelm{\pthelm#3}[#1][#2]}

\newcommand{\apset}{AP}
\newcommandx{\APSet}[3][1=, 2=, 3=]
	{\mthset{\apset#3}[#1][#2]}

\newcommand{\apsym}{p}
\newcommandx{\apSym}[3][1=, 2=, 3=]
	{\mthsym{\apsym#3}[#1][#2]}

\newcommand{\apelm}{p}
\newcommandx{\apElm}[3][1=, 2=, 3=]
	{\mthelm{\apelm#3}[#1][#2]}

\newcommand{\apfun}{ap}
\newcommandx{\apFun}[4][1=, 2=, 3=, 4=]
	{\mthargfun{\apfun#4}[#1][#2]{#3}}

\newcommand{\labgrp}{L\grp}
\newcommandx{\LabGrp}[5][1=, 2=, 3=, 4=, 5=]
	{\txtargname{\labgrp#5{\small\argint{$[$}{#1}{$]$}}}[#2][#3]{#4}\xspace}

\newcommand{\labgrpcls}{L\grpcls}
\newcommandx{\LabGrpCls}[5][1=, 2=, 3=, 4=, 5=]
	{\mthset[#5]{\labgrpcls#4\text{\small\txtname{\argint{$[$}{#1}{$]$}}}}[#2]%
	[#3]}

\newcommand{\trcset}{Trc}
\newcommandx{\TrcSet}[3][1=, 2=, 3=]
	{\mthset{\trcset#3}[#1][#2]}

\newcommand{\trcsym}{\varrho}
\newcommandx{\trcSym}[3][1=, 2=, 3=]
	{\mthsym{\trcsym#3}[#1][#2]}

\newcommand{\trcelm}{\varrho}
\newcommandx{\trcElm}[3][1=, 2=, 3=]
	{\mthelm{\trcelm#3}[#1][#2]}

\newcommand{\colset}{Cl}
\newcommandx{\ColSet}[3][1=, 2=, 3=]
	{\mthset{\colset#3}[#1][#2]}

\newcommand{\colsym}{c}
\newcommandx{\colSym}[3][1=, 2=, 3=]
	{\mthsym{\colsym#3}[#1][#2]}

\newcommand{\colelm}{c}
\newcommandx{\colElm}[3][1=, 2=, 3=]
	{\mthelm{\colelm#3}[#1][#2]}

\newcommand{\colfun}{cl}
\newcommandx{\colFun}[4][1=, 2=, 3=, 4=]
	{\mthargfun{\colfun#4}[#1][#2]{#3}}

\newcommand{\colgrp}{C\grp}
\newcommandx{\ColGrp}[5][1=, 2=, 3=, 4=, 5=]
	{\txtargname{\colgrp#5{\small\argint{$[$}{#1}{$]$}}}[#2][#3]{#4}\xspace}

\newcommand{\colgrpcls}{C\grpcls}
\newcommandx{\ColGrpCls}[5][1=, 2=, 3=, 4=, 5=]
	{\mthset[#5]{\colgrpcls#4\text{\small\txtname{\argint{$[$}{#1}{$]$}}}}[#2]%
	[#3]}

\newcommand{\wghset}{Wg}
\newcommandx{\WghSet}[3][1=, 2=, 3=]
	{\mthset{\wghset#3}[#1][#2]}

\newcommand{\wghsym}{w}
\newcommandx{\wghSym}[3][1=, 2=, 3=]
	{\mthsym{\wghsym#3}[#1][#2]}

\newcommand{\wghelm}{w}
\newcommandx{\wghElm}[3][1=, 2=, 3=]
	{\mthelm{\wghelm#3}[#1][#2]}

\newcommand{\wghfun}{wg}
\newcommandx{\wghFun}[4][1=, 2=, 3=, 4=]
	{\mthargfun{\wghfun#4}[#1][#2]{#3}}

\newcommand{\wghgrp}{W\grp}
\newcommandx{\WghGrp}[5][1=, 2=, 3=, 4=, 5=]
	{\txtargname{\wghgrp#5{\small\argint{$[$}{#1}{$]$}}}[#2][#3]{#4}\xspace}

\newcommand{\wghgrpcls}{W\grpcls}
\newcommandx{\WghGrpCls}[5][1=, 2=, 3=, 4=, 5=]
	{\mthset[#5]{\wghgrpcls#4\text{\small\txtname{\argint{$[$}{#1}{$]$}}}}[#2]%
	[#3]}

\newcommand{\gamkin}{2PT}

\newcommand{\plrset}{Pl}
\newcommandx{\PlrSet}[3][1=, 2=, 3=]
	{\mthset{\plrset#3}[#1][#2]}

\newcommand{\plrsym}{p}
\newcommandx{\plrSym}[3][1=, 2=, 3=]
	{\mthsym{\plrsym#3}[#1][#2]}

\newcommand{\plrelm}{p}
\newcommandx{\plrElm}[3][1=, 2=, 3=]
	{\mthelm{\plrelm#3}[#1][#2]}

\newcommand{\agnset}{Ag}
\newcommandx{\AgnSet}[3][1=, 2=, 3=]
	{\mthset{\agnset#3}[#1][#2]}

\newcommand{\agnsym}{a}
\newcommandx{\agnSym}[3][1=, 2=, 3=]
	{\mthsym{\agnsym#3}[#1][#2]}

\newcommand{\agnelm}{a}
\newcommandx{\agnElm}[3][1=, 2=, 3=]
	{\mthelm{\agnelm#3}[#1][#2]}

\newcommand{\movset}{Mv}
\newcommandx{\MovSet}[3][1=, 2=, 3=]
	{\mthset{\movset#3}[#1][#2]}

\newcommand{\movrel}{Mv}
\newcommandx{\MovRel}[3][1=, 2=, 3=]
	{\mthrel{\movrel#3}[#1][#2]}

\newcommand{\movsym}{m}
\newcommandx{\movSym}[3][1=, 2=, 3=]
	{\mthsym{\movsym#3}[#1][#2]}

\newcommand{\movelm}{m}
\newcommandx{\movElm}[3][1=, 2=, 3=]
	{\mthelm{\movelm#3}[#1][#2]}

\newcommand{\actset}{Ac}
\newcommandx{\ActSet}[3][1=, 2=, 3=]
	{\mthset{\actset#3}[#1][#2]}

\newcommand{\actrel}{Ac}
\newcommandx{\ActRel}[3][1=, 2=, 3=]
	{\mthrel{\actrel#3}[#1][#2]}

\newcommand{\actsym}{c}
\newcommandx{\actSym}[3][1=, 2=, 3=]
	{\mthsym{\actsym#3}[#1][#2]}

\newcommand{\actelm}{c}
\newcommandx{\actElm}[3][1=, 2=, 3=]
	{\mthelm{\actelm#3}[#1][#2]}

\newcommand{\decset}{Dc}
\newcommandx{\DecSet}[3][1=, 2=, 3=]
	{\mthset{\decset#3}[#1][#2]}

\newcommand{\decsym}{\delta}
\newcommandx{\decSym}[4][1=, 2=, 3=, 4=]
	{\mthargfun{\decsym#4}[#1][#2]{#3}}

\newcommand{\decelm}{\delta}
\newcommandx{\decElm}[4][1=, 2=, 3=, 4=]
	{\mthargfun{\decelm#4}[#1][#2]{#3}}

\newcommand{\posset}{Ps}
\newcommandx{\PosSet}[3][1=, 2=, 3=]
	{\mthset{\posset#3}[#1][#2]}
\newcommand{\fpossub}{0}
\newcommandx{\FPosSet}[3][1=, 2=, 3=]
	{\mthset{\posset#3}[\fpossub#1][#2]}
\newcommand{\spossub}{1}
\newcommandx{\SPosSet}[3][1=, 2=, 3=]
	{\mthset{\posset#3}[\spossub#1][#2]}

\newcommand{\possym}{v}
\newcommandx{\posSym}[3][1=, 2=, 3=]
	{\mthsym{\possym#3}[#1][#2]}
\newcommandx{\fposSym}[1][1=]
	{\posSym[\fpossub#1]}
\newcommandx{\sposSym}[1][1=]
	{\posSym[\spossub#1]}
\newcommand{\ipossub}{I}
\newcommandx{\iposSym}[1][1=]
	{\posSym[\ipossub#1]}

\newcommand{\poselm}{v}
\newcommandx{\posElm}[3][1=, 2=, 3=]
	{\mthelm{\poselm#3}[#1][#2]}
\newcommandx{\fposElm}[1][1=]
	{\posElm[\fpossub#1]}
\newcommandx{\sposElm}[1][1=]
	{\posElm[\spossub#1]}
\newcommandx{\iposElm}[1][1=]
	{\posElm[\ipossub#1]}

\newcommand{\sttset}{St}
\newcommandx{\SttSet}[3][1=, 2=, 3=]
	{\mthset{\sttset#3}[#1][#2]}
\newcommand{\fsttsub}{0}
\newcommandx{\FSttSet}[3][1=, 2=, 3=]
	{\mthset{\sttset#3}[\fsttsub#1][#2]}
\newcommand{\ssttsub}{1}
\newcommandx{\SSttSet}[3][1=, 2=, 3=]
	{\mthset{\sttset#3}[\ssttsub#1][#2]}

\newcommand{\sttsym}{s}
\newcommandx{\sttSym}[3][1=, 2=, 3=]
	{\mthsym{\sttsym#3}[#1][#2]}
\newcommandx{\fsttSym}[1][1=]
	{\sttSym[\fsttsub#1]}
\newcommandx{\ssttSym}[1][1=]
	{\sttSym[\ssttsub#1]}
\newcommand{\isttsub}{I}
\newcommandx{\isttSym}[1][1=]
	{\sttSym[\isttsub#1]}

\newcommand{\sttelm}{s}
\newcommandx{\sttElm}[3][1=, 2=, 3=]
	{\mthelm{\sttelm#3}[#1][#2]}
\newcommandx{\fsttElm}[1][1=]
	{\sttElm[\fsttsub#1]}
\newcommandx{\ssttElm}[1][1=]
	{\sttElm[\ssttsub#1]}
\newcommandx{\isttElm}[1][1=]
	{\sttElm[\isttsub#1]}

\newcommand{\plrfun}{pl}
\newcommandx{\plrFun}[4][1=, 2=, 3=, 4=]
	{\mthargfun{\plrfun#4}[#1][#2]{#3}}

\newcommand{\agnfun}{ag}
\newcommandx{\agnFun}[4][1=, 2=, 3=, 4=]
	{\mthargfun{\agnfun#4}[#1][#2]{#3}}

\newcommand{\movfun}{mv}
\newcommandx{\movFun}[4][1=, 2=, 3=, 4=]
	{\mthargfun{\movfun#4}[#1][#2]{#3}}

\newcommand{\actfun}{ac}
\newcommandx{\actFun}[4][1=, 2=, 3=, 4=]
	{\mthargfun{\actfun#4}[#1][#2]{#3}}

\newcommand{\decfun}{dc}
\newcommandx{\decFun}[4][1=, 2=, 3=, 4=]
	{\mthargfun{\decfun#4}[#1][#2]{#3}}

\newcommand{\trnfun}{tr}
\newcommandx{\trnFun}[4][1=, 2=, 3=, 4=]
	{\mthargfun{\trnfun#4}[#1][#2]{#3}}

\newcommand{\arn}{Ar}
\newcommandx{\Arn}[5][1=, 2=, 3=, 4=, 5=]
	{\txtargname{\arn#5{\small\argint{$[$}{#1}{$]$}}}[#2][#3]{#4}\xspace}

\newcommand{\arnname}{A}
\newcommand{\ArnName}
	{\mthname{\arnname}}

\newcommand{\arncls}{Ar}
\newcommandx{\ArnCls}[5][1=, 2=, 3=, 4=, 5=]
	{\mthset[#5]{\arncls#4\text{\small\txtname{\argint{$[$}{#1}{$]$}}}}[#2][#3]}

\newcommand{\ArnStr}[1][]
	{%
	\IfStrEqCase{\argdef{#1}{\gamkin}}
		{%
		{2PT}
			{\tuplecx{\FPosSet}{\SPosSet}{\MovRel}}%
		{MPC0}
			{\tupledx{\PlrSet}{\MovSet}{\PosSet}{\trnFun}}%
		{MPC1}
			{\tupleex{\PlrSet}{\MovSet}{\PosSet}{\decFun}{\trnFun}}%
		{MPC2}
			{\tuplefx{\PlrSet}{\MovSet}{\PosSet}{\plrFun}{\movFun}{\trnFun}}%
		{MPC3}
			{\tuplegx{\PlrSet}{\MovSet}{\PosSet}{\plrFun}{\movFun}{\decFun}{\trnFun}}%
		{2AT}
			{\tuplecx{\FSttSet}{\SSttSet}{\ActRel}}%
		{MAC0}
			{\tupledx{\AgnSet}{\ActSet}{\SttSet}{\trnFun}}%
		{MAC1}
			{\tupleex{\AgnSet}{\ActSet}{\SttSet}{\decFun}{\trnFun}}%
		{MAC2}
			{\tuplefx{\AgnSet}{\ActSet}{\SttSet}{\agnFun}{\actFun}{\trnFun}}%
		{MAC3}
			{\tuplegx{\AgnSet}{\ActSet}{\SttSet}{\agnFun}{\actFun}{\decFun}{\trnFun}}%
		}
		[\ensuremath{\clubsuit}]%
	}

\newcommand{\hstset}{Hst}
\newcommandx{\HstSet}[3][1=, 2=, 3=]
	{\mthset{\hstset#3}[#1][#2]}

\newcommand{\hstsym}{\rho}
\newcommandx{\hstSym}[3][1=, 2=, 3=]
	{\mthsym{\hstsym#3}[#1][#2]}

\newcommand{\hstelm}{\rho}
\newcommandx{\hstElm}[3][1=, 2=, 3=]
	{\mthelm{\hstelm#3}[#1][#2]}

\newcommand{\strset}{Str}
\newcommandx{\StrSet}[3][1=, 2=, 3=]
	{\mthset{\strset#3}[#1][#2]}

\newcommand{\strsym}{\sigma}
\newcommandx{\strSym}[4][1=, 2=, 3=, 4=]
	{\mthargfun{\strsym#4}[#1][#2]{#3}}

\newcommand{\strelm}{\sigma}
\newcommandx{\strElm}[4][1=, 2=, 3=, 4=]
	{\mthargfun{\strelm#4}[#1][#2]{#3}}

\newcommand{\prfset}{Prf}
\newcommandx{\PrfSet}[3][1=, 2=, 3=]
	{\mthset{\prfset#3}[#1][#2]}

\newcommand{\prfsym}{\xi}
\newcommandx{\prfSym}[4][1=, 2=, 3=, 4=]
	{\mthargfun{\prfsym#4}[#1][#2]{#3}}

\newcommandx{\prfElm}[4][1=, 2=, 3=, 4=]
	{\mthargfun{\prfsym#4}[#1][#2]{#3}}

\newcommand{\playfun}{play}
\newcommandx{\playFun}[4][1=, 2=, 3=, 4=]
	{\mthargfun{\playfun#4}[#1][#2]{#3}}

\newcommand{\labarn}{L\arn}
\newcommandx{\LabArn}[5][1=, 2=, 3=, 4=, 5=]
	{\txtargname{\labarn#5{\small\argint{$[$}{#1}{$]$}}}[#2][#3]{#4}\xspace}

\newcommand{\labarncls}{L\arncls}
\newcommandx{\LabArnCls}[5][1=, 2=, 3=, 4=, 5=]
	{\mthset[#5]{\labarncls#4\text{\small\txtname{\argint{$[$}{#1}{$]$}}}}[#2]%
	[#3]}

\newcommand{\colarn}{C\arn}
\newcommandx{\ColArn}[5][1=, 2=, 3=, 4=, 5=]
	{\txtargname{\colarn#5{\small\argint{$[$}{#1}{$]$}}}[#2][#3]{#4}\xspace}

\newcommand{\colarncls}{C\arncls}
\newcommandx{\ColArnCls}[5][1=, 2=, 3=, 4=, 5=]
	{\mthset[#5]{\colarncls#4\text{\small\txtname{\argint{$[$}{#1}{$]$}}}}[#2]%
	[#3]}

\newcommand{\wgharn}{W\arn}
\newcommandx{\WghArn}[5][1=, 2=, 3=, 4=, 5=]
	{\txtargname{\wgharn#5{\small\argint{$[$}{#1}{$]$}}}[#2][#3]{#4}\xspace}

\newcommand{\wgharncls}{W\arncls}
\newcommandx{\WghArnCls}[5][1=, 2=, 3=, 4=, 5=]
	{\mthset[#5]{\wgharncls#4\text{\small\txtname{\argint{$[$}{#1}{$]$}}}}[#2]%
	[#3]}

\newcommand{\winset}{Wn}
\newcommandx{\WinSet}[3][1=, 2=, 3=]
	{\mthset{\winset#3}[#1][#2]}

\newcommand{\prdset}{Pr}
\newcommandx{\PrdSet}[3][1=, 2=, 3=]
	{\mthset{\prdset#3}[#1][#2]}

\newcommand{\prdsym}{p}
\newcommandx{\prdSym}[3][1=, 2=, 3=]
	{\mthsym{\prdsym#3}[#1][#2]}

\newcommand{\prdelm}{p}
\newcommandx{\prdElm}[3][1=, 2=, 3=]
	{\mthelm{\prdelm#3}[#1][#2]}

\newcommand{\prdfun}{pr}
\newcommandx{\prdFun}[4][1=, 2=, 3=, 4=]
	{\mthargfun{\prdfun#4}[#1][#2]{#3}}

\newcommand{\extname}{E}
\newcommand{\ExtName}
	{\mthname{\extname}}

\newcommand{\extcls}{Ex}
\newcommandx{\ExtCls}[5][1=, 2=, 3=, 4=, 5=]
	{\mthset[#5]{\extcls#4\text{\small\txtname{\argint{$[$}{#1}{$]$}}}}[#2][#3]}

\newcommand{\conset}{Cn}
\newcommandx{\ConSet}[3][1=, 2=, 3=]
	{\mthset{\conset#3}[#1][#2]}

\newcommand{\consym}{\varphi}
\newcommandx{\conSym}[3][1=, 2=, 3=]
	{\mthsym{\consym#3}[#1][#2]}

\newcommand{\conelm}{\varphi}
\newcommandx{\conElm}[3][1=, 2=, 3=]
	{\mthelm{\conelm#3}[#1][#2]}

\newcommand{\schrel}{\models}
\newcommandx{\schRel}[4][1=, 2=, 3=, 4=]
	{\mthrel{\schrel#3}[#1][#2]}

\newcommand{\schcls}{Sc}
\newcommandx{\SchCls}[5][1=, 2=, 3=, 4=, 5=]
	{\mthset[#5]{\schcls#4\text{\small\txtname{\argint{$[$}{#1}{$]$}}}}[#2][#3]}

\newcommand{\gamcls}{Gm}
\newcommandx{\GamCls}[5][1=, 2=, 3=, 4=, 5=]
	{\mthset[#5]{\gamcls#4\text{\small\txtname{\argint{$[$}{#1}{$]$}}}}[#2][#3]}

\newcommandx{\GamStr}[1][1=]
	{%
	\StrLeft{\argdef{#1}{\gamkin}}{2}[\optgamkin]%
	\IfStrEqCase{\optgamkin}
		{%
		{2P}
			{\gamstrauxtp}%
		{MP}
			{\gamstrauxmp}%
		{2A}
			{\gamstrauxta}%
		{MA}
			{\gamstrauxma}%
		}
		[\ensuremath{\clubsuit}]%
	}
\newcommandx{\gamstrauxtp}[5][1=, 2=, 3=, 4=, 5=]
	{\tuplecx{\ArnName}{\iposElm}{\WinSet}[#3][#4][#5][#1][#2]}
\newcommandx{\gamstrauxmp}[5][1=, 2=, 3=, 4=, 5=]
	{\tuplecx{\ExtName}{\iposElm}{\conElm}[#3][#4][#5][#1][#2]}
\newcommandx{\gamstrauxta}[5][1=, 2=, 3=, 4=, 5=]
	{\tuplecx{\ArnName}{\isttElm}{\WinSet}[#3][#4][#5][#1][#2]}
\newcommandx{\gamstrauxma}[5][1=, 2=, 3=, 4=, 5=]
	{\tuplecx{\ExtName}{\isttElm}{\conElm}[#3][#4][#5][#1][#2]}

\newcommand{\worset}{W}
\newcommandx{\WorSet}[3][1=, 2=, 3=]
	{\mthset{\worset#3}[#1][#2]}

\newcommand{\worsym}{w}
\newcommandx{\worSym}[3][1=, 2=, 3=]
	{\mthsym{\worsym#3}[#1][#2]}

\newcommand{\worelm}{w}
\newcommandx{\worElm}[3][1=, 2=, 3=]
	{\mthelm{\worelm#3}[#1][#2]}

\newcommand{\trnrel}{R}
\newcommandx{\TrnRel}[3][1=, 2=, 3=]
	{\mthrel{\trnrel#3}[#1][#2]}

\newcommand{\trnsym}{r}
\newcommandx{\trnSym}[3][1=, 2=, 3=]
	{\mthsym{\trnsym#3}[#1][#2]}

\newcommand{\trnelm}{r}
\newcommandx{\trnElm}[3][1=, 2=, 3=]
	{\mthelm{\trnelm#3}[#1][#2]}

\newcommand{\labfun}{L}
\newcommandx{\labFun}[4][1=, 2=, 3=, 4=]
	{\mthargfun{\labfun#4}[#1][#2]{#3}}

\newcommand{\krpstr}{KS}
\newcommandx{\KrpStr}[5][1=, 2=, 3=, 4=, 5=]
	{\txtargname{\krpstr#5{\small\argint{$[$}{#1}{$]$}}}[#2][#3]{#4}\xspace}

\newcommand{\krpstrcls}{KS}
\newcommandx{\KrpStrCls}[5][1=, 2=, 3=, 4=, 5=]
	{\mthset[#5]{\krpstrcls#4\text{\small\txtname{\argint{$[$}{#1}{$]$}}}}[#2]%
	[#3]}

\newcommand{\trkset}{Trk}
\newcommandx{\TrkSet}[3][1=, 2=, 3=]
	{\mthset{\trkset#3}[#1][#2]}

\newcommand{\trksym}{\rho}
\newcommandx{\trkSym}[3][1=, 2=, 3=]
	{\mthsym{\trksym#3}[#1][#2]}

\newcommand{\trkelm}{\rho}
\newcommandx{\trkElm}[3][1=, 2=, 3=]
	{\mthelm{\trkelm#3}[#1][#2]}

\newcommand{\krptree}{KT}
\newcommandx{\KrpTree}[5][1=, 2=, 3=, 4=, 5=]
	{\txtargname{\krptree#5{\small\argint{$[$}{#1}{$]$}}}[#2][#3]{#4}\xspace}

\newcommand{\krptreecls}{KT}
\newcommandx{\KrpTreeCls}[5][1=, 2=, 3=, 4=, 5=]
	{\mthset[#5]{\krptreecls#4\text{\small\txtname{\argint{$[$}{#1}{$]$}}}}[#2]%
	[#3]}

\newcommand{\dirset}{Dir}
\newcommandx{\DirSet}[3][1=, 2=, 3=]
	{\mthset{\dirset#3}[#1][#2]}

\newcommand{\dirsym}{d}
\newcommandx{\dirSym}[3][1=, 2=, 3=]
	{\mthsym{\dirsym#3}[#1][#2]}

\newcommand{\direlm}{d}
\newcommandx{\dirElm}[3][1=, 2=, 3=]
	{\mthelm{\direlm#3}[#1][#2]}

\newcommand{\unwfun}{unw}
\newcommandx{\unwFun}[4][1=, 2=, 3=, 4=]
	{\mthargfun{\unwfun#4}[#1][#2]{#3}}

\newcommand{\congamstrkin}{MAC0}

\newcommand{\congamstr}{CGS}
\newcommandx{\ConGamStr}[5][1=, 2=, 3=, 4=, 5=]
	{\txtargname{\congamstr#5{\small\argint{$[$}{#1}{$]$}}}[#2][#3]{#4}\xspace}
\newcommand{\CGS}{\ConGamStr}

\newcommand{\epicongamstr}{ECGS}
\newcommandx{\EpiConGamStr}[5][1=, 2=, 3=, 4=, 5=]
	{\txtargname{\epicongamstr#5{\small\argint{$[$}{#1}{$]$}}}[#2][#3]{#4}\xspace}

\newcommand{\congamstrname}{G}
\newcommand{\ConGamStrName}
	{\mthname{\congamstrname}}
\newcommand{\CGSName}{\ConGamStrName}

\newcommandx{\ConGamStrCls}[5][1=, 2=, 3=, 4=, 5=]
	{\mthset[#5]{\arncls#4\text{\small\txtname{\argint{$[$}{#1}{$]$}}}}[#2][#3]}

\newcommandx{\ConGamStrStr}[1][1=]
	{%
	\IfStrEqCase{\argdef{#1}{\congamstrkin}}
		{%
		{IP}
			{\congamstrstrauxip}%
		{2PT}
			{\congamstrstrauxpt}%
		{MPC0}
			{\congamstrstrauxpca}%
		{MPC1}
			{\congamstrstrauxpcb}%
		{MPC2}
			{\congamstrstrauxpcc}%
		{MPC3}
			{\congamstrstrauxpcd}%
		{IA}
			{\congamstrstrauxia}%
		{2AT}
			{\congamstrstrauxat}%
		{MAC0}
			{\congamstrstrauxaca}%
		{MAC1}
			{\congamstrstrauxacb}%
		{MAC2}
			{\congamstrstrauxacc}%
		{MAC3}
			{\congamstrstrauxacd}%
		}
		[\ensuremath{\clubsuit}]%
	}
\newcommandx{\congamstrstrauxip}[3][1=, 2=, 3=]
	{%
	\def\defini{#1}%
	\def\defsubscr{#2}%
	\def\defsupscr{#3}%
	\congamstrstrauxxip%
	}
\newcommandx{\congamstrstrauxxip}[3][1=, 2=, 3=]
	{%
	\tupledx{\ArnName}{\APSet}{\apFun}{\iposElm}%
		[\defsubscr][\defsupscr][#1][#2][#3][\defini]%
	}
\newcommandx{\congamstrstrauxpt}[3][1=, 2=, 3=]
	{%
	\def\defini{#1}%
	\def\defsubscr{#2}%
	\def\defsupscr{#3}%
	\congamstrstrauxxpt%
	}
\newcommandx{\congamstrstrauxxpt}[5][1=, 2=, 3=, 4=, 5=]
	{%
	\tuplefx{\APSet}{\FPosSet}{\SPosSet}{\MovRel}{\apFun}{\iposElm}%
		[\defsubscr][\defsupscr][#1][#2][#3][#4][#5][\defini]%
	}
\newcommandx{\congamstrstrauxpca}[3][1=, 2=, 3=]
	{%
	\def\defini{#1}%
	\def\defsubscr{#2}%
	\def\defsupscr{#3}%
	\congamstrstrauxxpca%
	}
\newcommandx{\congamstrstrauxxpca}[6][1=, 2=, 3=, 4=, 5=, 6=]
	{%
	\tuplegx{\APSet}{\PlrSet}{\MovSet}{\PosSet}{\trnFun}{\apFun}{\iposElm}%
		[\defsubscr][\defsupscr][#1][#2][#3][#4][#5][#6][\defini]%
	}
\newcommandx{\congamstrstrauxpcb}[3][1=, 2=, 3=]
	{%
	\def\defini{#1}%
	\def\defsubscr{#2}%
	\def\defsupscr{#3}%
	\congamstrstrauxxpcb%
	}
\newcommandx{\congamstrstrauxxpcb}[7][1=, 2=, 3=, 4=, 5=, 6=, 7=]
	{%
	\tuplehx{\APSet}{\PlrSet}{\MovSet}{\PosSet}{\decFun}{\trnFun}{\apFun}%
		{\iposElm}%
		[\defsubscr][\defsupscr][#1][#2][#3][#4][#5][#6][#7][\defini]%
	}
\newcommandx{\congamstrstrauxpcc}[3][1=, 2=, 3=]
	{%
	\def\defini{#1}%
	\def\defsubscr{#2}%
	\def\defsupscr{#3}%
	\congamstrstrauxxpcc%
	}
\newcommandx{\congamstrstrauxxpcc}[8][1=, 2=, 3=, 4=, 5=, 6=, 7=, 8=]
	{%
	\tupleix{\APSet}{\PlrSet}{\MovSet}{\PosSet}{\plrFun}{\movFun}{\trnFun}%
		{\apFun}{\iposElm}%
		[\defsubscr][\defsupscr][#1][#2][#3][#4][#5][#6][#7][#8][\defini]%
	}
\newcommandx{\congamstrstrauxpcd}[3][1=, 2=, 3=]
	{%
	\def\defini{#1}%
	\def\defsubscr{#2}%
	\def\defsupscr{#3}%
	\congamstrstrauxxpcd%
	}
\newcommandx{\congamstrstrauxxpcd}[9][1=, 2=, 3=, 4=, 5=, 6=, 7=, 8=, 9=]
	{%
	\tuplejx{\APSet}{\PlrSet}{\MovSet}{\PosSet}{\plrFun}{\movFun}{\decFun}%
	{\trnFun}{\apFun}{\iposElm}%
		[\defsubscr][\defsupscr][#1][#2][#3][#4][#5][#6][#7][#8][#9][\defini]%
	}
\newcommandx{\congamstrstrauxia}[3][1=, 2=, 3=]
	{%
	\def\defini{#1}%
	\def\defsubscr{#2}%
	\def\defsupscr{#3}%
	\congamstrstrauxxia%
	}
\newcommandx{\congamstrstrauxxia}[3][1=, 2=, 3=]
	{%
	\tupledx{\ArnName}{\APSet}{\apFun}{\isttElm}%
		[\defsubscr][\defsupscr][#1][#2][#3][\defini]%
	}
\newcommandx{\congamstrstrauxat}[3][1=, 2=, 3=]
	{%
	\def\defini{#1}%
	\def\defsubscr{#2}%
	\def\defsupscr{#3}%
	\congamstrstrauxxat%
	}
\newcommandx{\congamstrstrauxxat}[5][1=, 2=, 3=, 4=, 5=]
	{%
	\tuplefx{\APSet}{\FSttSet}{\SSttSet}{\ActRel}{\apFun}{\isttElm}%
		[\defsubscr][\defsupscr][#1][#2][#3][#4][#5][\defini]%
	}
\newcommandx{\congamstrstrauxaca}[3][1=, 2=, 3=]
	{%
	\def\defini{#1}%
	\def\defsubscr{#2}%
	\def\defsupscr{#3}%
	\congamstrstrauxxaca%
	}
\newcommandx{\congamstrstrauxxaca}[6][1=, 2=, 3=, 4=, 5=, 6=]
	{%
	\tuplegx{\APSet}{\AgnSet}{\ActSet}{\SttSet}{\trnFun}{\apFun}{\isttElm}%
		[\defsubscr][\defsupscr][#1][#2][#3][#4][#5][#6][\defini]%
	}
\newcommandx{\congamstrstrauxacb}[3][1=, 2=, 3=]
	{%
	\def\defini{#1}%
	\def\defsubscr{#2}%
	\def\defsupscr{#3}%
	\congamstrstrauxxacb%
	}
\newcommandx{\congamstrstrauxxacb}[7][1=, 2=, 3=, 4=, 5=, 6=, 7=]
	{%
	\tuplehx{\APSet}{\AgnSet}{\ActSet}{\SttSet}{\decFun}{\trnFun}{\apFun}%
		{\isttElm}%
		[\defsubscr][\defsupscr][#1][#2][#3][#4][#5][#6][#7][\defini]%
	}
\newcommandx{\congamstrstrauxacc}[3][1=, 2=, 3=]
	{%
	\def\defini{#1}%
	\def\defsubscr{#2}%
	\def\defsupscr{#3}%
	\congamstrstrauxxacc%
	}
\newcommandx{\congamstrstrauxxacc}[8][1=, 2=, 3=, 4=, 5=, 6=, 7=, 8=]
	{%
	\tupleix{\APSet}{\AgnSet}{\ActSet}{\SttSet}{\agnFun}{\actFun}{\trnFun}%
		{\apFun}{\isttElm}%
		[\defsubscr][\defsupscr][#1][#2][#3][#4][#5][#6][#7][#8][\defini]%
	}
\newcommandx{\congamstrstrauxacd}[3][1=, 2=, 3=]
	{%
	\def\defini{#1}%
	\def\defsubscr{#2}%
	\def\defsupscr{#3}%
	\congamstrstrauxxacd%
	}
\newcommandx{\congamstrstrauxxacd}[9][1=, 2=, 3=, 4=, 5=, 6=, 7=, 8=, 9=]
	{%
	\tuplejx{\APSet}{\AgnSet}{\ActSet}{\SttSet}{\agnFun}{\actFun}{\decFun}%
		{\trnFun}{\apFun}{\isttElm}%
		[\defsubscr][\defsupscr][#1][#2][#3][#4][#5][#6][#7][#8][#9][\defini]%
	}

\newcommand{\turgamstr}{TGS}
\newcommandx{\TurGamStr}[5][1=, 2=, 3=, 4=, 5=]
	{\txtargname{\turgamstr#5{\small\argint{$[$}{#1}{$]$}}}[#2][#3]{#4}\xspace}

\newcommandx{\TurGamStrCls}[5][1=, 2=, 3=, 4=, 5=]
	{\mthset[#5]{\arncls#4\text{\small\txtname{\argint{$[$}{#1}{$]$}}}}[#2][#3]}

\newcommandx{\TurGamStrStr}[1][1=]
	{%
	\IfStrEqCase{\argdef{#1}{\turgamstrkin}}
		{%
		{IP}
			{\turgamstrstrauxip}%
		{2PT}
			{\turgamstrstrauxpt}%
		{MPC0}
			{\turgamstrstrauxpca}%
		{MPC1}
			{\turgamstrstrauxpcb}%
		{MPC2}
			{\turgamstrstrauxpcc}%
		{MPC3}
			{\turgamstrstrauxpcd}%
		{IA}
			{\turgamstrstrauxia}%
		{2AT}
			{\turgamstrstrauxat}%
		{MAC0}
			{\turgamstrstrauxaca}%
		{MAC1}
			{\turgamstrstrauxacb}%
		{MAC2}
			{\turgamstrstrauxacc}%
		{MAC3}
			{\turgamstrstrauxacd}%
		}
		[\ensuremath{\clubsuit}]%
	}
\newcommandx{\turgamstrstrauxip}[3][1=, 2=, 3=]
	{%
	\def\defini{#1}%
	\def\defsubscr{#2}%
	\def\defsupscr{#3}%
	\turgamstrstrauxxip%
	}
\newcommandx{\turgamstrstrauxxip}[3][1=, 2=, 3=]
	{%
	\tupledx{\ArnName}{\APSet}{\apFun}{\iposElm}%
		[\defsubscr][\defsupscr][#1][#2][#3][\defini]%
	}
\newcommandx{\turgamstrstrauxpt}[3][1=, 2=, 3=]
	{%
	\def\defini{#1}%
	\def\defsubscr{#2}%
	\def\defsupscr{#3}%
	\turgamstrstrauxxpt%
	}
\newcommandx{\turgamstrstrauxxpt}[5][1=, 2=, 3=, 4=, 5=]
	{%
	\tuplefx{\APSet}{\FPosSet}{\SPosSet}{\MovRel}{\apFun}{\iposElm}%
		[\defsubscr][\defsupscr][#1][#2][#3][#4][#5][\defini]%
	}
\newcommandx{\turgamstrstrauxpca}[3][1=, 2=, 3=]
	{%
	\def\defini{#1}%
	\def\defsubscr{#2}%
	\def\defsupscr{#3}%
	\turgamstrstrauxxpca%
	}
\newcommandx{\turgamstrstrauxxpca}[6][1=, 2=, 3=, 4=, 5=, 6=]
	{%
	\tuplegx{\APSet}{\PlrSet}{\MovSet}{\PosSet}{\trnFun}{\apFun}{\iposElm}%
		[\defsubscr][\defsupscr][#1][#2][#3][#4][#5][#6][\defini]%
	}
\newcommandx{\turgamstrstrauxpcb}[3][1=, 2=, 3=]
	{%
	\def\defini{#1}%
	\def\defsubscr{#2}%
	\def\defsupscr{#3}%
	\turgamstrstrauxxpcb%
	}
\newcommandx{\turgamstrstrauxxpcb}[7][1=, 2=, 3=, 4=, 5=, 6=, 7=]
	{%
	\tuplehx{\APSet}{\PlrSet}{\MovSet}{\PosSet}{\decFun}{\trnFun}{\apFun}%
		{\iposElm}%
		[\defsubscr][\defsupscr][#1][#2][#3][#4][#5][#6][#7][\defini]%
	}
\newcommandx{\turgamstrstrauxpcc}[3][1=, 2=, 3=]
	{%
	\def\defini{#1}%
	\def\defsubscr{#2}%
	\def\defsupscr{#3}%
	\turgamstrstrauxxpcc%
	}
\newcommandx{\turgamstrstrauxxpcc}[8][1=, 2=, 3=, 4=, 5=, 6=, 7=, 8=]
	{%
	\tupleix{\APSet}{\PlrSet}{\MovSet}{\PosSet}{\plrFun}{\movFun}{\trnFun}%
		{\apFun}{\iposElm}%
		[\defsubscr][\defsupscr][#1][#2][#3][#4][#5][#6][#7][#8][\defini]%
	}
\newcommandx{\turgamstrstrauxpcd}[3][1=, 2=, 3=]
	{%
	\def\defini{#1}%
	\def\defsubscr{#2}%
	\def\defsupscr{#3}%
	\turgamstrstrauxxpcd%
	}
\newcommandx{\turgamstrstrauxxpcd}[9][1=, 2=, 3=, 4=, 5=, 6=, 7=, 8=, 9=]
	{%
	\tuplejx{\APSet}{\PlrSet}{\MovSet}{\PosSet}{\plrFun}{\movFun}{\decFun}%
	{\trnFun}{\apFun}{\iposElm}%
		[\defsubscr][\defsupscr][#1][#2][#3][#4][#5][#6][#7][#8][#9][\defini]%
	}
\newcommandx{\turgamstrstrauxia}[3][1=, 2=, 3=]
	{%
	\def\defini{#1}%
	\def\defsubscr{#2}%
	\def\defsupscr{#3}%
	\turgamstrstrauxxia%
	}
\newcommandx{\turgamstrstrauxxia}[3][1=, 2=, 3=]
	{%
	\tupledx{\ArnName}{\APSet}{\apFun}{\isttElm}%
		[\defsubscr][\defsupscr][#1][#2][#3][\defini]%
	}
\newcommandx{\turgamstrstrauxat}[3][1=, 2=, 3=]
	{%
	\def\defini{#1}%
	\def\defsubscr{#2}%
	\def\defsupscr{#3}%
	\turgamstrstrauxxat%
	}
\newcommandx{\turgamstrstrauxxat}[5][1=, 2=, 3=, 4=, 5=]
	{%
	\tuplefx{\APSet}{\FSttSet}{\SSttSet}{\ActRel}{\apFun}{\isttElm}%
		[\defsubscr][\defsupscr][#1][#2][#3][#4][#5][\defini]%
	}
\newcommandx{\turgamstrstrauxaca}[3][1=, 2=, 3=]
	{%
	\def\defini{#1}%
	\def\defsubscr{#2}%
	\def\defsupscr{#3}%
	\turgamstrstrauxxaca%
	}
\newcommandx{\turgamstrstrauxxaca}[6][1=, 2=, 3=, 4=, 5=, 6=]
	{%
	\tuplegx{\APSet}{\AgnSet}{\ActSet}{\SttSet}{\trnFun}{\apFun}{\isttElm}%
		[\defsubscr][\defsupscr][#1][#2][#3][#4][#5][#6][\defini]%
	}
\newcommandx{\turgamstrstrauxacb}[3][1=, 2=, 3=]
	{%
	\def\defini{#1}%
	\def\defsubscr{#2}%
	\def\defsupscr{#3}%
	\turgamstrstrauxxacb%
	}
\newcommandx{\turgamstrstrauxxacb}[7][1=, 2=, 3=, 4=, 5=, 6=, 7=]
	{%
	\tuplehx{\APSet}{\AgnSet}{\ActSet}{\SttSet}{\decFun}{\trnFun}{\apFun}%
		{\isttElm}%
		[\defsubscr][\defsupscr][#1][#2][#3][#4][#5][#6][#7][\defini]%
	}
\newcommandx{\turgamstrstrauxacc}[3][1=, 2=, 3=]
	{%
	\def\defini{#1}%
	\def\defsubscr{#2}%
	\def\defsupscr{#3}%
	\turgamstrstrauxxacc%
	}
\newcommandx{\turgamstrstrauxxacc}[8][1=, 2=, 3=, 4=, 5=, 6=, 7=, 8=]
	{%
	\tupleix{\APSet}{\AgnSet}{\ActSet}{\SttSet}{\agnFun}{\actFun}{\trnFun}%
		{\apFun}{\isttElm}%
		[\defsubscr][\defsupscr][#1][#2][#3][#4][#5][#6][#7][#8][\defini]%
	}
\newcommandx{\turgamstrstrauxacd}[3][1=, 2=, 3=]
	{%
	\def\defini{#1}%
	\def\defsubscr{#2}%
	\def\defsupscr{#3}%
	\turgamstrstrauxxacd%
	}
\newcommandx{\turgamstrstrauxxacd}[9][1=, 2=, 3=, 4=, 5=, 6=, 7=, 8=, 9=]
	{%
	\tuplejx{\APSet}{\AgnSet}{\ActSet}{\SttSet}{\agnFun}{\actFun}{\decFun}%
		{\trnFun}{\apFun}{\isttElm}%
		[\defsubscr][\defsupscr][#1][#2][#3][#4][#5][#6][#7][#8][#9][\defini]%
	}

\newcommand{\trntabkin}{D}

\newcommand{\symset}{Sm}
\newcommandx{\SymSet}[3][1=, 2=, 3=]
	{\mthset{\symset#3}[#1][#2]}

\newcommand{\symsym}{\ell}
\newcommandx{\symSym}[3][1=, 2=, 3=]
	{\mthsym{\symsym#3}[#1][#2]}

\newcommand{\symelm}{\ell}
\newcommandx{\symElm}[3][1=, 2=, 3=]
	{\mthelm{\symelm#3}[#1][#2]}

\newcommand{\DSttSet}[1][]
	{\SttSet[\Delta#1]}

\newcommand{\ESttSet}[1][]
	{\SttSet[\exists#1]}

\newcommand{\ASttSet}[1][]
	{\SttSet[\forall#1]}

\newcommand{\trntab}{tt}
\newcommandx{\TrnTab}[5][1=, 2=, 3=, 4=, 5=]
	{\txtargname{\trntab#5{\small\argint{$[$}{#1}{$]$}}}[#2][#3]{#4}\xspace}

\newcommand{\trntabcls}{TT}
\newcommandx{\TrnTabCls}[5][1=, 2=, 3=, 4=, 5=]
	{\mthset[#5]{\trntabcls#4\text{\txtname{\small\argint{$[$}{#1}{$]$}}}}[#2]%
	[#3]}

\newcommand{\TrnTabStr}[1][]
	{%
	\IfStrEqCase{\argdef{#1}{\trntabkin}}
		{%
		{D}{\tuplecx{\SymSet}{\SttSet}{\trnFun}}%
		{N}{\tupledx{\SymSet}{\DSttSet}{\ESttSet}{\trnFun}}%
		{U}{\tupledx{\SymSet}{\DSttSet}{\ASttSet}{\trnFun}}%
		{A}{\tupleex{\SymSet}{\DSttSet}{\ESttSet}{\ASttSet}{\trnFun}}%
		}
		[\ensuremath{\clubsuit}]%
	}

\newcommandx{\PC}[5][1=, 2=, 3=, 4=, 5=]
	{\txtargname{PC#5{\small\argint{$[$}{#1}{$]$}}}[#2][#3]{#4}\xspace}

\newcommand{\bcset}{BC}
\newcommandx{\BCSet}[4][1=, 2=, 3=, 4=]
	{\mthset[3]{\bcset#4}[#1][#2]{#3}}

\newcommand{\bcelm}{\eta}
\newcommandx{\bcElm}[3][1=, 2=, 3=]
	{\mthelm{\bcelm#3}[#1][#2]}

\newcommand{\acset}{AC}
\newcommandx{\ACSet}[4][1=, 2=, 3=, 4=]
	{\mthset[3]{\acset#4}[#1][#2]{#3}}

\newcommand{\acelm}{\eta}
\newcommandx{\acElm}[3][1=, 2=, 3=]
	{\mthelm{\acelm#3}[#1][#2]}

\newcommandx{\QBF}[5][1=, 2=, 3=, 4=, 5=]
	{\txtargname{QBF#5{\small\argint{$[$}{#1}{$]$}}}[#2][#3]{#4}\xspace}

\newcommandx{\FOL}[5][1=, 2=, 3=, 4=, 5=]
	{\txtargname{FOL#5{\small\argint{$[$}{#1}{$]$}}}[#2][#3]{#4}\xspace}

\newcommand{\varset}{Vr}
\newcommandx{\VarSet}[3][1=, 2=, 3=]
	{\mthset{\varset#3}[#1][#2]}

\newcommand{\varsym}{x}
\newcommandx{\varSym}[3][1=, 2=, 3=]
	{\mthsym{\varsym#3}[#1][#2]}

\newcommand{\varelm}{x}
\newcommandx{\varElm}[3][1=, 2=, 3=]
	{\mthelm{\varelm#3}[#1][#2]}

\newcommand{\varfun}{vr}
\newcommandx{\varFun}[4][1=, 2=, 3=, 4=]
	{\mthargfun{\varfun#4}[#1][#2]{#3}}

\newcommand{\freefun}{free}
\newcommand{\freeFun}
	{\mthargfun{\freefun}}

\newcommand{\qntset}{Qn}
\newcommandx{\QntSet}[3][1=, 2=, 3=]
	{\mthset{\qntset#3}[#1][#2]}

\newcommand{\qntsym}{\wp}
\newcommandx{\qntSym}[3][1=, 2=, 3=]
	{\mthsym{\qntsym#3}[#1][#2]}

\newcommand{\qntelm}{\wp}
\newcommandx{\qntElm}[3][1=, 2=, 3=]
	{\mthelm{\qntelm#3}[#1][#2]}

\newcommand{\qntfun}{qnt}
\newcommandx{\qntFun}[4][1=, 2=, 3=, 4=]
	{\mthargfun{\qntfun#4}[#1][#2]{#3}}

\newcommand{\bndset}{Bn}
\newcommandx{\BndSet}[3][1=, 2=, 3=]
	{\mthset{\bndset#3}[#1][#2]}

\newcommand{\bndsym}{\flat}
\newcommandx{\bndSym}[3][1=, 2=, 3=]
	{\mthsym{\bndsym#3}[#1][#2]}

\newcommand{\bndelm}{\flat}
\newcommandx{\bndElm}[3][1=, 2=, 3=]
	{\mthelm{\bndelm#3}[#1][#2]}

\newcommand{\bndfun}{bnd}
\newcommandx{\bndFun}[4][1=, 2=, 3=, 4=]
	{\mthargfun{\bndfun#4}[#1][#2]{#3}}

\newcommand{\depset}{\Delta}
\newcommandx{\DepSet}[3][1=, 2=, 3=]
	{\mthset{\depset#3}[#1][#2]}

\newcommand{\denfun}{den}
\newcommandx{\denFun}[4][1=, 2=, 3=, 4=]
	{\mthargfun{\denfun#4}[#1][#2]{#3}}

\newcommand{\asgset}{Asg}
\newcommandx{\AsgSet}[3][1=, 2=, 3=]
	{\mthset{\asgset#3}[#1][#2]}

\newcommand{\asgfun}{\chi}
\newcommandx{\asgFun}[4][1=, 2=, 3=, 4=]
	{\mthargfun{\asgfun#4}[#1][#2]{#3}}

\newcommand{\smset}{SM}
\newcommandx{\SMSet}[3][1=, 2=, 3=]
	{\mthset{\smset#3}[#1][#2]}

\newcommand{\smfun}{\delta}
\newcommandx{\smFun}[4][1=, 2=, 3=, 4=]
	{\mthargfun{\smfun#4}[#1][#2]{#3}}

\newcommand{\cmset}{CM}
\newcommandx{\CMSet}[3][1=, 2=, 3=]
	{\mthset{\cmset#3}[#1][#2]}

\newcommand{\cmfun}{\gamma}
\newcommandx{\cmFun}[4][1=, 2=, 3=, 4=]
	{\mthargfun{\cmfun#4}[#1][#2]{#3}}

\newcommand{\schset}{Sch}
\newcommandx{\SchSet}[3][1=, 2=, 3=]
	{\mthset{\schset#3}[#1][#2]}

\newcommand{\schsym}{\sigma}
\newcommandx{\schSym}[3][1=, 2=, 3=]
	{\mthsym{\schsym#3}[#1][#2]}

\newcommand{\schelm}{\sigma}
\newcommandx{\schElm}[3][1=, 2=, 3=]
	{\mthelm{\schelm#3}[#1][#2]}

\newcommand{\entset}{Ent}
\newcommandx{\EntSet}[4][1=, 2=, 3=, 4=]
	{\mthset{\entset#4}[#1][#2]{#3}}

\newcommand{\entfun}{ent}
\newcommandx{\entFun}[4][1=, 2=, 3=, 4=]
	{\mthargfun{\entfun#4}[#1][#2]{#3}}

\newcommandx{\SOL}[5][1=, 2=, 3=, 4=, 5=]
	{\txtargname{SOL#5{\small\argint{$[$}{#1}{$]$}}}[#2][#3]{#4}\xspace}

\newcommandx{\TL}[5][1=, 2=, 3=, 4=, 5=]
	{\txtargname{TL#5{\small\argint{$[$}{#1}{$]$}}}[#2][#3]{#4}\xspace}

\newcommandx{\PL}[5][1=, 2=, 3=, 4=, 5=]
	{\txtargname{PL#5{\small\argint{$[$}{#1}{$]$}}}[#2][#3]{#4}\xspace}

\newcommand{\fvarset}{FVr}
\newcommandx{\FVarSet}[3][1=, 2=, 3=]
	{\mthset{\fvarset#3}[#1][#2]}

\newcommand{\fvarsym}{x}
\newcommandx{\fvarSym}[3][1=, 2=, 3=]
	{\mthsym{\fvarsym#3}[#1][#2]}

\newcommand{\fvarelm}{x}
\newcommandx{\fvarElm}[3][1=, 2=, 3=]
	{\mthelm{\fvarelm#3}[#1][#2]}

\newcommand{\fvarfun}{fvr}
\newcommandx{\fvarFun}[4][1=, 2=, 3=, 4=]
	{\mthargfun{\fvarfun#4}[#1][#2]{#3}}

\newcommand{\svarset}{SVr}
\newcommandx{\SVarSet}[3][1=, 2=, 3=]
	{\mthset{\svarset#3}[#1][#2]}

\newcommand{\svarsym}{X}
\newcommandx{\svarSym}[3][1=, 2=, 3=]
	{\mthsym{\svarsym#3}[#1][#2]}

\newcommand{\svarelm}{X}
\newcommandx{\svarElm}[3][1=, 2=, 3=]
	{\mthelm{\svarelm#3}[#1][#2]}

\newcommand{\svarfun}{svr}
\newcommandx{\svarFun}[4][1=, 2=, 3=, 4=]
	{\mthargfun{\svarfun#4}[#1][#2]{#3}}

\newcommandx{\ML}[5][1=, 2=, 3=, 4=, 5=]
	{\txtargname{ML#5{\small\argint{$[$}{#1}{$]$}}}[#2][#3]{#4}\xspace}

\newcommandx{\MC}[5][1=, 2=, 3=, 4=, 5=]
	{\txtargname{$\mu$-Calculus#5{\small\argint{$[$}{#1}{$]$}}}[#2][#3]{#4}\xspace}

\newcommandx{\LTL}[5][1=, 2=, 3=, 4=, 5=]
	{\txtargname{LTL#5{\small\argint{$[$}{#1}{$]$}}}[#2][#3]{#4}\xspace}

\newcommandx{\PTL}[5][1=, 2=, 3=, 4=, 5=]
	{\txtargname{PTL#5{\small\argint{$[$}{#1}{$]$}}}[#2][#3]{#4}\xspace}

\newcommand{\R}
	{\mthsym{R}}

\newcommand{\B}
	{\mthsym{B}}

\newcommandx{\CTL}[5][1=, 2=, 3=, 4=, 5=]
	{\txtargname{CTL#5{\small\argint{$[$}{#1}{$]$}}}[#2][#3]{#4}\xspace}

\newcommandx{\CTLP}[5][1=, 2=, 3=, 4=, 5=]
	{\txtargname{CTL$^{+}$#5{\small\argint{$[$}{#1}{$]$}}}[#2][#3]{#4}\xspace}

\newcommandx{\CTLS}[5][1=, 2=, 3=, 4=, 5=]
	{\txtargname{CTL$^{\star}$#5{\small\argint{$[$}{#1}{$]$}}}[#2][#3]{#4}\xspace}

\newcommandx{\STL}[5][1=, 2=, 3=, 4=, 5=]
	{\txtargname{STL#5{\small\argint{$[$}{#1}{$]$}}}[#2][#3]{#4}\xspace}

\newcommandx{\STLP}[5][1=, 2=, 3=, 4=, 5=]
	{\txtargname{STL$^{+}$#5{\small\argint{$[$}{#1}{$]$}}}[#2][#3]{#4}\xspace}

\newcommandx{\STLS}[5][1=, 2=, 3=, 4=, 5=]
	{\txtargname{STL$^{\star}$#5{\small\argint{$[$}{#1}{$]$}}}[#2][#3]{#4}\xspace}

\newcommandx{\ATL}[5][1=, 2=, 3=, 4=, 5=]
	{\txtargname{ATL#5{\small\argint{$[$}{#1}{$]$}}}[#2][#3]{#4}\xspace}

\newcommandx{\ATLP}[5][1=, 2=, 3=, 4=, 5=]
	{\txtargname{ATL$^{+}$#5{\small\argint{$[$}{#1}{$]$}}}[#2][#3]{#4}\xspace}

\newcommandx{\ATLS}[5][1=, 2=, 3=, 4=, 5=]
	{\txtargname{ATL$^{\star}$#5{\small\argint{$[$}{#1}{$]$}}}[#2][#3]{#4}\xspace}

\newcommandx{\SL}[5][1=, 2=, 3=, 4=, 5=]
	{\txtargname{SL#5{\small\argint{$[$}{#1}{$]$}}}[#2][#3]{#4}\xspace}

\newcommand{\OGSL}[1][]
	{\SL[\argb{1g}{#1}]}

\newcommand{\NGSL}[1][]
	{\SL[\argb{ng}{#1}]}

\newcommand{\EExs}[1]
	{\ensuremath{%
	\argint{\mbox{$\langle\!\langle$}}{#1}{\mbox{$\rangle\!\rangle$}}%
	}}

\newcommand{\AAll}[1]
	{\ensuremath{\argint{\mbox{$[\:\!\![$}}{#1}{\mbox{$]\:\!\!]$}}}}

\newcommand{\GSL}
	{\txtname{G}\SL}
	
	\newcommand{\GSLT}
		 {\txtname{Graded}\SL}
	
	\newcommand{\GSLfin}
		{\ensuremath{\txtname{Graded}_{\mathbb{N}}\SL}}

	\newcommand{\OGGSLT}
		 {\txtname{Graded}\OGSL}
	
	 \newcommand{\OGGSLfin}
		{\ensuremath{\txtname{Graded}_{\mathbb{N}}\OGSL}}

	\newcommand{\NGGSLT}
		 {\txtname{Graded}\NGSL}
	
	 \newcommand{\NGGSLfin}
		{\ensuremath{\txtname{Graded}_{\mathbb{N}}\NGSL}}

\newcommandx{\GSLK}[5][1=, 2=, 3=, 4=, 5=]
	{\txtargname{GSLK#5{\small\argint{$[$}{#1}{$]$}}}[#2][#3]{#4}\xspace}

\newcommandx{\EF}[5][1=, 2=, 3=, 4=, 5=]
	{\txtargname{EF#5{\small\argint{$[$}{#1}{$]$}}}[#2][#3]{#4}\xspace}

\newcommandx{\SG}[5][1=, 2=, 3=, 4=, 5=]
	{\txtargname{SG#5{\small\argint{$[$}{#1}{$]$}}}[#2][#3]{#4}\xspace}

\newcommandx{\LogTime}[4][1=, 2=, 3=, 4=]
	{\txtargname{LogTime#4}[#2][#3]{#1}\xspace}
\newcommandx{\LogTimeH}[4][1=, 2=, 3=, 4=]
	{\LogTime[#1][#2][#3][#4]-\HComp}
\newcommandx{\LogTimeE}[4][1=, 2=, 3=, 4=]
	{\LogTime[#1][#2][#3][#4]-\EComp}
\newcommandx{\LogTimeC}[4][1=, 2=, 3=, 4=]
	{\LogTime[#1][#2][#3][#4]-\CComp}

\newcommand{\NLogTime}
	{\txtname{N}\LogTime}
\newcommandx{\NLogTimeH}[4][1=, 2=, 3=, 4=]
	{\NLogTime[#1][#2][#3][#4]-\HComp}
\newcommandx{\NLogTimeE}[4][1=, 2=, 3=, 4=]
	{\NLogTime[#1][#2][#3][#4]-\EComp}
\newcommandx{\NLogTimeC}[4][1=, 2=, 3=, 4=]
	{\NLogTime[#1][#2][#3][#4]-\CComp}

\newcommand{\CoNLogTime}
	{\txtname{Co}\NLogTime}
\newcommandx{\CoNLogTimeH}[4][1=, 2=, 3=, 4=]
	{\CoNLogTime[#1][#2][#3][#4]-\HComp}
\newcommandx{\CoNLogTimeE}[4][1=, 2=, 3=, 4=]
	{\CoNLogTime[#1][#2][#3][#4]-\EComp}
\newcommandx{\CoNLogTimeC}[4][1=, 2=, 3=, 4=]
	{\CoNLogTime[#1][#2][#3][#4]-\CComp}

\newcommandx{\ALogTime}
	{\txtname{A}\LogTime}
\newcommandx{\ALogTimeH}[4][1=, 2=, 3=, 4=]
	{\ALogTime[#1][#2][#3][#4]-\HComp}
\newcommandx{\ALogTimeE}[4][1=, 2=, 3=, 4=]
	{\ALogTime[#1][#2][#3][#4]-\EComp}
\newcommandx{\ALogTimeC}[4][1=, 2=, 3=, 4=]
	{\ALogTime[#1][#2][#3][#4]-\CComp}

\newcommandx{\LogSpace}[4][1=, 2=, 3=, 4=]
	{\txtargname{LogSpace#4}[#2][#3]{#1}\xspace}
\newcommandx{\LogSpaceH}[4][1=, 2=, 3=, 4=]
	{\LogSpace[#1][#2][#3][#4]-\HComp}
\newcommandx{\LogSpaceE}[4][1=, 2=, 3=, 4=]
	{\LogSpace[#1][#2][#3][#4]-\EComp}
\newcommandx{\LogSpaceC}[4][1=, 2=, 3=, 4=]
	{\LogSpace[#1][#2][#3][#4]-\CComp}

\newcommandx{\NLogSpace}
	{\txtname{N}\LogSpace}
\newcommandx{\NLogSpaceH}[4][1=, 2=, 3=, 4=]
	{\NLogSpace[#1][#2][#3][#4]-\HComp}
\newcommandx{\NLogSpaceE}[4][1=, 2=, 3=, 4=]
	{\NLogSpace[#1][#2][#3][#4]-\EComp}
\newcommandx{\NLogSpaceC}[4][1=, 2=, 3=, 4=]
	{\NLogSpace[#1][#2][#3][#4]-\CComp}

\newcommandx{\CoNLogSpace}
	{\txtname{Co}\NLogSpace}
\newcommandx{\CoNLogSpaceH}[4][1=, 2=, 3=, 4=]
	{\CoNLogSpace[#1][#2][#3][#4]-\HComp}
\newcommandx{\CoNLogSpaceE}[4][1=, 2=, 3=, 4=]
	{\CoNLogSpace[#1][#2][#3][#4]-\EComp}
\newcommandx{\CoNLogSpaceC}[4][1=, 2=, 3=, 4=]
	{\CoNLogSpace[#1][#2][#3][#4]-\CComp}

\newcommandx{\ALogSpace}
	{\txtname{A}\LogSpace}
\newcommandx{\ALogSpaceH}[4][1=, 2=, 3=, 4=]
	{\ALogSpace[#1][#2][#3][#4]-\HComp}
\newcommandx{\ALogSpaceE}[4][1=, 2=, 3=, 4=]
	{\ALogSpace[#1][#2][#3][#4]-\EComp}
\newcommandx{\ALogSpaceC}[4][1=, 2=, 3=, 4=]
	{\ALogSpace[#1][#2][#3][#4]-\CComp}

\newcommandx{\PTime}[4][1=, 2=, 3=, 4=]
	{\txtargname{PTime#4}[#2][#3]{#1}\xspace}
\newcommandx{\PTimeH}[4][1=, 2=, 3=, 4=]
	{\PTime[#1][#2][#3][#4]-\HComp}
\newcommandx{\PTimeE}[4][1=, 2=, 3=, 4=]
	{\PTime[#1][#2][#3][#4]-\EComp}
\newcommandx{\PTimeC}[4][1=, 2=, 3=, 4=]
	{\PTime[#1][#2][#3][#4]-\CComp}

\newcommandx{\UPTime}
	{\txtname{U}\PTime}
\newcommandx{\UPTimeH}[4][1=, 2=, 3=, 4=]
	{\UPTime[#1][#2][#3][#4]-\HComp}
\newcommandx{\UPTimeE}[4][1=, 2=, 3=, 4=]
	{\UPTime[#1][#2][#3][#4]-\EComp}
\newcommandx{\UPTimeC}[4][1=, 2=, 3=, 4=]
	{\UPTime[#1][#2][#3][#4]-\CComp}

\newcommandx{\CoUPTime}
	{\txtname{Co}\UPTime}
\newcommandx{\CoUPTimeH}[4][1=, 2=, 3=, 4=]
	{\CoUPTime[#1][#2][#3][#4]-\HComp}
\newcommandx{\CoUPTimeE}[4][1=, 2=, 3=, 4=]
	{\CoUPTime[#1][#2][#3][#4]-\EComp}
\newcommandx{\CoUPTimeC}[4][1=, 2=, 3=, 4=]
	{\CoUPTime[#1][#2][#3][#4]-\CComp}

\newcommandx{\NPTime}
	{\txtname{N}\PTime}
\newcommandx{\NPTimeH}[4][1=, 2=, 3=, 4=]
	{\NPTime[#1][#2][#3][#4]-\HComp}
\newcommandx{\NPTimeE}[4][1=, 2=, 3=, 4=]
	{\NPTime[#1][#2][#3][#4]-\EComp}
\newcommandx{\NPTimeC}[4][1=, 2=, 3=, 4=]
	{\NPTime[#1][#2][#3][#4]-\CComp}

\newcommandx{\CoNPTime}
	{\txtname{Co}\NPTime}
\newcommandx{\CoNPTimeH}[4][1=, 2=, 3=, 4=]
	{\CoNPTime[#1][#2][#3][#4]-\HComp}
\newcommandx{\CoNPTimeE}[4][1=, 2=, 3=, 4=]
	{\CoNPTime[#1][#2][#3][#4]-\EComp}
\newcommandx{\CoNPTimeC}[4][1=, 2=, 3=, 4=]
	{\CoNPTime[#1][#2][#3][#4]-\CComp}

\newcommandx{\APTime}
	{\txtname{A}\PTime}
\newcommandx{\APTimeH}[4][1=, 2=, 3=, 4=]
	{\APTime[#1][#2][#3][#4]-\HComp}
\newcommandx{\APTimeE}[4][1=, 2=, 3=, 4=]
	{\APTime[#1][#2][#3][#4]-\EComp}
\newcommandx{\APTimeC}[4][1=, 2=, 3=, 4=]
	{\APTime[#1][#2][#3][#4]-\CComp}

\newcommandx{\PSpace}[4][1=, 2=, 3=, 4=]
	{\txtargname{PSpace#4}[#2][#3]{#1}\xspace}
\newcommandx{\PSpaceH}[4][1=, 2=, 3=, 4=]
	{\PSpace[#1][#2][#3][#4]-\HComp}
\newcommandx{\PSpaceE}[4][1=, 2=, 3=, 4=]
	{\PSpace[#1][#2][#3][#4]-\EComp}
\newcommandx{\PSpaceC}[4][1=, 2=, 3=, 4=]
	{\PSpace[#1][#2][#3][#4]-\CComp}

\newcommandx{\NPSpace}
	{\txtname{N}\PSpace}
\newcommandx{\NPSpaceH}[4][1=, 2=, 3=, 4=]
	{\NPSpace[#1][#2][#3][#4]-\HComp}
\newcommandx{\NPSpaceE}[4][1=, 2=, 3=, 4=]
	{\NPSpace[#1][#2][#3][#4]-\EComp}
\newcommandx{\NPSpaceC}[4][1=, 2=, 3=, 4=]
	{\NPSpace[#1][#2][#3][#4]-\CComp}

\newcommandx{\CoNPSpace}
	{\txtname{Co}\NPSpace}
\newcommandx{\CoNPSpaceH}[4][1=, 2=, 3=, 4=]
	{\CoNPSpace[#1][#2][#3][#4]-\HComp}
\newcommandx{\CoNPSpaceE}[4][1=, 2=, 3=, 4=]
	{\CoNPSpace[#1][#2][#3][#4]-\EComp}
\newcommandx{\CoNPSpaceC}[4][1=, 2=, 3=, 4=]
	{\CoNPSpace[#1][#2][#3][#4]-\CComp}

\newcommandx{\APSpace}
	{\txtname{A}\PSpace}
\newcommandx{\APSpaceH}[4][1=, 2=, 3=, 4=]
	{\APSpace[#1][#2][#3][#4]-\HComp}
\newcommandx{\APSpaceE}[4][1=, 2=, 3=, 4=]
	{\APSpace[#1][#2][#3][#4]-\EComp}
\newcommandx{\APSpaceC}[4][1=, 2=, 3=, 4=]
	{\APSpace[#1][#2][#3][#4]-\CComp}

\newcommandx{\ExpTime}[4][1=, 2=, 3=, 4=]
	{\txtargname{ExpTime#4}[#2][#3]{#1}\xspace}
\newcommandx{\ExpTimeH}[4][1=, 2=, 3=, 4=]
	{\ExpTime[#1][#2][#3][#4]-\HComp}
\newcommandx{\ExpTimeE}[4][1=, 2=, 3=, 4=]
	{\ExpTime[#1][#2][#3][#4]-\EComp}
\newcommandx{\ExpTimeC}[4][1=, 2=, 3=, 4=]
	{\ExpTime[#1][#2][#3][#4]-\CComp}

\newcommandx{\NExpTime}
	{\txtname{N}\ExpTime}
\newcommandx{\NExpTimeH}[4][1=, 2=, 3=, 4=]
	{\NExpTime[#1][#2][#3][#4]-\HComp}
\newcommandx{\NExpTimeE}[4][1=, 2=, 3=, 4=]
	{\NExpTime[#1][#2][#3][#4]-\EComp}
\newcommandx{\NExpTimeC}[4][1=, 2=, 3=, 4=]
	{\NExpTime[#1][#2][#3][#4]-\CComp}

\newcommandx{\CoNExpTime}
	{\txtname{Co}\NExpTime}
\newcommandx{\CoNExpTimeH}[4][1=, 2=, 3=, 4=]
	{\CoNExpTime[#1][#2][#3][#4]-\HComp}
\newcommandx{\CoNExpTimeE}[4][1=, 2=, 3=, 4=]
	{\CoNExpTime[#1][#2][#3][#4]-\EComp}
\newcommandx{\CoNExpTimeC}[4][1=, 2=, 3=, 4=]
	{\CoNExpTime[#1][#2][#3][#4]-\CComp}

\newcommandx{\AExpTime}
	{\txtname{A}\ExpTime}
\newcommandx{\AExpTimeH}[4][1=, 2=, 3=, 4=]
	{\AExpTime[#1][#2][#3][#4]-\HComp}
\newcommandx{\AExpTimeE}[4][1=, 2=, 3=, 4=]
	{\AExpTime[#1][#2][#3][#4]-\EComp}
\newcommandx{\AExpTimeC}[4][1=, 2=, 3=, 4=]
	{\AExpTime[#1][#2][#3][#4]-\CComp}

\newcommandx{\ExpSpace}[4][1=, 2=, 3=, 4=]
	{\txtargname{ExpSpace#4}[#2][#3]{#1}\xspace}
\newcommandx{\ExpSpaceH}[4][1=, 2=, 3=, 4=]
	{\ExpSpace[#1][#2][#3][#4]-\HComp}
\newcommandx{\ExpSpaceE}[4][1=, 2=, 3=, 4=]
	{\ExpSpace[#1][#2][#3][#4]-\EComp}
\newcommandx{\ExpSpaceC}[4][1=, 2=, 3=, 4=]
	{\ExpSpace[#1][#2][#3][#4]-\CComp}

\newcommandx{\NExpSpace}
	{\txtname{N}\ExpSpace}
\newcommandx{\NExpSpaceH}[4][1=, 2=, 3=, 4=]
	{\NExpSpace[#1][#2][#3][#4]-\HComp}
\newcommandx{\NExpSpaceE}[4][1=, 2=, 3=, 4=]
	{\NExpSpace[#1][#2][#3][#4]-\EComp}
\newcommandx{\NExpSpaceC}[4][1=, 2=, 3=, 4=]
	{\NExpSpace[#1][#2][#3][#4]-\CComp}

\newcommandx{\CoNExpSpace}
	{\txtname{Co}\NExpSpace}
\newcommandx{\CoNExpSpaceH}[4][1=, 2=, 3=, 4=]
	{\CoNExpSpace[#1][#2][#3][#4]-\HComp}
\newcommandx{\CoNExpSpaceE}[4][1=, 2=, 3=, 4=]
	{\CoNExpSpace[#1][#2][#3][#4]-\EComp}
\newcommandx{\CoNExpSpaceC}[4][1=, 2=, 3=, 4=]
	{\CoNExpSpace[#1][#2][#3][#4]-\CComp}

\newcommandx{\AExpSpace}
	{\txtname{A}\ExpSpace}
\newcommandx{\AExpSpaceH}[4][1=, 2=, 3=, 4=]
	{\AExpSpace[#1][#2][#3][#4]-\HComp}
\newcommandx{\AExpSpaceE}[4][1=, 2=, 3=, 4=]
	{\AExpSpace[#1][#2][#3][#4]-\EComp}
\newcommandx{\AExpSpaceC}[4][1=, 2=, 3=, 4=]
	{\AExpSpace[#1][#2][#3][#4]-\CComp}

\newcommandx{\NonElm}[4][1=, 2=, 3=, 4=]
	{\txtargname{NonElementary#4}[#2][#3]{#1}\xspace}
\newcommandx{\NonElmH}[4][1=, 2=, 3=, 4=]
	{\NonElm[#1][#2][#3][#4]-\HComp}
\newcommandx{\NonElmE}[4][1=, 2=, 3=, 4=]
	{\NonElm[#1][#2][#3][#4]-\EComp}
\newcommandx{\NonElmC}[4][1=, 2=, 3=, 4=]
	{\NonElm[#1][#2][#3][#4]-\CComp}

\newcommandx{\NonElmTime}[4][1=, 2=, 3=, 4=]
	{\txtargname{NonElementaryTime#4}[#2][#3]{#1}\xspace}
\newcommandx{\NonElmTimeH}[4][1=, 2=, 3=, 4=]
	{\NonElmTime[#1][#2][#3][#4]-\HComp}
\newcommandx{\NonElmTimeE}[4][1=, 2=, 3=, 4=]
	{\NonElmTime[#1][#2][#3][#4]-\EComp}
\newcommandx{\NonElmTimeC}[4][1=, 2=, 3=, 4=]
	{\NonElmTime[#1][#2][#3][#4]-\CComp}

\newcommandx{\NonElmSpace}[4][1=, 2=, 3=, 4=]
	{\txtargname{NonElementarySpace#4}[#2][#3]{#1}\xspace}
\newcommandx{\NonElmSpaceH}[4][1=, 2=, 3=, 4=]
	{\NonElmSpace[#1][#2][#3][#4]-\HComp}
\newcommandx{\NonElmSpaceE}[4][1=, 2=, 3=, 4=]
	{\NonElmSpace[#1][#2][#3][#4]-\EComp}
\newcommandx{\NonElmSpaceC}[4][1=, 2=, 3=, 4=]
	{\NonElmSpace[#1][#2][#3][#4]-\CComp}

\newcommandx{\DLHier}[4][2=, 3=, 4=]
	{\mthargset[0]{\Delta#4}[#1][#3]{#2}\xspace}
\newcommandx{\DLHierH}[4][2=, 3=, 4=]
	{\DLHier{#1}[#2][#3][#4]-\HComp}
\newcommandx{\DLHierE}[4][2=, 3=, 4=]
	{\DLHier{#1}[#2][#3][#4]-\EComp}
\newcommandx{\DLHierC}[4][2=, 3=, 4=]
	{\DLHier{#1}[#2][#3][#4]-\CComp}

\newcommandx{\ELHier}[4][2=, 3=, 4=]
	{\mthargset[0]{\Sigma#4}[#1][#3]{#2}\xspace}
\newcommandx{\ELHierH}[4][2=, 3=, 4=]
	{\ELHier{#1}[#2][#3][#4]-\HComp}
\newcommandx{\ELHierE}[4][2=, 3=, 4=]
	{\ELHier{#1}[#2][#3][#4]-\EComp}
\newcommandx{\ELHierC}[4][2=, 3=, 4=]
	{\ELHier{#1}[#2][#3][#4]-\CComp}

\newcommandx{\ULHier}[4][2=, 3=, 4=]
	{\mthargset[0]{\Pi#4}[#1][#3]{#2}\xspace}
\newcommandx{\ULHierH}[4][2=, 3=, 4=]
	{\ULHier{#1}[#2][#3][#4]-\HComp}
\newcommandx{\ULHierE}[4][2=, 3=, 4=]
	{\ULHier{#1}[#2][#3][#4]-\EComp}
\newcommandx{\ULHierC}[4][2=, 3=, 4=]
	{\ULHier{#1}[#2][#3][#4]-\CComp}

\newcommandx{\DBHier}[4][2=, 3=, 4=]
	{\mthargset[3]{\Delta#4}[#1][#3]{#2}\xspace}
\newcommandx{\DBHierH}[4][2=, 3=, 4=]
	{\DBHier{#1}[#2][#3][#4]-\HComp}
\newcommandx{\DBHierE}[4][2=, 3=, 4=]
	{\DBHier{#1}[#2][#3][#4]-\EComp}
\newcommandx{\DBHierC}[4][2=, 3=, 4=]
	{\DBHier{#1}[#2][#3][#4]-\CComp}

\newcommandx{\EBHier}[4][2=, 3=, 4=]
	{\mthargset[3]{\Sigma#4}[#1][#3]{#2}\xspace}
\newcommandx{\EBHierH}[4][2=, 3=, 4=]
	{\EBHier{#1}[#2][#3][#4]-\HComp}
\newcommandx{\EBHierE}[4][2=, 3=, 4=]
	{\EBHier{#1}[#2][#3][#4]-\EComp}
\newcommandx{\EBHierC}[4][2=, 3=, 4=]
	{\EBHier{#1}[#2][#3][#4]-\CComp}

\newcommandx{\UBHier}[4][2=, 3=, 4=]
	{\mthargset[3]{\Pi#4}[#1][#3]{#2}\xspace}
\newcommandx{\UBHierH}[4][2=, 3=, 4=]
	{\UBHier{#1}[#2][#3][#4]-\HComp}
\newcommandx{\UBHierE}[4][2=, 3=, 4=]
	{\UBHier{#1}[#2][#3][#4]-\EComp}
\newcommandx{\UBHierC}[4][2=, 3=, 4=]
	{\UBHier{#1}[#2][#3][#4]-\CComp}

\newcommandx{\DPolHier}[4][2=, 3=, 4=]
	{\DLHier{#1}[#2][\argb{\mathrm{P}}{#3}][#4]}
\newcommandx{\DPolHierH}[4][2=, 3=, 4=]
	{\DPolHier{#1}[#2][#3][#4]-\HComp}
\newcommandx{\DPolHierE}[4][2=, 3=, 4=]
	{\DPolHier{#1}[#2][#3][#4]-\EComp}
\newcommandx{\DPolHierC}[4][2=, 3=, 4=]
	{\DPolHier{#1}[#2][#3][#4]-\CComp}

\newcommandx{\EPolHier}[4][2=, 3=, 4=]
	{\ELHier{#1}[#2][\argb{\mathrm{P}}{#3}][#4]}
\newcommandx{\EPolHierH}[4][2=, 3=, 4=]
	{\EPolHier{#1}[#2][#3][#4]-\HComp}
\newcommandx{\EPolHierE}[4][2=, 3=, 4=]
	{\EPolHier{#1}[#2][#3][#4]-\EComp}
\newcommandx{\EPolHierC}[4][2=, 3=, 4=]
	{\EPolHier{#1}[#2][#3][#4]-\CComp}

\newcommandx{\UPolHier}[4][2=, 3=, 4=]
	{\ULHier{#1}[#2][\argb{\mathrm{P}}{#3}][#4]}
\newcommandx{\UPolHierH}[4][2=, 3=, 4=]
	{\UPolHier{#1}[#2][#3][#4]-\HComp}
\newcommandx{\UPolHierE}[4][2=, 3=, 4=]
	{\UPolHier{#1}[#2][#3][#4]-\EComp}
\newcommandx{\UPolHierC}[4][2=, 3=, 4=]
	{\UPolHier{#1}[#2][#3][#4]-\CComp}

\newcommandx{\DAriHier}[4][2=, 3=, 4=]
	{\DLHier{#1}[#2][\argb{0}{#3}][#4]}
\newcommandx{\DAriHierH}[4][2=, 3=, 4=]
	{\DAriHier{#1}[#2][#3][#4]-\HComp}
\newcommandx{\DAriHierE}[4][2=, 3=, 4=]
	{\DAriHier{#1}[#2][#3][#4]-\EComp}
\newcommandx{\DAriHierC}[4][2=, 3=, 4=]
	{\DAriHier{#1}[#2][#3][#4]-\CComp}

\newcommandx{\EAriHier}[4][2=, 3=, 4=]
	{\ELHier{#1}[#2][\argb{0}{#3}][#4]}
\newcommandx{\EAriHierH}[4][2=, 3=, 4=]
	{\EAriHier{#1}[#2][#3][#4]-\HComp}
\newcommandx{\EAriHierE}[4][2=, 3=, 4=]
	{\EAriHier{#1}[#2][#3][#4]-\EComp}
\newcommandx{\EAriHierC}[4][2=, 3=, 4=]
	{\EAriHier{#1}[#2][#3][#4]-\CComp}

\newcommandx{\UAriHier}[4][2=, 3=, 4=]
	{\ULHier{#1}[#2][\argb{0}{#3}][#4]}
\newcommandx{\UAriHierH}[4][2=, 3=, 4=]
	{\UAriHier{#1}[#2][#3][#4]-\HComp}
\newcommandx{\UAriHierE}[4][2=, 3=, 4=]
	{\UAriHier{#1}[#2][#3][#4]-\EComp}
\newcommandx{\UAriHierC}[4][2=, 3=, 4=]
	{\UAriHier{#1}[#2][#3][#4]-\CComp}

\newcommandx{\DAnaHier}[4][2=, 3=, 4=]
	{\DLHier{#1}[#2][\argb{1}{#3}][#4]}
\newcommandx{\DAnaHierH}[4][2=, 3=, 4=]
	{\DAnaHier{#1}[#2][#3][#4]-\HComp}
\newcommandx{\DAnaHierE}[4][2=, 3=, 4=]
	{\DAnaHier{#1}[#2][#3][#4]-\EComp}
\newcommandx{\DAnaHierC}[4][2=, 3=, 4=]
	{\DAnaHier{#1}[#2][#3][#4]-\CComp}

\newcommandx{\EAnaHier}[4][2=, 3=, 4=]
	{\ELHier{#1}[#2][\argb{1}{#3}][#4]}
\newcommandx{\EAnaHierH}[4][2=, 3=, 4=]
	{\EAnaHier{#1}[#2][#3][#4]-\HComp}
\newcommandx{\EAnaHierE}[4][2=, 3=, 4=]
	{\EAnaHier{#1}[#2][#3][#4]-\EComp}
\newcommandx{\EAnaHierC}[4][2=, 3=, 4=]
	{\EAnaHier{#1}[#2][#3][#4]-\CComp}

\newcommandx{\UAnaHier}[4][2=, 3=, 4=]
	{\ULHier{#1}[#2][\argb{1}{#3}][#4]}
\newcommandx{\UAnaHierH}[4][2=, 3=, 4=]
	{\UAnaHier{#1}[#2][#3][#4]-\HComp}
\newcommandx{\UAnaHierE}[4][2=, 3=, 4=]
	{\UAnaHier{#1}[#2][#3][#4]-\EComp}
\newcommandx{\UAnaHierC}[4][2=, 3=, 4=]
	{\UAnaHier{#1}[#2][#3][#4]-\CComp}

\newcommandx{\DBorHier}[4][2=, 3=, 4=]
	{\DBHier{#1}[#2][\argb{\mathrm{B}}{#3}][#4]}
\newcommandx{\DBorHierH}[4][2=, 3=, 4=]
	{\DBorHier{#1}[#2][#3][#4]-\HComp}
\newcommandx{\DBorHierE}[4][2=, 3=, 4=]
	{\DBorHier{#1}[#2][#3][#4]-\EComp}
\newcommandx{\DBorHierC}[4][2=, 3=, 4=]
	{\DBorHier{#1}[#2][#3][#4]-\CComp}

\newcommandx{\EBorHier}[4][2=, 3=, 4=]
	{\EBHier{#1}[#2][\argb{\mathrm{B}}{#3}][#4]}
\newcommandx{\EBorHierH}[4][2=, 3=, 4=]
	{\EBorHier{#1}[#2][#3][#4]-\HComp}
\newcommandx{\EBorHierE}[4][2=, 3=, 4=]
	{\EBorHier{#1}[#2][#3][#4]-\EComp}
\newcommandx{\EBorHierC}[4][2=, 3=, 4=]
	{\EBorHier{#1}[#2][#3][#4]-\CComp}

\newcommandx{\UBorHier}[4][2=, 3=, 4=]
	{\UBHier{#1}[#2][\argb{\mathrm{B}}{#3}][#4]}
\newcommandx{\UBorHierH}[4][2=, 3=, 4=]
	{\UBorHier{#1}[#2][#3][#4]-\HComp}
\newcommandx{\UBorHierE}[4][2=, 3=, 4=]
	{\UBorHier{#1}[#2][#3][#4]-\EComp}
\newcommandx{\UBorHierC}[4][2=, 3=, 4=]
	{\UBorHier{#1}[#2][#3][#4]-\CComp}

\newcommand{\HComp}
	{\txtname{hard}\xspace}

\newcommand{\EComp}
	{\txtname{easy}\xspace}

\newcommand{\CComp}
	{\txtname{complete}\xspace}

\providecommand{\SymSet}[1][]{\mthset[#1]{\Sigma}}

\providecommand{\QSet}[1][]{\mthset[#1]{Q}}

\providecommand{\PSet}[1][]{\mthset[#1]{P}}

\newcommand{\WATuple}[5]
	{
	\ifx&#5&
		\tupled{#1}{#2}{#3}{#4}
	\else
		\tuplee{#1}{#2}{#3}{#4}{#5}
	\fi
	}

\newcommand{\WMTuple}[7]
	{
	\ifx&#7&
		\tuplef{#1}{#2}{#3}{#4}{#5}{#6}
	\else
		\tupleg{#1}{#2}{#3}{#4}{#5}{#6}{#7}
	\fi
	}

\newcommand{\NTA}{\txtname{Nta}}

\newcommand{\ATA}{\txtname{Ata}}

\newcommand{\NPT}{\txtname{Npt}}

\newcommand{\APT}{\txtname{Apt}}

\providecommand{\SymSet}[1][]{\mthset[#1]{\Sigma}}

\providecommand{\QSet}[1][]{\mthset[#1]{Q}}

\providecommand{\PSet}[1][]{\mthset[#1]{P}}

\newcommand{\TATuple}[6]
	{
	\ifx&#2&
		\ifx&#6&
			\tupled{#1}{#3}{#4}{#5}
		\else
			\tuplee{#1}{#3}{#4}{#5}{#6}
		\fi
	\else
		\ifx&#6&
			\tuplee{#1}{#2}{#3}{#4}{#5}
		\else
			\tuplef{#1}{#2}{#3}{#4}{#5}{#6}
		\fi
	\fi
	}

\newcommand{\TMTuple}[8]
	{
	\ifx&#3&
		\ifx&#8&
			\tuplef{#1}{#2}{#4}{#5}{#6}{#7}
		\else
			\tupleg{#1}{#2}{#4}{#5}{#6}{#7}{#8}
		\fi
	\else
		\ifx&#8&
			\tupleg{#1}{#2}{#3}{#4}{#5}{#6}{#7}
		\else
			\tupleh{#1}{#2}{#3}{#4}{#5}{#6}{#7}{#8}
		\fi
	\fi
	}

\providecommand{\LangSet}[1][]{\mthset[#1]{L}}

\providecommand{\infFun}[1][]{\mthfun[#1]{inf}}

\newtheorem{definition}{Definition}[section]

\newtheorem{lemma}{Lemma}[section]
\newtheorem{theorem}{Theorem}[section]

\newtheorem{proof}{Proof}[section]

\newcounter{flushenumerate}
	{\end{list}}

% ...

\newcommand{\sucfun}{suc}
\newcommandx{\sucFun}[4][1=, 2=, 3=, 4=]
	{\mthargfun{\sucfun#4}[#1][#2]{#3}}

\newcommand{\plyset}{P}
\newcommandx{\PlySet}[3][1=, 2=, 3=]
	{\mthset{\plyset#3}[#1][#2]}

\newcommand{\plysym}{Player}
\newcommandx{\plySym}[3][1=, 2=, 3=]
	{\mthset{\plysym#3}[#1][#2]}

\def\it{\begin{itemize} }
\def\-{\item }
\def\ti{\end{itemize} }
\def\en{\begin{enumerate} }
\def\ne{\end{enumerate} }

\usepackage{changebar}

\usepackage{breakurl}             % Not needed if you use pdflatex only.

\usepackage{wasysym}

\newcommand{\Nat}{\mathbb{N}}

\newcommand{\tpl}[1]{\langle {#1} \rangle }
\def\ValSet{\textsc{Val}}

\DeclareMathOperator{\nextt}{\mathsf{X}}

\DeclareMathOperator{\until}{\mathbin{\mathsf{U}}}

\DeclareMathOperator{\always}{\mathsf{G}}

\DeclareMathOperator{\eventually}{\mathsf{F}}

\newcommand{\true}{\mathsf{true}}
\newcommand{\false}{\mathsf{false}}

\newcommand{\T}{\mathsf{T}}
\renewcommand{\R}{\mathsf{R}}

\title{Extended Graded Modalities in Strategy Logic}
\author{
Benjamin Aminof\\
\institute{Technische Universitat Wien, Austria}\\
\email{benj@forsyte.tuwien.ac.at}
\and
Vadim Malvone, \ \ Aniello Murano, \ \ Sasha Rubin \\
\institute{Universit\`a degli Studi di Napoli Federico II, Italy}\\
\email{\{malvone,murano,rubin\}@unina.it}
}

\def\titlerunning{Extended Graded Modalities in Strategy Logic}
\def\authorrunning{B. Aminof, V. Malvone, A. Murano, S. Rubin}

\begin{document}
	
 	%\nolinenumbers

	\maketitle

	\begin{abstract}
Strategy Logic (\SL) is a logical formalism for strategic reasoning in
multi-agent systems. Its main feature is that it has variables for strategies that are associated to specific agents with a binding operator. We introduce Graded Strategy
Logic\,({\GSLT}), an extension of \SL\ by graded quantifiers over
tuples of strategy variables, i.e., 
``there exist at least $g$ different tuples $(x_1,...,x_n)$ of strategies'' where $g$ is a cardinal from the set 
$\SetN \cup \{\aleph_0, \aleph_1, 2^{\aleph_0}\}$.
We prove that the model-checking problem of \GSLT\ is decidable. We then turn to the complexity of fragments of \GSLT. When the $g$'s are restricted to finite cardinals, written $\GSLfin$,
the complexity of model-checking is no harder
than for \SL, i.e., it is non-elementary in the quantifier rank. We illustrate our formalism by showing how to count the number of different strategy
profiles that are Nash equilibria (NE), or subgame-perfect equilibria (SPE).
%, in games with a fixed number of outcomes.
By analyzing the structure of the specific formulas involved, we conclude
that the important problems of checking for the existence of a unique NE or SPE
can both be solved in 2\ExpTime, which
is not harder than merely checking for the existence of such equilibria.
\end{abstract}

	% Begin of file Introduction.tex
\begin{section}{Introduction}

Strategy Logic (\SL) is a powerful formalism for reasoning about strategies in
multi-agent systems~\cite{MMV10b,MMPV14}.  Strategies tell an agent what to do
--- they are functions that prescribe an action based on the history. The key
idea in \SL\ is to treat strategies as first-order object~\cite{CHP10}.  A strategy $\xElm$
can be quantified existentially $\EExs{\xElm}$ (read: there exists a strategy
$x$) and, dually, universally $\AAll{\xElm}$ (read: for all strategies $x$).  Strategies
are not intrinsically glued to specific agents: the \emph{binding} operator
$(\alpha, \xElm)$ allows one to bind an agent $\alpha$ to the strategy $\xElm$.
\SL\ is built over the structure of the linear-time temporal-logic \LTL~\cite{Pnu77} and strictly subsumes several other well-known logics for the strategic reasoning including \ATLS~\cite{AHK02} and the like~\cite{HJW05,WHW07,WHY11,HLW13,LLM10,BLLM09,AGJ07,FMP08}.
	
We extend \SL by replacing the quantification
$\EExs{\xElm}$ and $\AAll{\xElm}$ over strategy variables with \emph{graded
quantification over tuples of strategy variables}: $\EExs{\varElm_{1},\ldots,
\varElm_{n}}^{\geq g}$ (read $\EExs{\varElm_{1},\ldots, \varElm_{n}}^{\geq g}$
as ``there exist at least $g$ different tuples $(\varElm_{1},\ldots,
\varElm_{n})$ of strategies'') and its dual $\AAll{\varElm_{1},\ldots,
\varElm_{n}}^{< g}$, where $g \in \SetN \cup \{\aleph_0, \aleph_1,
2^{\aleph_0}\}$.	Here, two tuples are different if they are different in some component, and two strategies are different if they disagree on
some history.  That is, we count strategies syntactically in a way similar as it is usually done in graded
extensions of modal and description logics~\cite{BLMV08,KSV02,CGL99,Gradel99,SSM08,BL14,BGLS11,DL10,DL03}.
	
	We address the model-checking problem for \GSLT\ and, by means of an automata-theoretic approach, we prove that it is decidable. Our algorithm is not elementary, i.e., it is not bounded by any tower of exponentials. Let \GSLfin\ denote the fragment in which all grades are finite (i.e., $g \in \SetN$). We studied this fragment in \cite{AMMR16} and report the results here. The model-checking problem has
the same complexity as \SL. That is, model checking \GSLfin\ formulas with a
nesting depth $k>0$ of blocks of quantifiers\footnote{A \emph{block} of quantifiers is a
maximally-consecutive sequence of quantifiers of the same type, i.e., either
all existential, or all universal.} is in $(k+1)$\ExpTime, and that for the
special case where the formula starts with a block of quantifiers, it is in
$k$\ExpTime. 
	We also study natural fragments of \GSLT, in line with the fragments of \SL. In particular, the	Nested-Goal fragment of \SL requires that bindings and 
	quantifications appear in exhaustive blocks (see Section~\ref{sec:gradedSL} for details), and 
	can express Nash Equilibrium (NE)~\cite{MMPV14}.
	Similarly, we define {Nested-Goal \GSLT}, a syntactic fragment of \GSLT.
	We show that Nested-Goal \GSLfin\ has the same
	model-checking
	complexity as Nested-Goal \SL, i.e.,  non-elementary in the {alternation number} of the
	quantifiers appearing in the formula.\footnote{The \emph{alternation number} of a Nested-Goal formula is, roughly speaking, the maximum number of
	existential/universal quantifier switches in a consecutive sequence of quantifiers.} 
	
	Since many natural formulas have a small number of quantifiers, and even smaller nesting depth of blocks of quantifiers, the complexity of the model-checking problem is not as bad as it seems. Several solution concepts can be expressed as \SL\ formulas with a small number of quantifiers\cite{CHP10,MMPV14,GHW14,KPV14,KPV16,Bel15}. We illustrate this by expressing \emph{uniqueness} of various solution concepts, including winning-strategies in
	two-player zero-sum games and equilibria in multi-player non zero-sum games.
	It is important to observe that, when dealing with NE, to establish whether the game admits a \emph{unique} equilibrium is a very challenging and important question~\cite{AKH02,PC79,CHS99}. This problem has an impact on the predictive power of NE since, in case there are multiple equilibria, the outcome of the game cannot be pinned
	down~\cite{SCB13,ZG11,Pavel12}. Before our works involving Strategy Logic, there was no uniform framework to
	deal with this problem. So far, this question has mainly been addressed algorithmically and for very restrictive game topologies~\cite{ORS93a}. 
	By means of \GSLT, in this paper we show how to address and efficiently solve this problem and in a general way. In particular, thanks to the small nesting depth of the blocks of quantifiers of the
	formulas involved, we show that checking the uniqueness of winning strategies, NE, or subgame-perfect equilibria (SPE) can be solved in $2$\ExpTime, which is no worse than checking their existence.
	
	\noindent {\bf Related work.}
	The work closest to ours is~\cite{AMMR16} that also considers a graded extension of Strategy Logic, aimed at counting strategies in solution concepts. Precisely, strategies in~\cite{AMMR16} are counted syntactically, as we do here, but graded quantifiers numbers are limited to finite cardinals.
	This explains the words ``extended graded modalities'' in the title of our work.
	In this paper, for the sake of completeness, we have decided to report in summary the main results from~\cite{AMMR16} as an introductory material. Also, most of the definitions we provide extend those reported in~\cite{AMMR16} and we refer to that paper for additional comments and observations.
	
 	Another work close to ours is~\cite{MMMS15}. It also introduces a graded extension of \SL, called \GSL. However, in contrast with our work and~\cite{AMMR16} (and by extending an idea of counting paths in~\cite{AMR15,BMM12}), \GSL\ provides a quite intricate way of counting strategies: it gives a semantic definition of equivalence of strategies, and the graded modalities count non-equivalent strategies. While this approach is sound, it heavily complicates the model-checking problem. Indeed, only a very weak fragment of \GSL\ has been solved in~\cite{MMMS15} by exploiting an \emph{ad hoc} solution that does not seem to be scalable to (all of) \GSL. Such a fragment is orthogonal to \ATLS and, as far as we know, not enough powerful to represent NE neither SPE.
 	
	\noindent {\bf Outline.}
	The sequel of the paper is organized as follows.
	In Section~\ref{sec:gradedSL} we introduce \GSLT and provide some preliminary related concepts.
	In Section~\ref{sec:illustration} we recall the notion of objective and illustrate how to express some solution concepts and their uniqueness.
	In Section~\ref{sec:MC} we address the model-checking problem for \GSLT and its fragments.
	We conclude with Section~\ref{conclusion} in which we have a discussion and suggestions for future work.

\end{section}

% End of file Introduction.tex

	% Begin of file SectionI.tex

\begin{section}{Graded Strategy Logic}
	\label{sec:gradedSL}
	In this section we introduce Graded Strategy Logic, which we call $\GSLT$ for
	short.
	\begin{subsection}{Models}
		Sentences of \GSLT\ are interpreted over \emph{concurrent game
		structures}, just as for \ATL\ and \SL~\cite{AHK02,MMPV14}.
		
		\begin{definition}
			\label{CGS}
			A \emph{concurrent game structure} (\emph{$\CGS$})
			is a tuple $\CGSName \defeq \tpl{\APSet, \AgnSet, \ActSet,
			\SttSet, \isttElm, \apFun, \trnFun}$,
			where 
			\it
			\- $\APSet$ is a finite set \emph{atomic propositions}, 
			\- $\AgnSet$ is a finite set of 			\emph{agents},
			\- $\ActSet$ is a finite set of \emph{actions},
			\- $\SttSet$ is a 	finite set of		\emph{states},
			\- $\isttElm \in \SttSet$ is the \emph{initial
			state},
			\- $\apFun : \SttSet \to \pow{\APSet}$ is the \emph{labeling
			function} mapping each state to the set of atomic propositions true in
			that state, and
			\- let $\DecSet \!\defeq\! \AgnSet \!\to\! \ActSet$ be the
			set of \emph{decisions}, \ie, functions describing the choice
			of an action by every agent. Then, $\trnFun : \DecSet \!\to\! (\SttSet
			\!\to\! \SttSet)$, a \emph{transition function}, maps every
			decision $\decElm \!\in\! \DecSet$ to a function
			$\trnFun(\decElm): \SttSet \to \SttSet$.
			\ti
		\end{definition}
		
		We will usually take the set $\AgnSet$ of agents to be $\{\alpha_1, \dots,
		\alpha_n\}$.
		A \emph{path (from $s$)} is a finite or infinite non-empty sequence of states $s_1
		s_2 \dots$ such that $s = s_1$ and for every $i$ there exists a decision $\delta$
		with $\trnFun(\delta)(s_i) = s_{i+1}$. The set of paths starting with $s$ is denoted
		$\PthSet(s)$.
		The set of finite paths from $s$, called the \emph{histories (from $s$)}, is denoted
		$\HstSet(s)$.
		A \emph{strategy (from $s$)} is a function $\strElm \!\in\! \StrSet(s) \!\defeq\! \HstSet(s)
		\!\to\! \ActSet$ that prescribes which action has to be performed given a
		 history.
		We write $\PthSet, \HstSet, \StrSet$ for the set of all paths, histories, and strategies (no
		matter where they start).
		We use the standard notion of equality between strategies, \cite{LB08}, i.e.,
		$\strElm_1 = \strElm_2$ iff for all $\rho \in \HstSet$, $\strElm_{1}(\rho) = \strElm_{2}(\rho)$.
		This extends to equality between two $n$-tuples of strategies in the
		natural way, i.e., coordinate-wise.
		
	\end{subsection}
	
	\begin{subsection}{Syntax}
	
		\GSLT extends \SL\ by replacing the singleton strategy quantifiers
		$\EExs{\varElm}$ and $\AAll{\varElm}$ with the graded
		(tupled) quantifiers $\EExs{\varElm_{1},\ldots, \varElm_{n}}^{\geq g}$ and
		$\AAll{\varElm_{1},\ldots, \varElm_{n}}^{< g}$, respectively, where each
		$\varElm_{i}$ belongs to a countable set of variables $\VarSet$
		and $g \in \SetN \cup \{\aleph_0, \aleph_1, 2^{\aleph_0}\}$ is called the \emph{grade} of the quantifier.
		Intuitively, these are read as ``there
		exist at least $g$ tuples of strategies $(\varElm_{1},\ldots,
		\varElm_{n})$'' and ``all but less than $g$ many tuples of strategies'',
		respectively. The syntax $(\alpha, \varElm)$ denotes a \emph{binding} of the agent $\alpha$ to the strategy $\varElm$.
		
		\begin{definition}
			\GSLT \emph{formulas} are built inductively by means of the following
			grammar, where $\apElm \in \APSet$, $\alpha \in \AgnSet$,
			$\varElm,\varElm_1, \dots \varElm_n \in \VarSet$ such that $x_i \neq x_j$ for $i \neq j$,
			$n \in \SetN$, and $g \in \SetN \cup \{\aleph_0, \aleph_1, 2^{\aleph_0}\}$:
			\begin{center}
				$\varphi \seteq
				\apElm \mid
				\neg \varphi
				\mid \varphi \vee \varphi
				\mid \nextt \varphi
				\mid \varphi \until \varphi
				\mid \EExs{\varElm_{1},\ldots, \varElm_{n}}^{\geq g} \varphi
				\mid (\alpha, \varElm) \varphi$.
			\end{center}
		\end{definition}

		{\bf Notation.} Whenever we write $\EExs{\varElm_{1},\ldots, \varElm_{n}}^{\geq g}$
		we mean that $\varElm_i \neq \varElm_j$ for $i \neq j$, i.e., the variables in a tuple are distinct. Shorthands are derived as usual.
		Specifically, $\true \defeq p \vee \neg p$, $\false \defeq \neg \true$,
$\eventually \varphi \defeq \true \until \varphi$, and
		$\always \varphi \defeq
		\neg \eventually \neg \varphi$. % $\varphi \R \psi \defeq \neg (\neg \varphi \until \neg		\psi)$.
		Also, we have that $\AAll{\varElm_{1},\ldots, \varElm_{n}}^{< g}
		\varphi \defeq \neg \EExs{\varElm_{1},\ldots, \varElm_{n}}^{\geq g}
		\neg\varphi$.
		
		In order to define the semantics, we first define the concept of
		\emph{free placeholders} in a formula, which refer to agents and variables.
		Intuitively, an agent or variable is free in $\varphi$ if it does
		not have a strategy associated with it (either by quantification or binding) but one is required in order
		to ascertain if $\varphi$ is true or not.
								The set of \emph{free agents} and \emph{free variables} of a \GSLT\
								formula $\varphi$ is given by
								the function $\mthfun{free} : \GSLT \to \pow{\AgnSet \cup \VarSet}$
								defined as follows:
								\begin{itemize}
									\item
										$\mthfun{free}(\apElm) \defeq \emptyset$, where $\apElm \in \APSet$;
									\vspace{-0.35em}
									\item\label{def:sl(freeagvar:neg)}
										$\freeFun{\neg \varphi} \defeq \freeFun{\varphi}$;
									\vspace{-0.35em}
									\item\label{def:sl(freeagvar:conjdisj)}
										$\freeFun{\varphi_{1} \vee \varphi_{2}} \defeq \freeFun{\varphi_{1}}
										\cup \freeFun{\varphi_{2}}$;
									\vspace{-0.35em}
									\item\label{def:sl(freeagvar:next)}
										$\freeFun{\nextt \varphi} \defeq \AgnSet \cup \freeFun{\varphi}$;
									\vspace{-0.35em}
									\item\label{def:sl(freeagvar:untilrelease)}
										$\freeFun{\varphi_{1} \until \varphi_{2}} \defeq \AgnSet \cup
										\freeFun{\varphi_{1}} \cup \freeFun{\varphi_{2}}$;
									\vspace{-0.35em}
									\item\label{def:sl(freeagvar:qnt)}
										$\freeFun{\EExs{\varElm_{1},\ldots, \varElm_{n}}^{\geq g} \varphi}
										\defeq \freeFun{\varphi} \setminus \{ \varElm_{1},\ldots,
										\varElm_{n} \}$;
									\vspace{-0.35em}
									\item\label{def:sl(freeagvar:bndprs)}
										$\freeFun{(\alpha, \xElm) \varphi} \defeq \freeFun{\varphi}$, if $\alpha
										\not\in \freeFun{\varphi}$, where $\alpha \in \AgnSet$ and $\xElm \in
										\VarSet$;
									\vspace{-0.35em}
									\item\label{def:sl(freeagvar:bndrem)}
										$\freeFun{(\alpha, \xElm) \varphi} \defeq (\freeFun{\varphi} \setminus
										\{ \alpha \}) \cup \{ \xElm \}$, if $\alpha \in \freeFun{\varphi}$,
										where $\alpha \in \AgnSet$ and $\xElm \in \VarSet$.
								\end{itemize}
								A formula $\varphi$ without free agents (resp., variables), i.e., with
								$\freeFun{\varphi} \cap \AgnSet = \emptyset$ (resp., $\freeFun{\varphi}
								\cap \VarSet = \emptyset$), is called \emph{agent-closed} (resp.,
								\emph{variable-closed}).
								If $\varphi$ is both agent- and variable-closed, it is called a
								\emph{sentence}.

		\SL\ has a few natural syntactic fragments, the most powerful of which is
		called Nested-Goal \SL.
		Similarly, we define \emph{Nested-Goal \GSLT} (abbreviated \NGGSLT), as a syntactic fragment of
		\GSLT.
		As in  \NGSL, in \NGGSLT we require that bindings and 
		quantifications appear in exhaustive blocks. I.e., whenever there is a 
quantification over a variable in a formula $\psi$ it is part of a consecutive 
sequence of quantifiers that covers all of the free variables that appear in 
$\psi$, and whenever an agent is bound to a strategy then it is part of a 
consecutive sequence of bindings of all agents to strategies. Finally, formulas 
with free agents are not allowed.
		To formalize \NGGSLT we first introduce some notions.
		A \emph{quantification prefix} over a finite set $\VSet \!\subseteq\! \VarSet$ of 
		variables is a sequence $\qntElm \!\in\! \set{ \EExs{\varElm_{1},\ldots,
		\varElm_{n}}^{\geq g}, \AAll{\varElm_{1},\ldots, \varElm_{n}}^{< g} }{n \in \SetN,
		\varElm_{1},\ldots, \varElm_{n} \!\in\! \VSet \land g \!\in\! 
		\SetN \cup \{\aleph_0, \aleph_1, 2^{\aleph_0}\}
		}^{*}$ such that each $\varElm \!\in\!
		\VSet$ occurs exactly once in $\qntElm$.
		%\cbend
		A \emph{binding prefix} is a sequence
		$\bndElm \!\in\! \set{ (\agnElm, \varElm) }{ \alpha \!\in\! \AgnSet \land
		\varElm \!\in\! \VarSet }^{*}$ such that each $\alpha \!\in\!
		\AgnSet$ occurs exactly once in $\bndElm$.
		We denote the set of binding prefixes by $\BndSet$, and the set
		of quantification prefixes over $\VSet$ by $\QntSet(\VSet)$.

		\begin{definition}
			\NGGSLT formulas are built inductively
			using the following grammar, with $\pElm \in \APSet$, $\qntElm \in
			 \QntSet(\VSet)$ $(\VSet \subseteq \VarSet)$,
			 and $\bndElm \in
			\BndSet$:
			\begin{center}
				$\varphi ::= \pElm \mid \neg \varphi \mid
				\varphi \vee \varphi \mid \nextt \varphi \mid \varphi \until \varphi \mid
				 \qntElm \varphi \mid \bndElm \varphi$,
			\end{center}
			where in the rule $\qntElm \varphi$ we require that $\varphi$
			is agent-closed and $\qntElm \in \QntSet(\freeFun{\varphi})$.
		\end{definition}

		Formulas of \NGGSLT\ can be classified according to their \emph{alternation 
		number}, i.e., the maximum number of quantifier switches in a quantification prefix. Formally:

		                \begin{definition}
		                \label{def:sl(freeagvar)}
		                        The 
		                        \emph{alternation number} of a \NGGSLT\
		                        formula is given by:
		                        \begin{itemize}
		                                \item
		                                $\mthfun{alt}(\apElm) \defeq 0$, where $\apElm \in \APSet$;
		                                \vspace{-0.35em}
		                                \item
		                                $\mthfun{alt}(\mthfun{OP}\varphi) \defeq \mthfun{alt}(\varphi)$,
		                                where $\mthfun{OP} \in \{ \neg, \nextt, \bndElm \}$;
		                                \vspace{-0.35em}
		                                \item
		                                $\mthfun{alt}(\varphi_{1} \mthfun{OP} \varphi_{2}) \defeq 
		                                \mthfun{max}(\mthfun{alt}(\varphi_{1}), \mthfun{alt}(\varphi_{2}))$
		                                where $\mthfun{OP} \in \{ \vee, \until \}$;
		                                \vspace{-0.35em}
		                                \item
		                                $\mthfun{alt}(\qntElm \varphi) \defeq 
		                                \mthfun{max}(\mthfun{alt}(\varphi), \mthfun{alt}(\qntElm))$ where $\qntElm$ is a quantification prefix;
		                                \vspace{-0.35em}
		                                \item
		                                $\mthfun{alt}(\qntElm) \defeq 
		                                \sum_{i = 1}^{\card{\qntElm}-1} \mthfun{switch}(\qntElm_i,
		                                \qntElm_{i+1})$, where $\mthfun{switch}(Q,Q') = 
		                                0$ if $Q$ and $Q'$ are either both universal or both existential quantifiers, and $1$ otherwise.
		                        \end{itemize}
		                \end{definition}

		The \emph{quantifier rank} of $\varphi$ is the maximum nesting of 
quantifiers in $\varphi$,
		e.g., $\EExs{\varElm_{1},\ldots, \varElm_{n}}^{\geq g} (\alpha_{1},  
\varElm[1]) \ldots \allowbreak
	(\alpha_{n}, \varElm[n]) \bigwedge_{i = 1}^{n} (\EExs{\ySym} (\alpha_{i},
	\ySym) \psi_{i}) \rightarrow \psi_{i}$ has quantifier rank $2$ if each 
$\psi_i$ is quantifier free.
		Moreover, a \emph{quantifier-block of $\varphi$} is a maximally-consecutive
		sequence of quantifiers in $\varphi$ of the same type (\ie, either all 
existential, or
		all universal). The \emph{quantifier-block rank of $\varphi$} is exactly 
like the quantifier rank
		except that a quantifier block of $j$ quantifiers contributes $1$ instead of 
$j$ to the count.

	We conclude this subsection by introducing \emph{One-Goal \GSLT}, written \OGGSLT.
	The importance of this fragment in \SL stems from the fact that it strictly includes \ATLS\ while maintaining the same complexity for
	both the model checking and the satisfiability problems, \ie
	2\ExpTimeC~\cite{MMPV14}. However, it is commonly believed that Nash
	Equilibrium cannot be expressed in this fragment.
	The definition of \OGGSLT follows.
	\begin{definition}
		\OGGSLT formulas are built inductively using the following grammar, with
		$\pElm \in \APSet$, $\qntElm \in
		\QntSet(\VSet)$ $(\VSet \subseteq \VarSet)$,
		and $\bndElm \in \BndSet$:
			\begin{center}
				$\varphi ::= \pElm \mid \neg \varphi \mid
				\varphi \vee \varphi \mid \nextt \varphi \mid \varphi \until \varphi
				\mid \qntElm \bndElm \varphi$,
			\end{center}
			where $\qntElm$ is a quantification prefix over $\freeFun{\bndElm \varphi}$.
		\end{definition}

	\end{subsection}

Finally, we subscript with $\SetN$ and write \GSLfin, \OGGSLfin, and \NGGSLfin\ for the fragments in which all grades are from the set $\SetN$. 

	\begin{subsection}{Semantics}

		As for \SL, the interpretation of a \GSLT formula requires a
		valuation of its free placeholders.
		
		\begin{definition}
		An \emph{assignment (from $s$)} is a function $\asgFun \! \in
		\AsgSet(s) \!\defeq \!(\VarSet \cup \AgnSet) \!\to\! \StrSet(s)$ mapping
		variables and agents to strategies.
		\end{definition}
				We denote by ${\asgFun}[\eElm \mapsto \strElm]$, with $\eElm \in \VarSet
				\cup \AgnSet$ and $\strElm \in \StrSet(s)$, the assignment
				that differs from $\asgFun$ only in the fact that $\eElm$ maps to $\strElm$.
		Extend this definition to tuples: for $\overline{\eElm} = (e_1, \dots,
		e_n)$ with $e_i \neq e_j$ for $i \neq j$, define
		${\asgFun}[\overline{\eElm} \mapsto \overline{\strElm}]$ to be
		the assignment that differs from $\asgFun$ only in the fact that $\eElm_i$
		maps to $\strElm_i$ (for each $i$).

		Since an assignment ensures that all free variables are associated with strategies, it induces a play.
		\begin{definition}
		For an assignment $\asgFun \in \AsgSet(\sElm)$ the \emph{$(\asgFun, \sElm)$-play}
		denotes the path $\pi \in \PthSet(s)$ such that for all $i \in \SetN$, it holds that $\pi_{i + 1} =
		\trnFun(\decFun)(\pi_{i})$, where $\decFun(\alpha) \defeq
		\asgFun(\alpha)(\pi_{\leq i})$ for $\alpha \in
		\AgnSet$. 				The function $\playFun : \AsgSet \times \SttSet \to \PthSet$, with
				$\dom{\playFun} \defeq \set{ (\asgFun, \sttElm) }{ \asgFun \in \AsgSet(\sttElm)
				}$, maps  $(\asgFun, \sttElm)$ to the $(\asgFun,
				\sttElm)$-play $\playFun(\asgFun, \sttElm) \allowbreak \in
				\PthSet(\sttElm)$.
		\end{definition}

		The notation $\pi_{\leq i}$ (resp. $\pi_{< i}$) denotes the prefix of the sequence $\pi$ of length $i$ (resp. $i-1$). Similarly, the notation
		$\pi_i$ denotes the $i$th symbol of $\pi$. Thus, 
$\playFun(\asgFun,\sttElm)_i$ is the $i$th state on the play determined by 
$\asgFun$ from $\sttElm$.

 		The following definition of $\asgFun_i$ says how to interpret an assignment $\asgFun$ starting from a point $i$ along the play, i.e.,
 		for each placeholder $e$, take the action the strategy $\asgFun(e)$ would do if it were given the prefix of the play up to $i$ followed by
 		the current history.
 		\begin{definition}
		For $\asgFun \in \AsgSet(\sElm)$ and $i \in \Nat$, writing
		$\rho \defeq \playFun(\asgFun,\sElm)_{\leq i}$ (the prefix of the play up to $i$) and $\tElm \defeq \playFun(\asgFun,\sElm)_i$ (the last state of $\rho$)
		define $\asgFun_i \in \AsgSet(\tElm)$ to be the assignment from $\tElm$ that maps
		$e \in \VarSet \cup \AgnSet$ to the strategy that maps $h \in \HstSet(\tElm)$ to the action $\asgFun(e)(\rho_{< i} \cdot h)$.
 		\end{definition}
		
			The semantics of \GSLT\ mimics the one for \SL as given in~\cite{MMPV14}.
      Given a $\CGS$ $\CGSName$, for all
			states $\sttElm \in \SttSet$ and assignments $\asgFun \!\in\!
			\AsgSet(s)$, we now define the relation $\CGSName\!, \asgFun\!, \sttElm \!\models
			\!\varphi$, read \emph{$\varphi$ holds at $\sttElm$ in $\CGSName$ under $\asgFun$}.
			                \begin{definition}
			                        \label{}
			                        Fix a $\CGS$ $\CGSName$. For all states $\sttElm \in \SttSet$ and assignments $\asgFun \in \AsgSet(s)$,
			                        the relation $\CGSName, \asgFun, \sttElm \models \varphi$ is defined inductively on the structure of $\varphi$:
			                        \begin{itemize}
			                                \item
			                                        $\CGSName, \asgFun, \sttElm \models \apElm$ iff $\apElm \in
			                                        \apFun(\sttElm)$;
			                                \item
			                                        $\CGSName, \asgFun, \sttElm \models \neg \varphi$ iff
			                                        $\CGSName, \asgFun, \sttElm \not\models \varphi$;
			                                \item
			                                        $\CGSName, \asgFun, \sttElm \models \varphi_{1} \vee \varphi_{2}$ iff
			                                        $\CGSName, \asgFun, \sttElm \models \varphi_{1}$ or
			                                        $\CGSName, \asgFun, \sttElm \models \varphi_{2}$;
			                                \item
			                                        $\GName, \asgFun, \sttElm \models \nextt \varphi$ iff $\GName, \asgFun_1,
			                                        \playFun(\asgFun, \sttElm)_{1} \models \varphi$;
			                                \item
			                                        $\GName, \asgFun, \sttElm \models \varphi_{1} \until \varphi_{2}$ iff
			                                        there is an index $i \in \SetN$ such that $\GName, \asgFun_i,
			                                        \playFun(\asgFun, \sttElm)_{i} \models \varphi_{2}$ and, for all indexes $j
			                                        \in \SetN$ with $j < i$, it holds that $\GName, \asgFun_j,
			                                        \playFun(\asgFun,\sttElm)_{j} \models \varphi_{1}$;
			                                \item $\CGSName, \asgFun, \sttElm \models (\alpha, \varElm) \varphi$ iff
			                                        $\CGSName, {\asgFun}[\alpha \mapsto \asgFun(\varElm)], \sttElm \models
			                                        \varphi$;
			                        \item 
			                 			$\CGSName, \asgFun,\sttElm \!\models \!\EExs{\varElm_{1},\ldots,
			           					\varElm_{n}}^{\geq g} \varphi\!$                         					iff there exist $g$ many $n$-tuples of strategies $\overline{\strElm_{i}}$ ($0 \leq i < g$) such that:
			                            \it
			                            \- $\overline{\strElm_{i}} \neq \overline{\strElm_{j}}$ for $i \neq
		    								j$, and
			                                						
			                            \- $\CGSName, {\asgFun}[\overline{\varElm} \mapsto
    								\overline{\strElm_{i}}],
			                         \sttElm  \models \varphi$
			                         for $0\leq i < g$.
			                         \ti

			                        \end{itemize}
			                \end{definition}

			Intuitively, $\EExs{\varElm_{1},\ldots, \varElm_{n}}^{
			\geq g} \varphi\!$ expresses that the number of
			distinct tuples of strategies that satisfy $\varphi$ is at least $g$.

		As usual, if $\asgFun$ and $\asgFun'$ agree on $\freeFun{\varphi}$, then
		$\CGSName, \asgFun,\sttElm \models \varphi$ if and only if $\CGSName, \asgFun',\sttElm \models \varphi$, i.e., the
		truth of $\varphi$ does not depend on the values the assignment takes on placeholders that are not free.
		Thus, for a sentence $\varphi$, we write $\CGSName \models \varphi$ to mean
		that $\CGSName, \asgFun, \isttElm \models \varphi$ for some (equivalently, for all)
		assignments $\asgFun$, and where $\isttElm$ is the initial state of $\CGSName$.
	\end{subsection}
	
\end{section}

% End of file SectionI.tex

		\begin{section}{Illustrating \GSLT: uniqueness of solutions} \label{sec:illustration}

	In game theory, players have objectives that are summarized in a payoff
	function that maps plays to real numbers.
	In order to specify such payoffs with formulas, we follow a formalization from~\cite{KPV14}
	called \emph{objective \LTL}.
	We then discuss appropriate solution concepts, and show how to express these in
	\GSLT.
		
	Let $\CGSName$ be a \CGS with $n$ agents.
	Let $m \in \Nat$ and fix, for each agent $\alpha_i \in \AgnSet$, an \emph{objective} tuple
	$S_i \defeq \tpl{\fFun_i,\varphi_i^1,\dots,\varphi_i^m}$, where $\fFun_i:\{0,1\}^m \to
	Z$, and each $\varphi_i^j$ is an \LTL\ formula over $\APSet$.
	If $\pi$ is a play, then agent $\alpha_i$ receives payoff $\fFun_i(\bar{h}) \in
	\Nat$ where $\bar{h}_j$, the $j$th bit of $\bar{h}$, is $1$ if and only if $\pi \models \varphi_i^j$. We assume agents are trying to maximize their payoffs. 
	
	In case each $f_i:\{0,1\}^m \to \{-1,1\}$ we say that the game is \emph{win/lose}.
	If $\sum_{1 \leq i \leq n} \fFun_i(\bar{h}) = 0$ for all $\pi$, then $\CGSName$ is a \emph{zero-sum game}, otherwise it is a \emph{non zero-sum game}. 
			
	\noindent {\bf Two player, Win/Lose,  Zero-sum games.}
	In these types of games the main solution concept is the \emph{winning strategy}. In general, a strategy is winning for agent $\alpha_i$ if and only if for all strategies of adversarial agents the resulting induced play has payoff $1$ for agent $i$. In~\cite{MMS15} the authors describe a two-player game named ``Cop and the Robber'', played in a maze, in which the objective of the Robber is to reach an exit (and thus the objective of the Cop is to ensure the Robber never reaches the exit). The authors describe two closely related mazes in which the Robber has, respectively, exactly one and exactly two winning strategies. Both these properties can be easily expressed by \GSL. For instance, the Robber has a single \LTL\ objective $\varphi \defeq \eventually exit$, and the following formula of \GSLfin\ expresses that the Robber has exactly two winning strategies: 
	 \[
	 \EExs{\varElm}^{\geq 2} \AAll{\ySym} (\text{Robber},\varElm)(\text{Cop}, \ySym)\varphi \land 
	 \neg \EExs{\varElm}^{\geq 3} \AAll{\ySym} (\text{Robber},\varElm)(\text{Cop}, \ySym)\varphi
	 \]
	
	\noindent {\bf Non zero-sum games.}
	The central solution concept in these type of games is the Nash Equilibrium.
	A tuple of strategies, one for each player, is called a \emph{strategy profile}.
	A strategy profile is a \emph{Nash equilibrium ($N\!E$)} if
	no agent can increase his payoff by unilaterally choosing a different strategy.
	A game may have zero, one, or many NE. Suppose $\phi_{N\!E}(\overline{x})$ expresses that the profile $\overline{x} \defeq (\varElm[1] 
	\ldots \varElm[n])$  is a NE. Then the following formula expresses that there is a unique NE:
		\begin{equation} \label{eqn:uniqueNE}
				\EExs{\varElm[1], \ldots, \varElm[n]}^{\geq 1}
				\phi_{N\!E}(\overline{\varElm}) \wedge \neg \EExs{\varElm[1], \ldots,
				\varElm[n]}^{\geq 2} \phi_{N\!E}(\overline{\varElm}).
				\end{equation}
	
	We now describe $\phi_{N\!E}$, first in the win/lose setting, and then in the general case.
	
First, consider a win/lose setting in which the objective of agent $\alpha_i \in \AgnSet$ is a single \LTL\ formula $\varphi_i$ and the payoff is $1$ if the formula is true and $-1$ otherwise. The 
following formula expresses that $\overline{x} \defeq (\varElm[1] 
\ldots \varElm[n])$ is a Nash Equilibrium:
	\[
		\AAll{\ySym_1} \ldots \AAll{\ySym_n}
	\bigwedge_{i = 1}^{n} ( \bndElm_{i}  \varphi_{i})
	\rightarrow  \bndElm \varphi_{i}
	\]
	where $\bndElm = (\alpha_{1}, \varElm[1]) \ldots (\alpha_{n},
	\varElm[n])$, and $\bndElm_{i} = (\alpha_{1},  \varElm[1]) \ldots
	(\alpha_{i-1} , \allowbreak \varSym[i-1])(\alpha_{i}, \ySym_{i})
	(\alpha_{i+1}, \varSym[i+1])\ldots (\alpha_{n}, \varElm[n])$. Although this is not in the nested-goal fragment, (the quantifiers don't bind the $x$s), plugging it into Formula~(\ref{eqn:uniqueNE}) results in a formula of \NGGSLfin.

Next, consider the case that each agent $\alpha_i$ has a general objective 
tuple $S_i \defeq \tpl{\fFun_i,\varphi_i^1,\dots,\varphi_i^m}$.
Given a vector $\bar{h} \in \{0,1\}^m$, let $gd_i(\bar{h}) \defeq 
\{ \bar{h}' \in \{0,1\}^m \mid \fFun_i(\bar{h}') \geq 
\fFun_i(\bar{h}) \}$ be the set of vectors $\bar{h}'$ for which the 
payoff for agent 
$\alpha_i$ is not worse than for $\bar{h}$. Also, let 
$\eta_i^{\bar{h}}$ be
the formula obtained by taking a conjunction of the formulas $\varphi_i^1,
\dots,\varphi_i^m$ or their negations according to $\bar{h}$, i.e., by 
taking $\varphi_a^j$ if the $j$th bit in $\bar{h}$ is $1$, and otherwise 
taking $\neg \varphi_a^j$.
Formally, $\eta_i^{\bar{h}} \defeq \wedge_{j \in \{1 \leq j \leq m \mid 
\bar{h}_j=1\}} \varphi_i^j \bigwedge \allowbreak \wedge_{j \in \{1 \leq j \leq m \mid 
\bar{h}_j=0\}} \neg \varphi_i^j$. Observe that the following formula says that 
$\overline{x} \defeq (\varElm[1] \ldots \varElm[n])$ is a Nash Equilibrium:
\[    
\AAll{\ySym_1} \ldots
\AAll{\ySym_n} \bigwedge_{i = 1}^{n} \bigwedge_{\bar{h} \in \{0,1\}^m} 
(\bndElm_{i} \eta_i^{\bar{h}}) \rightarrow \bigvee_{\bar{h}' \in 
gd_i(\bar{h})} \bndElm \eta_i^{\bar{h}'}
\]
Again, plugging it into Formula~(\ref{eqn:uniqueNE}) results in a formula of \NGGSLfin\ expressing there is a unique NE.

It has been argued (in \cite{Ummels06,KPV14}) that
NE may be implausible when used for sequential games (of which iterated one
shot games are central examples), and that a more robust notion is
subgame-perfect equilibrium~\cite{Sel65}.
Given a game $\GName$, a strategy profile is a \emph{subgame-perfect
equilibrium (SPE)} if starting at any reachable subgame, the profile is a NE.
The following formula expresses that $\overline{x} \defeq (\alpha_{1},
\varElm[1]) \ldots (\alpha_{n}, \varElm[n])$ is an SPE:
	\[
	\phi_{S\!P\!E}(\overline{\varElm}) \!\defeq\!
	\AAll{\zElm[1], \ldots, \zElm[n]} (\alpha_{1},  \zElm[1]) \ldots
	(\alpha_{n}, \zElm[n]) \always
	\phi_{N\!E}(\overline{\varElm})
	\]
Using graded modalities, we can thus express there is a unique SPE using the 
following \NGGSLfin\ formula:
	\[
	\EExs{\varElm[1], \ldots, \varElm[n]}^{\geq 1}
	\phi_{{S\!P\!E}}(\overline{\varElm}) \wedge \neg \EExs{\varElm[1], \ldots,
	\varElm[n]}^{\geq 2} \phi_{{S\!P\!E}}(\overline{\varElm}).
	\]
	\end{section}

%\cbend
%\end{section}

	% Begin of file SectionII.tex

\begin{section}{Model-checking \GSLT}
	\label{sec:MC}
	In this section we study the model-checking problem for \GSLT\ and show
	that it is decidable with a time-complexity that is non-elementary (i.e., not bounded by any fixed
	tower of exponentials). However, it is elementary if the number of blocks of quantifiers is fixed.
	For the algorithmic procedures, we follow an \emph{automata-theoretic
	approach}~\cite{KVW00}, reducing the decision problem for the logic to the
	emptiness problem of an automaton.
	The procedure we propose here extends that used for \SL\ in~\cite{MMPV14}. The
	only case that is different is the new graded quantifier over tuples of strategies.

	We start with the central notions of automata theory, and then show how to convert a
	\GSLT\ sentence $\varphi$ into an automaton that accepts exactly the (tree encodings) of the concurrent game structures
	that satisfy $\varphi$. This is used to prove the main result about \GSLT\ model checking.

	{\bf Automata Theory.}
		A \emph{$\Sigma$-labeled $\Upsilon$-tree} $\T$ is a pair $\tpl{T,\vFun}$ where 
		$T \subseteq \Upsilon^+$ is prefix-closed (i.e., if $t \in T$ and $s \in 
\Upsilon^+$ is a prefix of $t$ then also $s \in T$),
		and $\vFun:T \to \Sigma$ is a labeling function. Note that every word $w \in 
		\Upsilon^+ \cup \Upsilon^\omega$ with the property that every prefix of $w$ 
is in $T$,
		can be thought of as a path in $\T$. Infinite paths are called \emph{branches}.
		
		\emph{Nondeterministic tree automata} (\emph{\NTA}) are a
		generalization to infinite trees of the classical automata on words~\cite{thomas1990automata}.
		\emph{Alternating tree automata} (\emph{\ATA}) are a further
		generalization of nondeterministic tree automata~\cite{EJ91}.
		Intuitively, on visiting a node of the input tree, while an \NTA\ sends
		exactly one copy of itself to each of the successors of the node, an \ATA\
		can send several copies to the same successor. We use the parity acceptance
		condition~\cite{KVW00}.

		 For a set $X$, let $\B^+(X)$ be the set of positive Boolean formulas over $X$,
		 including the constants \textbf{true} and \textbf{false}.
		 A set $Y \subseteq X$ satisfies a formula $\theta \in \B^+(X)$, written $Y
		 \models \theta$, if assigning \textbf{true} to elements in $Y$ and
		 \textbf{false} to elements in $X \setminus Y$ makes $\theta$ true.

		 		\begin{definition}
		 			\label{def:ata}
		 			An Alternating Parity Tree-Automaton (\APT) is a tuple $\AName \defeq
		 			\tpl{\Sigma,\Delta,\QSet,\delta,\qElm[0],\mathcal{F}}$, where 
		 			\it
		 			\- $\Sigma$ is the \emph{input alphabet}, 
		 			\- $\Delta$ is a set of \emph{directions}, 
		 			\- $\QSet$ is a finite set of
		 			\emph{states}, 
		 			\- $\qElm[0] \in \QSet$ is an \emph{initial state}, 
		 			\- $\delta : \QSet \times \Sigma \to \B^{+}(\Delta
		 			\times \QSet)$ is an \emph{alternating transition function}, and
		 			\-
		 			 $\mathcal{F}$, an \emph{acceptance condition}, is of the form
		 			 $(\FSet_{1},
		 		\ldots, \FSet_{k}) \in (\pow{\QSet})^{+}$ where $\FSet_{1} \subseteq \FSet_{2} \ldots
		 		\subseteq \FSet_{k} = \QSet$.
		 			\ti
		 		\end{definition}
		 		The set $\Delta \times \QSet$ is called
		 		the set of \emph{moves}.
		 		An \NTA\ is an \ATA\ in which each conjunction in
		 		the transition function $\delta$ has exactly one move $(\dElm, \qElm)$
		 		associated with each direction $\dElm$.
		 
		 		An \emph{input tree} for an \APT\ is a $\Sigma$-labeled $\Delta$-tree $\T =
		 		\tpl{\TSet, \vFun}$.
		 		A \emph{run} of an \APT\ on an input tree $\T =
		 		\tpl{\TSet, \vFun}$ is a $(\Delta \times \QSet)$-tree $\R$ such that, for
		 		all
		 		nodes $\xElm \in \RSet$, where $\xElm = (\dElm[1], \qElm[1]) \dots (\dElm[n], \qElm[n])$ (for some $n \in \Nat$),
		 		it holds that \emph{(i)}		$\yElm \defeq (\dElm[1],\dots,\dElm[n]) \in \TSet$ and
		 		\emph{(ii)} there is a set of moves $\SSet \subseteq \Delta \times \QSet$
		 		with $\SSet \models \delta(\qElm[n], \vFun(\yElm))$ such that $\xElm \cdot
		 		(\dElm, \qElm) \in \RSet$ for all $(\dElm, \qElm) \in \SSet$.

		 		The acceptance condition allows us to say when a run is successful.
		 		Let $\R$ be a run of an \APT\ $\AName$ on an input tree $\T$ and $u \in (\Delta \times \QSet)^\omega$
		 		one
		 		of its branches.
		 		Let $\infFun(u) \subseteq \QSet$
		 		denote the set of states that occur in infinitely many moves of $u$.
		 		Say that $u$ \emph{satisfies the parity acceptance condition} $\mathcal{F}
		 		\!=\! (\FSet_{1}, \ldots, \FSet_{k})$ if the least index $i \!\in\!
		 		\numcc{1}{k}$ for which $\infFun(u) \cap \FSet_{i} \neq \emptyset$ is even.
		 		A run is \emph{successful} if all its branches satisfy the parity acceptance condition $\mathcal{F}$. 
		 		An \APT\ \emph{accepts} an input tree $\T$ iff
		 				there exists a successful run $\R$ of $\AName$ on $\T$.

		The \emph{language} $\LangSet(\AName)$ of the \APT\ $\AName$ is the set  of trees $\T$ accepted by $\AName$.
		Two automata are \emph{equivalent} if they
		have the same language.
		The \emph{emptiness problem} for alternating parity tree-automata is to decide, given $\AName$, whether
		$\LangSet(\AName) = \emptyset$. The \emph{universality problem} is to decide whether \AName\ accepts all trees.

	\begin{subsection}{From Logic to Automata}
		\label{subsec:FLtA}
		Following an automata-theoretic approach, we reduce the model-checking problem of \GSLT\ to the
		emptiness problem for alternating parity tree automata \cite{MMPV14}. The
main step is to translate every \GSLT\ formula $\varphi$ (i.e.,
		$\varphi$ may have free placeholders), concurrent-game structure $\GName$, and state $\sttElm$, into an \APT\ that accepts a tree
		if and only if the tree encodes an assignment $\asgFun$ such that $\GName,\asgFun,\sttElm \models \varphi$.
		
		We first describe the encoding, following \cite{MMPV14}. Informally, the
\CGS\ $\GName$ is encoded by its ``tree-unwinding starting from $\sttElm$''
whose
		nodes represent histories, i.e., the $\SttSet$-labeled $\SttSet$-tree $\T \defeq \tpl{\HstSet(s),\uFun}$ such that $\uFun(h)$ is the last symbol of $h$.
		Then, every strategy $\asgFun(e)$ with $e \in \freeFun{\varphi}$
		is encoded as an
		$\ActSet$-labeled tree over the unwinding. The unwinding and these strategies $\asgFun(e)$ are viewed as a single $(\ValSet \times \SttSet)$-labeled tree
		where $\ValSet \defeq \freeFun{\varphi} \to \ActSet$.
		
		\begin{definition} \label{dfn:encoding}
		The \emph{encoding of $\asgFun$ (w.r.t.\ $\varphi,\GName,\sttElm$)} is the $(\ValSet \times \SttSet)$-labeled $\SttSet$-tree
		$\T \defeq \tpl{\TSet,\uFun}$ such that $\TSet$ is the set of histories $h$ of $\GName$ starting with $\sttElm$
		and $\uFun(h) \defeq (\fFun,q)$ where $q$ is the last symbol in $h$ and $\fFun:\freeFun{\varphi} \to \ActSet$ is defined by $\fFun(e) \defeq \asgFun(e)(h)$ for all
		$e \in \freeFun{\varphi}$.\footnote{In case $\freeFun{\varphi} = \emptyset$, then $\fFun$ is the (unique) empty function.
		In this case, the encoding of every $\asgFun$ may be viewed as the tree-unwinding from $s$.}
		\end{definition}

		The following lemma is proved by induction on the structure of the formula $\varphi$, as in \cite{MMPV14}. The idea for handling the new case, i.e., the graded quantifier $\EExs{\varElm_{1},\ldots,\varElm_{n}}^{\geq g} \psi$, is to build an APT that is a projection of an APT that itself checks that each of the $g$ tuples of strategies satisfies $\psi$ and that each pair of $g$ tuples is distinct.

		\begin{lemma}
			\label{lem:apt}
			For every \GSLT\ formula $\varphi$, \CGS\ $\GName$, and state $\sElm \in \SttSet$,
			there exists an \APT\ $\AName_\varphi$ such that for all assignments $\asgFun$, if
			$\T$ is the encoding of $\asgFun$ (w.r.t.\ $\varphi,\GName,\sttElm$), then
			$\GName, \asgFun, \sElm \models \varphi$ iff $\T \in
			\LangSet(\AName_\varphi)$.

		\end{lemma}

\begin{proof}

		 			As in \cite{MMPV14} we induct on the structure of the formula $\varphi$ to construct the corresponding automaton $\AName_\varphi$.
		 			The Boolean operations are easily dealt with using the fact that disjunction corresponds to non-determinism,
		 			and negation corresponds to dualizing the automaton.
		 			Note $(\dagger)$ that thus also conjunction is dealt with due to 			De Morgan's laws.
		 			The temporal operators
		 			are dealt with by following the unique play (determined by the given assignment) and verifying the required subformulas, e.g., for $\nextt \psi$
		 			the automaton, after taking one step along the play, launches a copy of the automaton for $\psi$.
		 			All of these operations incur a linear blowup in the size of the automaton.
		 			The only case that differs from \cite{MMPV14}
		 			is the quantification, i.e., we need to handle the case that  $\varphi = \EExs{\varElm_{1},\ldots,\varElm_{n}}^{\geq g} \psi$.
		 						Recall that
		 					$\CGSName, \asgFun,\sttElm \!\models \!\EExs{\varElm_{1},\ldots,
		 					\varElm_{n}}^{\geq g} \psi\!$
		 					iff there exists $g$ many tuples $\overline{\strElm_{i}}$
		 					of strategies such that:
		 						$\overline{\strElm_{a}} \neq \overline{\strElm_{b}}$ for $a \neq b$, and
		 						$\CGSName, {\asgFun}[\overline{\varElm} \mapsto
		 						\overline{\strElm_{i}}],\sttElm  \models \psi$
		 						for $0 \leq i < g$.
		 						
		 			There are two cases.

		 			{\bf Case $g \in \{\aleph_0, \aleph_1, 2^{\aleph_0}\}$.} 
		 			
		 			Write MSOL for monadic second-order logic in the signature of trees. We use the following two results.
		 			
		 			\begin{enumerate}
		 			\item For every MSOL formula $\alpha(Y)$ there exists an \APT\ accepting the set of trees $Y$ such that $\alpha(Y)$ holds. Conversely, for every \APT\ there is an MSOL formula $\alpha(Y)$ that holds on those $Y$ that are accepted by the \APT, see \cite{thomas1990automata}.
		 			\item 
		 			For every MSOL formula $\alpha(\overline{X},Y)$ and $\kappa \in \{\aleph_0,\aleph_1,2^{\aleph_0}\}$ there exists an MSOL formula $\beta(\overline{X})$ equivalent to ``there exist $\kappa$ many trees $Y$ such that $\alpha(\overline{X},Y)$''~\cite{BaranyKR10}.
		 			\end{enumerate}

		 			Thus, consider $\EExs{\varElm_{1},\ldots,\varElm_{n}}^{\geq g} \psi$. By induction, there is an \APT\ $\DName$ for $\psi$. Compile $\DName$ into an MSOL formula $\alpha$, and then produce an MSOL formula $\beta$ that holds iff ``there exist $\kappa$ many tuples of trees such that $\alpha$'' (recall that a tuple of trees is coded as a single tree). Finally, convert $\beta$ back into an \APT.
		 				
		 			Note that the blowup in the translations (MSOL to \APT, and closure under ``there exists $\kappa$ many trees'') is non-elementary.

		 			{\bf Case $g \in \Nat$.}
		 			
		 			We show how to build an \NPT\ for $\varphi$ that mimics this definition: it will be a projection of an \APT\ $\DName_\psi$, which itself is the
		 			intersection of two automata, one checking that each of
		 			the $g$ tuples of strategies satisfies $\psi$, and the other checking that each pair of the $g$ tuples of strategies is distinct.
		 
		 			In more detail, introduce a set of fresh variables $X \defeq \{ x_i^j \in \VarSet : i \leq n, j \leq g\}$, and consider the formulas
		 			$\psi^j$ (for $j \leq g$) formed from $\psi$ by renaming $x_i$ (for $i \leq n$) to $x_i^j$. Define $\psi' \defeq \wedge_{j \leq g} \psi^j$.
		 			Note that, by induction, each $\psi^j$ has a corresponding \APT, and thus, using the conjunction-case $(\dagger)$ above,
		 			there is an \APT\ $\BName$ for $\psi'$. Note that the input alphabet for $\BName$ is $(\freeFun{\psi'} \to \ActSet) \times \SttSet$ and that
		 			$X \subseteq \freeFun{\psi'}$.
		 			
		 			On the other hand, let $\CName$ be an \APT\ with input alphabet $(\freeFun{\psi'} \to \ActSet) \times \SttSet$ that accepts a tree $\T = \tpl{\TSet,\vFun}$
		 			if and only if for every $a \neq b \leq g$ there exists $i \leq n$ and $h \in \TSet$ such that $\vFun(h) = (f,q)$ and $f(x_i^a) \neq f(x_i^b)$.

		 			Form the \APT\ $\DName_\psi$ for the intersection of $\BName$ and $\CName$.
		 			
		 			Now, using the classic transformation~\cite{MS95},
		 			we remove alternation from the \APT\ $\DName_\psi$
		 			to get an equivalent \NPT\  $\NName$ (note that this step costs an exponential).
		 			Finally, use the fact that \NPT s are closed under projection (with no blowup) to get an \NPT\ for the language
		 			$proj_X(L(\NName))$ of trees that encode assignments $\asgFun$ satisfying $\varphi$.

		 			 For completeness we recall this last step.
		 			 If $L$ is a language of $\Sigma$-labeled trees with
		 			$\Sigma \defeq \ASet \to \BSet$, and $X \subset \ASet$, then the \emph{$\XSet$-projection of $L$}, written $proj_{\XSet}(L)$, is the language of
		 			$\Sigma'$-labeled trees with $\Sigma' \defeq \ASet \setminus \XSet \to \BSet$ such that $\T \defeq \tpl{\TSet,\vFun} \in proj_{\XSet}(L)$ if and only if
		 			there exists an $\XSet$-labeled tree $\tpl{\TSet,\wFun}$ such that the language $L$ contains the tree $\tpl{\TSet,\uFun}$
		 			where $\uFun: \TSet \to (\ASet \to \BSet)$ maps 			$t \in \TSet$ to $\vFun(t) \cup \wFun(t)$.
		 			%
		 %			the function that maps $x \in \ASet \setminus \XSet$ to $\vFun(x)$ and maps $x \in \XSet$ to $\wFun(x)$,
		 			Now, if $\NName$ is an \NPT\  with input alphabet $\Sigma \defeq \ASet \to \BSet$, and if $\XSet \subset \ASet$, then there is an \NPT\  with input alphabet
		 			$\Sigma' \defeq \ASet \setminus \XSet \to \BSet$ with language $proj_{\XSet}(L(\NName))$.
		 %			\NB{give transition relation for projection}

		 			The proof that the construction is correct is immediate. 
		 			\hfill  $\Square$
		 			\end{proof}

We make some remarks about this lemma. First, all the cases in the induction 
incur a linear blowup except for the quantification case. For the quantification case, if $g \in \Nat$ then the translation incurs an exponential blowup (the translation from an \APT\ to an \NPT\ results in an exponentially larger automaton \cite{KVW00}); in case $g \in \{\aleph_0,\aleph_1,2^{\aleph_0}\}$ the blowup is non-elementary. In case all grades are from $\Nat$, we can say a
little more. Note that a block of $k$ identical quantifiers only costs, in the
worst case, a single exponential (and not $k$ many exponentials) because we can
extend the proof above to deal with a block of quantifiers at once. Thus, we get
that the size of the \APT\ for $\varphi$ is a tower of exponentials whose height is the
quantifier-block rank of $\varphi$.

		\begin{theorem}
			\label{thm:MCSL}
			The model-checking problem for $\GSLT$ is decidable. Regarding complexity:
			\en
			\- The complexity is not bounded by any fixed tower of exponentials.
			
			\- The complexity is	\PTimeC\ w.r.t.\ the size of the			model.
			
			\- If all grades are restricted to be in $\Nat$, then the complexity is $(k+1)$\ExpTime\ if $k \geq 1$ is the quantifier-block rank of $\varphi$.
			Moreover, if $\varphi$ is the form $\qntElm \psi$, where $\qntElm$ is a quantifier-block, and $\psi$ is of  quantifier-block rank $k-1$, then the complexity is $k$\ExpTime.
			\ne
			\end{theorem}

	\begin{proof}
	The lower-bounds already holds for \SL\ \cite{MMPV14}.
	For decidability, use Lemma~\ref{lem:apt} to transform the $\CGS$ and $\varphi$ into an \APT\ and test its emptiness. For the upper bound in item 2., use the fact that the membership problem for \APT\ is in \PTime. For the upper bound in item 3., proceeds as follows.
	The complexity of checking emptiness (or indeed, universality) of an \APT\ is in \ExpTime\cite{KVW00}.
	As discussed after Lemma~\ref{lem:apt}, for the case that all grades are in $\SetN$, the size of the \APT\ is a tower of 
	exponentials whose height is the quantifier-block rank of $\varphi$.
	This gives the $(k+1)$\ExpTime\ upper bound.
			\end{proof}
		\begin{theorem}
			\label{thm:MCNG}
			The model-checking problem for $\NGGSLT$ is
			\PTimeC\ w.r.t.\ the size of the model, and $(k + 1)$-\ExpTime\ when
			restricted to formulas of maximum alternation number $k$ and grades in $\tt$.
		\end{theorem}

		\begin{proof}
			The lower bound already holds for \NGSL~\cite{MMPV14}, and the \PTime upper bound is inherited from Theorem~\ref{thm:MCSL}.
			The upper bound for $\NGGSLfin$ is obtained by following the same reasoning for \NGSL 			of the singleton existential quantifier~\cite{MMPV14} but using the automaton
			construction as	in Theorem~\ref{thm:MCSL}.		
		\end{proof}
	
Directly from the statements reported above, we get the following results:
		
\begin{theorem}
Checking the uniqueness of NE, and checking the uniqueness of SPE,
can be done in $2$\ExpTime.
\end{theorem}

		We conclude this section with the complexity of the model
		checking problem for \OGGSLfin.
		Also in this case one can derive the lower bound from the one holding for
		the corresponding sub-logic in \SL (\ie, \OGSL) and the upper bound by using
		the same algorithm for \OGSL but plugging a (yet no more complex)
		different automata construction for the new existential quantifier
		modality. Indeed the model checking problem for \OGGSLfin\ is 2\ExpTimeC.
		It is worth recalling that \OGSL strictly subsumes \ATLS~\cite{MMPV14}.
		It is quite immediate to see that this also holds in the
		graded setting (note that \ATLS already allows quantifying over tuples of agents' (bound) strategies).
		As the model checking for \ATLS is already 2\ExpTimeH, we get that
		also for the graded extension for this logic, which we name
		$\txtname{G}\ATLS$, the model checking problem is 2\ExpTimeC.
		The model checking results for both $\txtname{G}\ATLS$ and \OGGSLT are
		reported in the following theorem.
		
		\begin{theorem}
			\label{thmII}
			The model-checking problem for $\txtname{G}\ATLS$ and \OGGSLfin\ is
			\PTimeC\ \wrt the size of the model and 2\ExpTimeC\ in the size of
			formula.
		\end{theorem}

		\end{subsection}

\end{section}

% End of file SectionII.tex

% Begin of file Conclusion.tex

\begin{section}{Conclusion}\label{conclusion}

We introduced \GSLT, a logic for strategic reasoning that has the ability to count tuples of strategies, \ie, partial strategy-profiles. The logic is elegant, simple, and powerful: indeed, we can use it to solve the unique NE problem for \LTL\ objectives in $2$\ExpTime, and thus at the same complexity that is required to merely decide if a NE exists. 

The positive results presented in this paper open several directions for future work:

\en
\- Study the relative expressive power of fragments of \GSL.

\- In order to express uniqueness of solution concepts, we added grades to the strategy quantifiers. One may, instead, study the expressive power of \SL with the atomic expression $x = y$ (saying that strategies $x$ and $y$ are the same), just as one does for first-order logic with and without equality. Note that equality cannot replace the infinite grades. However, if one restricts to finite grades, are there advantages of doing it one way rather than the other? 

\- We showed how to compile quantified formulas into automata. Other generalized quantifiers have been studied in the context of automata and monadic second-order logic~\cite{DBLP:conf/csl/Bojanczyk04, DBLP:journals/bsl/Rubin08,bojanczyk_et_al:LIPIcs:2012:3427}, including the boundedness quantifier and the modulo-counting quantifiers. Thus, one might try understand which generalized quantifiers are useful for strategic reasoning.

\- One might implement \GSLT\ and its model-checking procedure in a formal verification tool such as SLK-MCMAS~\cite{CLMM14,CLM15}. 
\ne

\end{section}

%%****************************************************************************%%
%%                                                                            %%
%% Article Title                                                              %%
%%                                                                            %%
%% Acknowledgments.tex                                                        %%
%%                                                                            %%
%% Revision 0                                                                 %%
%%                                                                            %%
%% Copyright (C) 20xx, Fabio Mogavero.                                        %%
%% All rights reserved.                                                       %%
%%                                                                            %%
%%****************************************************************************%%

% Begin of file Acknowledgments.tex

\begin{section}*{Acknowledgments}

We thank Michael Wooldridge for suggesting uniqueness
of Nash Equilibria as an application of graded strategy logic.
Benjamin Aminof is supported  by  the  Austrian  National  Research Network 
S11403-N23 (RiSE) of the Austrian Science Fund (FWF)  and  by  the  Vienna  
Science  and  Technology  Fund (WWTF) through grant ICT12-059. Sasha Rubin is a 
Marie Curie fellow of the Istituto Nazionale di Alta Matematica.
Aniello  Murano  is partially supported  by  the GNCS  2016 project: Logica, 
Automi e Giochi per Sistemi Auto-adattivi.

\end{section}

% End of file Acknowledgments.tex

%
%
	\bibliographystyle{eptcs}
	\bibliography{References}

\begin{thebibliography}{10}
\providecommand{\bibitemdeclare}[2]{}
\providecommand{\surnamestart}{}
\providecommand{\surnameend}{}
\providecommand{\urlprefix}{Available at }
\providecommand{\url}[1]{\texttt{#1}}
\providecommand{\href}[2]{\texttt{#2}}
\providecommand{\urlalt}[2]{\href{#1}{#2}}
\providecommand{\doi}[1]{doi:\urlalt{http://dx.doi.org/#1}{#1}}
\providecommand{\bibinfo}[2]{#2}

\bibitemdeclare{inproceedings}{AGJ07}
\bibitem{AGJ07}
\bibinfo{author}{T.~\surnamestart {\AA}gotnes\surnameend},
  \bibinfo{author}{V.~\surnamestart Goranko\surnameend} \&
  \bibinfo{author}{W.~\surnamestart Jamroga\surnameend} (\bibinfo{year}{2007}):
  \emph{\bibinfo{title}{Alternating-time temporal logics with irrevocable
  strategies}}.
\newblock In: {\sl \bibinfo{booktitle}{TARK-2007}}, pp.
  \bibinfo{pages}{15--24}, \doi{10.1145/1324249.1324256}.

\bibitemdeclare{article}{AKH02}
\bibitem{AKH02}
\bibinfo{author}{E.~\surnamestart Altman\surnameend},
  \bibinfo{author}{H.~\surnamestart Kameda\surnameend} \&
  \bibinfo{author}{Y.~\surnamestart Hosokawa\surnameend}
  (\bibinfo{year}{2002}): \emph{\bibinfo{title}{Nash Equilibria in Load
  Balancing in Distributed Computer Systems}}.
\newblock {\sl \bibinfo{journal}{IGTR}}
  \bibinfo{volume}{4}(\bibinfo{number}{2}), pp. \bibinfo{pages}{91--100},
  \doi{10.1142/S0219198902000574}.

\bibitemdeclare{article}{AHK02}
\bibitem{AHK02}
\bibinfo{author}{R.~\surnamestart Alur\surnameend}, \bibinfo{author}{T.A.
  \surnamestart Henzinger\surnameend} \& \bibinfo{author}{O.~\surnamestart
  Kupferman\surnameend} (\bibinfo{year}{2002}):
  \emph{\bibinfo{title}{{Alternating-Time Temporal Logic.}}}
\newblock {\sl \bibinfo{journal}{JACM}}
  \bibinfo{volume}{49}(\bibinfo{number}{5}), pp. \bibinfo{pages}{672--713},
  \doi{10.1145/585265.585270}.

\bibitemdeclare{inproceedings}{AMMR16}
\bibitem{AMMR16}
\bibinfo{author}{B.~\surnamestart Aminof\surnameend},
  \bibinfo{author}{V.~\surnamestart Malvone\surnameend},
  \bibinfo{author}{A.~\surnamestart Murano\surnameend} \&
  \bibinfo{author}{S.~\surnamestart Rubin\surnameend} (\bibinfo{year}{2016}):
  \emph{\bibinfo{title}{Graded Strategy Logic: Reasoning about Uniqueness of
  Nash Equilibria}}.
\newblock In: {\sl \bibinfo{booktitle}{{AAMAS 2016}}},
  \bibinfo{publisher}{{IFAAMAS}}, pp. \bibinfo{pages}{698--706}.

\bibitemdeclare{inproceedings}{AMR15}
\bibitem{AMR15}
\bibinfo{author}{B.~\surnamestart Aminof\surnameend},
  \bibinfo{author}{A.~\surnamestart Murano\surnameend} \&
  \bibinfo{author}{S.~\surnamestart Rubin\surnameend} (\bibinfo{year}{2015}):
  \emph{\bibinfo{title}{On CTL* with Graded Path Modalities}}.
\newblock In: {\sl \bibinfo{booktitle}{LPAR-20 2016}}, pp.
  \bibinfo{pages}{281--296}, \doi{10.1007/978-3-662-48899-7\_20}.

\bibitemdeclare{article}{BaranyKR10}
\bibitem{BaranyKR10}
\bibinfo{author}{V.~\surnamestart B{\'{a}}r{\'{a}}ny\surnameend},
  \bibinfo{author}{L.~\surnamestart Kaiser\surnameend} \&
  \bibinfo{author}{A.~M. \surnamestart Rabinovich\surnameend}
  (\bibinfo{year}{2010}): \emph{\bibinfo{title}{Expressing Cardinality
  Quantifiers in Monadic Second-Order Logic over Trees}}.
\newblock {\sl \bibinfo{journal}{Fundam. Inform.}}
  \bibinfo{volume}{100}(\bibinfo{number}{1-4}), pp. \bibinfo{pages}{1--17},
  \doi{10.3233/FI-2010-260}.

\bibitemdeclare{inproceedings}{BGLS11}
\bibitem{BGLS11}
\bibinfo{author}{E.~\surnamestart B{\'{a}}rcenas\surnameend},
  \bibinfo{author}{P.~\surnamestart Genev{\`{e}}s\surnameend},
  \bibinfo{author}{N.~\surnamestart Laya{\"{\i}}da\surnameend} \&
  \bibinfo{author}{A.~\surnamestart Schmitt\surnameend} (\bibinfo{year}{2011}):
  \emph{\bibinfo{title}{Query Reasoning on Trees with Types, Interleaving, and
  Counting}}.
\newblock In: {\sl \bibinfo{booktitle}{{IJCAI} 2011}}, pp.
  \bibinfo{pages}{718--723}, \doi{10.5591/978-1-57735-516-8/IJCAI11-127}.

\bibitemdeclare{article}{BL14}
\bibitem{BL14}
\bibinfo{author}{E.~\surnamestart B{\'{a}}rcenas\surnameend} \&
  \bibinfo{author}{J.~\surnamestart Lavalle\surnameend} (\bibinfo{year}{2014}):
  \emph{\bibinfo{title}{Global Numerical Constraints on Trees}}.
\newblock {\sl \bibinfo{journal}{Logical Methods in Computer Science}}
  \bibinfo{volume}{10}(\bibinfo{number}{2}), \doi{10.2168/LMCS-10(2:10)2014}.

\bibitemdeclare{inproceedings}{Bel15}
\bibitem{Bel15}
\bibinfo{author}{F.~\surnamestart Belardinelli\surnameend}
  (\bibinfo{year}{2015}): \emph{\bibinfo{title}{A Logic of Knowledge and
  Strategies with Imperfect Information}}.
\newblock In: {\sl \bibinfo{booktitle}{LAMAS 15}}.

\bibitemdeclare{article}{BMM12}
\bibitem{BMM12}
\bibinfo{author}{A.~\surnamestart Bianco\surnameend},
  \bibinfo{author}{F.~\surnamestart Mogavero\surnameend} \&
  \bibinfo{author}{A.~\surnamestart Murano\surnameend} (\bibinfo{year}{2012}):
  \emph{\bibinfo{title}{{Graded Computation Tree Logic.}}}
\newblock {\sl \bibinfo{journal}{TOCL}}
  \bibinfo{volume}{13}(\bibinfo{number}{3}), pp. \bibinfo{pages}{25:1--53},
  \doi{10.1145/2287718.2287725}.

\bibitemdeclare{inproceedings}{DBLP:conf/csl/Bojanczyk04}
\bibitem{DBLP:conf/csl/Bojanczyk04}
\bibinfo{author}{M.~\surnamestart Boja{\'{n}}czyk\surnameend}
  (\bibinfo{year}{2004}): \emph{\bibinfo{title}{A Bounding Quantifier}}.
\newblock In: {\sl \bibinfo{booktitle}{{CSL 2004}}}, {\sl \bibinfo{series}{LNCS
  3210}} \bibinfo{volume}{3210}, \bibinfo{publisher}{Springer}, pp.
  \bibinfo{pages}{41--55}, \doi{10.1007/978-3-540-30124-0\_7}.

\bibitemdeclare{inproceedings}{bojanczyk_et_al:LIPIcs:2012:3427}
\bibitem{bojanczyk_et_al:LIPIcs:2012:3427}
\bibinfo{author}{M.~\surnamestart Bojanczyk\surnameend} \&
  \bibinfo{author}{S.~\surnamestart Torunczyk\surnameend}
  (\bibinfo{year}{2012}): \emph{\bibinfo{title}{{Weak MSO+U over infinite
  trees}}}.
\newblock In \bibinfo{editor}{Christoph \surnamestart D{\"u}rr\surnameend} \&
  \bibinfo{editor}{Thomas \surnamestart Wilke\surnameend}, editors: {\sl
  \bibinfo{booktitle}{STACS 2012}}, {\sl
  \bibinfo{series}{LIPIcs}}~\bibinfo{volume}{14}, \bibinfo{publisher}{Schloss
  Dagstuhl--Leibniz-Zentrum fuer Informatik}, pp. \bibinfo{pages}{648--660},
  \doi{10.4230/LIPIcs.STACS.2012.648}.

\bibitemdeclare{article}{BLMV08}
\bibitem{BLMV08}
\bibinfo{author}{P.A. \surnamestart Bonatti\surnameend},
  \bibinfo{author}{C.~\surnamestart Lutz\surnameend},
  \bibinfo{author}{A.~\surnamestart Murano\surnameend} \& \bibinfo{author}{M.Y.
  \surnamestart Vardi\surnameend} (\bibinfo{year}{2008}):
  \emph{\bibinfo{title}{{The Complexity of Enriched muCalculi.}}}
\newblock {\sl \bibinfo{journal}{LMCS}}
  \bibinfo{volume}{4}(\bibinfo{number}{3}), pp. \bibinfo{pages}{1--27},
  \doi{10.2168/LMCS-4(3:11)2008}.

\bibitemdeclare{inproceedings}{BLLM09}
\bibitem{BLLM09}
\bibinfo{author}{T.~\surnamestart Brihaye\surnameend},
  \bibinfo{author}{A.~Da~Costa \surnamestart Lopes\surnameend},
  \bibinfo{author}{F.~\surnamestart Laroussinie\surnameend} \&
  \bibinfo{author}{N.~\surnamestart Markey\surnameend} (\bibinfo{year}{2009}):
  \emph{\bibinfo{title}{{ATL} with Strategy Contexts and Bounded Memory}}.
\newblock In: {\sl \bibinfo{booktitle}{LFCS 2009}}, pp.
  \bibinfo{pages}{92--106}, \doi{10.1007/978-3-540-92687-0\_7}.

\bibitemdeclare{inproceedings}{CGL99}
\bibitem{CGL99}
\bibinfo{author}{D.~\surnamestart Calvanese\surnameend},
  \bibinfo{author}{G.~\surnamestart {De Giacomo}\surnameend} \&
  \bibinfo{author}{M.~\surnamestart Lenzerini\surnameend}
  (\bibinfo{year}{1999}): \emph{\bibinfo{title}{Reasoning in Expressive
  Description Logics with Fixpoints based on Automata on Infinite Trees}}.
\newblock In: {\sl \bibinfo{booktitle}{IJCAI 99}}, pp. \bibinfo{pages}{84--89}.

\bibitemdeclare{inproceedings}{CLMM14}
\bibitem{CLMM14}
\bibinfo{author}{P.~\surnamestart {\v{C}}erm{\'a}k\surnameend},
  \bibinfo{author}{A.~\surnamestart Lomuscio\surnameend},
  \bibinfo{author}{F.~\surnamestart Mogavero\surnameend} \&
  \bibinfo{author}{A.~\surnamestart Murano\surnameend} (\bibinfo{year}{2014}):
  \emph{\bibinfo{title}{{MCMAS-SLK: A Model Checker for the Verification of
  Strategy Logic Specifications.}}}
\newblock In: {\sl \bibinfo{booktitle}{CAV'14}}, \bibinfo{series}{LNCS 8559},
  \bibinfo{publisher}{Springer}, pp. \bibinfo{pages}{524--531},
  \doi{10.1007/978-3-319-08867-9\_34}.

\bibitemdeclare{inproceedings}{CLM15}
\bibitem{CLM15}
\bibinfo{author}{P.~\surnamestart Cerm{\'{a}}k\surnameend},
  \bibinfo{author}{A.~\surnamestart Lomuscio\surnameend} \&
  \bibinfo{author}{A.~\surnamestart Murano\surnameend} (\bibinfo{year}{2015}):
  \emph{\bibinfo{title}{Verifying and Synthesising Multi-Agent Systems against
  One-Goal Strategy Logic Specifications}}.
\newblock In: {\sl \bibinfo{booktitle}{AAAI 2015}}, pp.
  \bibinfo{pages}{2038--2044}.

\bibitemdeclare{article}{CHP10}
\bibitem{CHP10}
\bibinfo{author}{K.~\surnamestart Chatterjee\surnameend}, \bibinfo{author}{T.A.
  \surnamestart Henzinger\surnameend} \& \bibinfo{author}{N.~\surnamestart
  Piterman\surnameend} (\bibinfo{year}{2010}): \emph{\bibinfo{title}{{Strategy
  Logic.}}}
\newblock {\sl \bibinfo{journal}{IC}}
  \bibinfo{volume}{208}(\bibinfo{number}{6}), pp. \bibinfo{pages}{677--693},
  \doi{10.1016/j.ic.2009.07.004}.

\bibitemdeclare{article}{CHS99}
\bibitem{CHS99}
\bibinfo{author}{R.~\surnamestart Cornes\surnameend},
  \bibinfo{author}{R.~\surnamestart Hartley\surnameend} \&
  \bibinfo{author}{T.~\surnamestart Sandler\surnameend} (\bibinfo{year}{1999}):
  \emph{\bibinfo{title}{An Elementary Proof via Contraction}}.
\newblock {\sl \bibinfo{journal}{Journal of Public Economic Theory}}
  \bibinfo{volume}{1}(\bibinfo{number}{4}), pp. \bibinfo{pages}{499--509},
  \doi{10.1111/1097-3923.00023}.

\bibitemdeclare{book}{SCB13}
\bibitem{SCB13}
\bibinfo{author}{J.~Bramel \surnamestart D.~Simchi-Levi\surnameend, X.~Chen}
  (\bibinfo{year}{2013}): \emph{\bibinfo{title}{The Logic of Logistics: Theory,
  Algorithms, and Applications for Logistics Management}}.
\newblock \bibinfo{series}{Science and Business Media},
  \bibinfo{publisher}{Springer}.

\bibitemdeclare{inproceedings}{DL03}
\bibitem{DL03}
\bibinfo{author}{S.~\surnamestart Dal{-}Zilio\surnameend} \&
  \bibinfo{author}{D.~\surnamestart Lugiez\surnameend} (\bibinfo{year}{2003}):
  \emph{\bibinfo{title}{{XML} Schema, Tree Logic and Sheaves Automata}}.
\newblock In: {\sl \bibinfo{booktitle}{RTA 2003}}, pp.
  \bibinfo{pages}{246--263}, \doi{10.1007/3-540-44881-0\_18}.

\bibitemdeclare{article}{DL10}
\bibitem{DL10}
\bibinfo{author}{S.~\surnamestart Demri\surnameend} \&
  \bibinfo{author}{D.~\surnamestart Lugiez\surnameend} (\bibinfo{year}{2010}):
  \emph{\bibinfo{title}{Complexity of modal logics with Presburger
  constraints}}.
\newblock {\sl \bibinfo{journal}{J. Applied Logic}}
  \bibinfo{volume}{8}(\bibinfo{number}{3}), pp. \bibinfo{pages}{233--252},
  \doi{10.1016/j.jal.2010.03.001}.

\bibitemdeclare{inproceedings}{EJ91}
\bibitem{EJ91}
\bibinfo{author}{E.~A. \surnamestart Emerson\surnameend} \&
  \bibinfo{author}{C.~S. \surnamestart Jutla\surnameend}
  (\bibinfo{year}{1991}): \emph{\bibinfo{title}{Tree Automata, Mu-Calculus and
  Determinacy (Extended Abstract)}}.
\newblock In: {\sl \bibinfo{booktitle}{ASFCS 1991}}, pp.
  \bibinfo{pages}{368--377}, \doi{10.1109/SFCS.1991.185392}.

\bibitemdeclare{article}{FMP08}
\bibitem{FMP08}
\bibinfo{author}{A.~\surnamestart Ferrante\surnameend},
  \bibinfo{author}{A.~\surnamestart Murano\surnameend} \&
  \bibinfo{author}{M.~\surnamestart Parente\surnameend} (\bibinfo{year}{2008}):
  \emph{\bibinfo{title}{{Enriched Mu-Calculi Module Checking.}}}
\newblock {\sl \bibinfo{journal}{LMCS}}
  \bibinfo{volume}{4}(\bibinfo{number}{3}), pp. \bibinfo{pages}{1--21},
  \doi{10.2168/LMCS-4(3:1)2008}.

\bibitemdeclare{article}{Gradel99}
\bibitem{Gradel99}
\bibinfo{author}{Erich \surnamestart Gr{\"{a}}del\surnameend}
  (\bibinfo{year}{1999}): \emph{\bibinfo{title}{On The Restraining Power of
  Guards}}.
\newblock {\sl \bibinfo{journal}{J. Symb. Log.}}
  \bibinfo{volume}{64}(\bibinfo{number}{4}), pp. \bibinfo{pages}{1719--1742},
  \doi{10.2307/2586808}.

\bibitemdeclare{inproceedings}{GHW14}
\bibitem{GHW14}
\bibinfo{author}{J.~\surnamestart Gutierrez\surnameend},
  \bibinfo{author}{P.~\surnamestart Harrenstein\surnameend} \&
  \bibinfo{author}{M.~\surnamestart Wooldridge\surnameend}
  (\bibinfo{year}{2014}): \emph{\bibinfo{title}{Reasoning about Equilibria in
  Game-Like Concurrent Systems}}.
\newblock In: {\sl \bibinfo{booktitle}{KR 2014}}, \bibinfo{publisher}{AAAI}.

\bibitemdeclare{inproceedings}{HLW13}
\bibitem{HLW13}
\bibinfo{author}{A.~\surnamestart Herzig\surnameend},
  \bibinfo{author}{E.~\surnamestart Lorini\surnameend} \&
  \bibinfo{author}{D.~\surnamestart Walther\surnameend} (\bibinfo{year}{2013}):
  \emph{\bibinfo{title}{Reasoning about Actions Meets Strategic Logics}}.
\newblock In: {\sl \bibinfo{booktitle}{LORI 2013}}, pp.
  \bibinfo{pages}{162--175}, \doi{10.1007/978-3-642-40948-6\_13}.

\bibitemdeclare{inproceedings}{HJW05}
\bibitem{HJW05}
\bibinfo{author}{W.~\surnamestart van~der Hoek\surnameend},
  \bibinfo{author}{W.~\surnamestart Jamroga\surnameend} \&
  \bibinfo{author}{M.~\surnamestart Wooldridge\surnameend}
  (\bibinfo{year}{2005}): \emph{\bibinfo{title}{A logic for strategic
  reasoning}}.
\newblock In: {\sl \bibinfo{booktitle}{AAMAS 2005}}, pp.
  \bibinfo{pages}{157--164}, \doi{10.1145/1082473.1082497}.

\bibitemdeclare{article}{KPV16}
\bibitem{KPV16}
\bibinfo{author}{O.~\surnamestart Kupferman\surnameend},
  \bibinfo{author}{G.~\surnamestart Perelli\surnameend} \&
  \bibinfo{author}{M.~\surnamestart Vardi\surnameend} (\bibinfo{year}{2016}):
  \emph{\bibinfo{title}{Synthesis with rational environments}}.
\newblock {\sl \bibinfo{journal}{Annals of Mathematics and Artificial
  Intelligence}}, pp. \bibinfo{pages}{1--18}, \doi{10.1007/s10472-016-9508-8}.

\bibitemdeclare{inproceedings}{KPV14}
\bibitem{KPV14}
\bibinfo{author}{O.~\surnamestart Kupferman\surnameend},
  \bibinfo{author}{G.~\surnamestart Perelli\surnameend} \&
  \bibinfo{author}{M.~Y. \surnamestart Vardi\surnameend}
  (\bibinfo{year}{2014}): \emph{\bibinfo{title}{Synthesis with Rational
  Environments}}.
\newblock In: {\sl \bibinfo{booktitle}{EUMAS 2014}}, pp.
  \bibinfo{pages}{219--235}, \doi{10.1007/978-3-319-17130-2\_15}.

\bibitemdeclare{inproceedings}{KSV02}
\bibitem{KSV02}
\bibinfo{author}{O.~\surnamestart Kupferman\surnameend},
  \bibinfo{author}{U.~\surnamestart Sattler\surnameend} \&
  \bibinfo{author}{M.Y. \surnamestart Vardi\surnameend} (\bibinfo{year}{2002}):
  \emph{\bibinfo{title}{{The Complexity of the Graded muCalculus.}}}
\newblock In: {\sl \bibinfo{booktitle}{CADE'02}}, \bibinfo{series}{LNCS 2392},
  \bibinfo{publisher}{Springer}, pp. \bibinfo{pages}{423--437},
  \doi{10.1007/3-540-45620-1\_34}.

\bibitemdeclare{article}{KVW00}
\bibitem{KVW00}
\bibinfo{author}{O.~\surnamestart Kupferman\surnameend}, \bibinfo{author}{M.Y.
  \surnamestart Vardi\surnameend} \& \bibinfo{author}{P.~\surnamestart
  Wolper\surnameend} (\bibinfo{year}{2000}): \emph{\bibinfo{title}{{An Automata
  Theoretic Approach to Branching-Time Model Checking.}}}
\newblock {\sl \bibinfo{journal}{JACM}}
  \bibinfo{volume}{47}(\bibinfo{number}{2}), pp. \bibinfo{pages}{312--360},
  \doi{10.1145/333979.333987}.

\bibitemdeclare{book}{LB08}
\bibitem{LB08}
\bibinfo{author}{K.~\surnamestart Leyton-Brown\surnameend} \&
  \bibinfo{author}{Y.~\surnamestart Shoham\surnameend} (\bibinfo{year}{2008}):
  \emph{\bibinfo{title}{Essentials of Game Theory: A Concise, Multidisciplinary
  Introduction (Synthesis Lectures on Artificial Intelligence and Machine
  Learning)}}.
\newblock \bibinfo{publisher}{M\&C}, \doi{10.2200/S00108ED1V01Y200802AIM003}.

\bibitemdeclare{inproceedings}{LLM10}
\bibitem{LLM10}
\bibinfo{author}{A.~Da~Costa \surnamestart Lopes\surnameend},
  \bibinfo{author}{F.~\surnamestart Laroussinie\surnameend} \&
  \bibinfo{author}{N.~\surnamestart Markey\surnameend} (\bibinfo{year}{2010}):
  \emph{\bibinfo{title}{{ATL} with Strategy Contexts: Expressiveness and Model
  Checking}}.
\newblock In: {\sl \bibinfo{booktitle}{FSTTCS 2010}}, pp.
  \bibinfo{pages}{120--132}, \doi{10.4230/LIPIcs.FSTTCS.2010.120}.

\bibitemdeclare{inproceedings}{MMMS15}
\bibitem{MMMS15}
\bibinfo{author}{V.~\surnamestart Malvone\surnameend},
  \bibinfo{author}{F.~\surnamestart Mogavero\surnameend},
  \bibinfo{author}{A.~\surnamestart Murano\surnameend} \&
  \bibinfo{author}{L.~\surnamestart Sorrentino\surnameend}
  (\bibinfo{year}{2015}): \emph{\bibinfo{title}{On the Counting of
  Strategies}}.
\newblock In: {\sl \bibinfo{booktitle}{TIME 2015}}, pp.
  \bibinfo{pages}{170--179}, \doi{10.1109/TIME.2015.19}.

\bibitemdeclare{inproceedings}{MMS15}
\bibitem{MMS15}
\bibinfo{author}{V.~\surnamestart Malvone\surnameend},
  \bibinfo{author}{A.~\surnamestart Murano\surnameend} \&
  \bibinfo{author}{L.~\surnamestart Sorrentino\surnameend}:
  \emph{\bibinfo{title}{Games with additional winning strategies}}.
\newblock In: {\sl \bibinfo{booktitle}{CILC, 2015}}, \bibinfo{publisher}{CEUR}.

\bibitemdeclare{article}{MMPV14}
\bibitem{MMPV14}
\bibinfo{author}{F.~\surnamestart Mogavero\surnameend},
  \bibinfo{author}{A.~\surnamestart Murano\surnameend},
  \bibinfo{author}{G.~\surnamestart Perelli\surnameend} \&
  \bibinfo{author}{M.Y. \surnamestart Vardi\surnameend} (\bibinfo{year}{2014}):
  \emph{\bibinfo{title}{{Reasoning About Strategies: On the Model-Checking
  Problem.}}}
\newblock {\sl \bibinfo{journal}{TOCL}}
  \bibinfo{volume}{15}(\bibinfo{number}{4}), pp. \bibinfo{pages}{34:1--42},
  \doi{10.1145/2631917}.

\bibitemdeclare{inproceedings}{MMV10b}
\bibitem{MMV10b}
\bibinfo{author}{F.~\surnamestart Mogavero\surnameend},
  \bibinfo{author}{A.~\surnamestart Murano\surnameend} \& \bibinfo{author}{M.Y.
  \surnamestart Vardi\surnameend} (\bibinfo{year}{2010}):
  \emph{\bibinfo{title}{{Reasoning About Strategies.}}}
\newblock In: {\sl \bibinfo{booktitle}{FSTTCS'10}}, \bibinfo{series}{LIPIcs 8},
  \bibinfo{publisher}{Leibniz-Zentrum fuer Informatik}, pp.
  \bibinfo{pages}{133--144}, \doi{10.4230/LIPIcs.FSTTCS.2010.133}.

\bibitemdeclare{article}{MS95}
\bibitem{MS95}
\bibinfo{author}{D.~E. \surnamestart Muller\surnameend} \&
  \bibinfo{author}{P.~E. \surnamestart Schupp\surnameend}
  (\bibinfo{year}{1995}): \emph{\bibinfo{title}{Simulating Alternating Tree
  Automata by Nondeterministic Automata: New Results and New Proofs of the
  Theorems of Rabin, McNaughton and Safra}}.
\newblock {\sl \bibinfo{journal}{Theor. Comput. Sci.}}
  \bibinfo{volume}{141}(\bibinfo{number}{1{\&}2}), pp.
  \bibinfo{pages}{69--107}, \doi{10.1016/0304-3975(94)00214-4}.

\bibitemdeclare{article}{ORS93a}
\bibitem{ORS93a}
\bibinfo{author}{A.~\surnamestart Orda\surnameend},
  \bibinfo{author}{R.~\surnamestart Rom\surnameend} \&
  \bibinfo{author}{N.~\surnamestart Shimkin\surnameend} (\bibinfo{year}{1993}):
  \emph{\bibinfo{title}{Competitive routing in multiuser communication
  networks}}.
\newblock {\sl \bibinfo{journal}{IEEE/ACM Trans. Netw.}}
  \bibinfo{volume}{1}(\bibinfo{number}{5}), pp. \bibinfo{pages}{510--521},
  \doi{10.1109/90.251910}.

\bibitemdeclare{article}{PC79}
\bibitem{PC79}
\bibinfo{author}{G.P. \surnamestart Papavassilopoulos\surnameend} \&
  \bibinfo{author}{J.~B. \surnamestart Cruz\surnameend} (\bibinfo{year}{1979}):
  \emph{\bibinfo{title}{On the Uniqueness of Nash Strategies for a Class of
  Analytic Differential Games}}.
\newblock {\sl \bibinfo{journal}{Journal of Optimization Theory and
  Applications}} \bibinfo{volume}{27}(\bibinfo{number}{2}), pp.
  \bibinfo{pages}{309--314}, \doi{10.1007/BF00933234}.

\bibitemdeclare{book}{Pavel12}
\bibitem{Pavel12}
\bibinfo{author}{L.~\surnamestart Pavel\surnameend} (\bibinfo{year}{2012}):
  \emph{\bibinfo{title}{Game Theory for Control of Optical Networks}}.
\newblock \bibinfo{series}{Science and Business Media},
  \bibinfo{publisher}{Springer}, \doi{10.1007/978-0-8176-8322-1}.

\bibitemdeclare{inproceedings}{Pnu77}
\bibitem{Pnu77}
\bibinfo{author}{A.~\surnamestart Pnueli\surnameend} (\bibinfo{year}{1977}):
  \emph{\bibinfo{title}{{The Temporal Logic of Programs.}}}
\newblock In: {\sl \bibinfo{booktitle}{FOCS'77}}, \bibinfo{publisher}{IEEE
  Computer Society}, pp. \bibinfo{pages}{46--57}, \doi{10.1109/SFCS.1977.32}.

\bibitemdeclare{article}{DBLP:journals/bsl/Rubin08}
\bibitem{DBLP:journals/bsl/Rubin08}
\bibinfo{author}{S.~\surnamestart Rubin\surnameend} (\bibinfo{year}{2008}):
  \emph{\bibinfo{title}{Automata Presenting Structures: {A} Survey of the
  Finite String Case}}.
\newblock {\sl \bibinfo{journal}{Bulletin of Symbolic Logic}}
  \bibinfo{volume}{14}(\bibinfo{number}{2}), pp. \bibinfo{pages}{169--209},
  \doi{10.2178/bsl/1208442827}.

\bibitemdeclare{inproceedings}{SSM08}
\bibitem{SSM08}
\bibinfo{author}{H.~\surnamestart Seidl\surnameend},
  \bibinfo{author}{T.~\surnamestart Schwentick\surnameend} \&
  \bibinfo{author}{A.~\surnamestart Muscholl\surnameend}
  (\bibinfo{year}{2008}): \emph{\bibinfo{title}{Counting in trees}}.
\newblock In: {\sl \bibinfo{booktitle}{Logic and Automata: History and
  Perspectives}}, pp. \bibinfo{pages}{575--612}.

\bibitemdeclare{article}{Sel65}
\bibitem{Sel65}
\bibinfo{author}{R.~\surnamestart Selten\surnameend} (\bibinfo{year}{1965}):
  \emph{\bibinfo{title}{Spieltheoretische Behandlung eines Oligopolmodells mit
  Nachfragetragheit.}}
\newblock {\sl \bibinfo{journal}{Zeitschrift fur die gesamte
  Staatswissenschaft}} \bibinfo{volume}{121}, pp. \bibinfo{pages}{301--324}.

\bibitemdeclare{article}{thomas1990automata}
\bibitem{thomas1990automata}
\bibinfo{author}{W.~\surnamestart Thomas\surnameend} (\bibinfo{year}{1990}):
  \emph{\bibinfo{title}{Automata on infinite objects}}.
\newblock {\sl \bibinfo{journal}{Handbook of theoretical computer science,
  Volume B}}, pp. \bibinfo{pages}{133--191}.

\bibitemdeclare{inproceedings}{Ummels06}
\bibitem{Ummels06}
\bibinfo{author}{M.~\surnamestart Ummels\surnameend} (\bibinfo{year}{2006}):
  \emph{\bibinfo{title}{Rational Behaviour and Strategy Construction in
  Infinite Multiplayer Games}}.
\newblock In: {\sl \bibinfo{booktitle}{FSTTCS}}, pp. \bibinfo{pages}{212--223},
  \doi{10.1007/11944836\_21}.

\bibitemdeclare{inproceedings}{WHW07}
\bibitem{WHW07}
\bibinfo{author}{D.~\surnamestart Walther\surnameend},
  \bibinfo{author}{W.~\surnamestart van~der Hoek\surnameend} \&
  \bibinfo{author}{M.~\surnamestart Wooldridge\surnameend}
  (\bibinfo{year}{2007}): \emph{\bibinfo{title}{Alternating-time temporal logic
  with explicit strategies}}.
\newblock In: {\sl \bibinfo{booktitle}{TARK 2007}}, pp.
  \bibinfo{pages}{269--278}, \doi{10.1145/1324249.1324285}.

\bibitemdeclare{inproceedings}{WHY11}
\bibitem{WHY11}
\bibinfo{author}{F.~\surnamestart Wang\surnameend},
  \bibinfo{author}{C.~\surnamestart Huang\surnameend} \&
  \bibinfo{author}{F.~\surnamestart Yu\surnameend} (\bibinfo{year}{2011}):
  \emph{\bibinfo{title}{A Temporal Logic for the Interaction of Strategies}}.
\newblock In: {\sl \bibinfo{booktitle}{{CONCUR} 2011}}, \bibinfo{series}{LNCS
  6901}, \bibinfo{publisher}{Springer}, pp. \bibinfo{pages}{466--481},
  \doi{10.1007/978-3-642-23217-6\_31}.

\bibitemdeclare{book}{ZG11}
\bibitem{ZG11}
\bibinfo{author}{Y.~\surnamestart Zhang\surnameend} \&
  \bibinfo{author}{M.~\surnamestart Guizani\surnameend} (\bibinfo{year}{2011}):
  \emph{\bibinfo{title}{Game Theory for Wireless Communications and
  Networking}}.
\newblock \bibinfo{publisher}{CRC Press}.

\end{thebibliography}

\end{document}